\documentclass[aps,prd,reprint,superscriptaddress]{revtex4-2}

\usepackage{comment}
\usepackage[utf8]{inputenc} 
\usepackage{graphicx,color,overpic,mathtools}
\usepackage{amsthm,amsmath,amssymb,mathrsfs}
\usepackage[draft]{hyperref}
\usepackage{braket,bm,bbm,setspace}
\PassOptionsToPackage{normalem}{ulem}
\usepackage{ulem} 
\usepackage{physics}
\usepackage{rotating}
\usepackage{float}
\usepackage[makeroom]{cancel}
\usepackage[english]{babel}
\usepackage{graphicx}
\graphicspath{ {images/} }
\addto\captionsspanish{}
\hypersetup{
    colorlinks=false,
    pdfborder={0 0 0},
}
\definecolor{emerald}{rgb}{0.31, 0.60, 0.95}

\begin{document}

\title{Semiclassical constant-density spheres in a regularized Polyakov approximation}

\author{Julio Arrechea}
\affiliation{Instituto de Astrof\'isica de Andaluc\'ia (IAA-CSIC),
Glorieta de la Astronom\'ia, 18008 Granada, Spain}
\author{Carlos Barcel\'o} 
\affiliation{Instituto de Astrof\'isica de Andaluc\'ia (IAA-CSIC),
Glorieta de la Astronom\'ia, 18008 Granada, Spain}
\author{Ra\'ul Carballo-Rubio}
\affiliation{Florida Space Institute, University of Central Florida, 12354 Research Parkway, Partnership 1, 32826 Orlando, FL, USA}
\author{Luis J. Garay} 
\affiliation{Departamento de F\'{\i}sica Te\'orica and IPARCOS, Universidad Complutense de Madrid, 28040 Madrid, Spain}  
\affiliation{Instituto de Estructura de la Materia (IEM-CSIC), Serrano 121, 28006 Madrid, Spain}

\begin{abstract}

We provide an exhaustive analysis of the complete set of solutions of the equations of stellar equilibrium under semiclassical effects. As classical matter we use a perfect fluid of constant density; as the semiclassical source we use the renormalized stress-energy tensor (RSET) of a minimally coupled massless scalar field in the Boulware vacuum (the only vacuum consistent with asymptotic flatness and staticity). For the RSET we use a regularized version of the Polyakov approximation. We present a complete catalogue of the semiclassical self-consistent solutions which incorporates regular as well as singular solutions, showing that the semiclassical corrections are highly relevant in scenarios of high compactness. Semiclassical corrections allow the existence of ultra-compact equilibrium configurations which have bounded pressures and masses up to a central core of Planckian radius, precisely where the regularized Polyakov approximation is not accurate. Our analysis strongly suggests the absence of a Buchdahl limit in semiclasical gravity, while indicating that the regularized Polyakov approximation used here must be improved to describe equilibrium configurations of arbitrary compactness that remain regular at the center of spherical symmetry.

\end{abstract}


\maketitle

\section{Introduction}

The study of spherically symmetric models of relativistic stars with isotropic pressures is a well-know subject and has provided some of the most important insights into the nature of stellar configurations~\cite{Wheeler1955}.
These configurations are the relativistic version of the Newtonian fluid spheres in hydrostatic equilibrium. 
In the Newtonian theory, there are no bounds to how compact fluid spheres can be as, given a regular density profile, there is always a regular pressure profile able to withstand the gravitational pull.
However, in general relativity, where pressure acts itself as a source of spacetime curvature, gravitational collapse is unavoidable for bodies that surpass the Buchdahl limit \cite{Buchdahl1959}. Buchdahl's theorem states that, for fluid spheres satisfying reasonable regularity conditions, the compactness $C$ (the quotient between twice their Misner-Sharp mass at the surface and their radius) must be smaller than $8/9$ in geometrical units. The constant-density relativistic star, derived by Schwarzschild in 1916 \cite{Schwarzschild1916}, saturates this limit, being the stellar configuration that has the smallest central pressure for a given compactness. Therefore, standard classical general relativity predicts that black holes should be formed at some point as a matter of principle.

The idea that behind the astrophysical black-hole-like objects there are indeed entities with a structure very close to that of relativistic black holes is supported by a mixture of theoretical and observational arguments. On the one hand, there are constraints to the compactness and brightness of these objects~\cite{Eckart2017,Cardoso2019}; on the other hand, the black holes predicted by general relativity are arguably the simplest and better motivated model consistent with these observations. Nonetheless, the new observational capacity into astrophysical black holes (mainly, gravitational waves and Event Horizon Telescope observations), together with some somewhat stalled theoretical tensions when extending the classical model of black holes into the quantum regime (think e.g. on the information loss problem~\cite{Hawking1976}) motivates a renewed interest in analyzing alternatives to black holes. The search for alternatives to black holes comprises an active research field, with proposals that vary from exotic equilibrium configurations still below the Buchdahl limit (e.g. boson stars \cite{Colpi1986}) to ultracompact objects whose surface lies extremely close to their gravitational radius. Known proposals typically involve exotic effective matter contents, such as anisotropic fluids \cite{Raposo2018}, quark stars \cite{Freedman1978}, or gravitational-vacuum-condensate stars \cite{Mazur2015, Cattoen2005} (see \cite{Cardoso2019,Olmo2019} for a more detailed list of proposals). Those that allude to semiclassical or quantum theories of gravity \cite{Barcelo2008,Mathur2005} typically modify the standard black hole picture only from a radius extremely close to the gravitational radius inwards. In the next years, electromagnetic and gravitational observations will improve the testability of these models by putting constraints to the various observational parameters that identify these alternatives \cite{Carballo-Rubio2018}. 

This work is motivated by the search of ultracompact objects within the realm of semiclassical gravity \cite{Barcelo2009, Visser2009,Carballo-Rubio2018a}. 
These hypothetical ultracompact and horizonless equilibrium configurations (so-called black stars) would be supported by vacuum polarization, i.e. the contribution of the vacuum energy of quantum fields to spacetime curvature, in the form of a renormalized stress-energy tensor (RSET). It is the subject of semiclassical gravity to account for the way spacetime responds to such effects. 
While in flat spacetime the contribution of zero-point energies of fields to curvature can be fully subtracted, in curved spacetimes it must undergo a covariant renormalization procedure \cite{FullingDavies1977, BirrellDavies1982}, rendering a finite contribution. The resulting object is the RSET, whose expectation value  $\langle\hat{T}_{\mu\nu}\rangle$ in some vacuum state enters the right-hand side of the semiclassical field equations,
\begin{equation}\label{eq:semieinstein}
G_{\mu\nu}=8\pi \left(T_{\mu\nu}+\hbar \langle\hat{T}_{\mu\nu}\rangle\right).
\end{equation}
Here, Greek indices take $4$ spacetime values and we have chosen units $c=G=1$. The RSET is a function of the components of the metric and their derivatives and, as such, responds to the geometry of spacetime and backreacts on it.

The contribution of vacuum energy to curvature, being proportional to $\hbar$, is negligible in most astrophysical scenarios. However, vacuum polarization becomes relevant in the presence of high spacetime curvatures \cite{Brout1995,Hiscock1997,Chakraborty2015}
(close to spacetime singularities), in the early universe \cite{Fischetti1979,Grib1981,Simon1992,Parker1993}, and in the vicinity of horizons \cite{Banerjee2009,Barcelo2019}. The RSET can violate the (pointwise) energy conditions \cite{Visser1996,Barcelo2002,Kontou2020} being able to, tentatively, serve as a source of quantum repulsion on matter and so allowing for equilibrium in situations forbidden in the classical theory \cite{Carballo-Rubio2018a}. In fact, it is known that a configuration with its surface hovering just above its gravitational radius would experience important semiclassical deviations from its classical dynamics \cite{Barcelo2008,Barcelo2015,Carballo-Rubio2018a,Harada2018,Barcelo2019}, opening the possibility of reaching stability.

This paper belongs to a series of investigations put forward to analyze systematically whether qualitatively new equilibrium configurations are naturally possible within semiclassical gravity. For reasons explained in the next subsection, our investigations have started using as quantum field a single minimally-coupled massless scalar field whose RSET is a regularized version of the Polyakov approximation. Before turning to more refined analyses, we decided to exhaust this framework to clearly see its scope and limitations. In a previous paper we analyzed the form of the self-consistent vacuum solutions of semiclassical gravity~\cite{Arrechea2020}. Here, we add a classical perfect fluid of constant energy density to the semiclassical vacuum. This matter content is specially interesting for its simplicity, but moreover because it leaves the pressure term free to evolve in the precise form needed to attain equilibrium. During these analyses we have realized the importance of understanding not just the regular solutions to the gravitational equations, but also the different non-regular solutions that appear. This paper specifically focuses in presenting the complete set of self-consistent solutions of the semiclassical equation \eqref{eq:semieinstein} for constant-density spheres, both regular and non-regular, and comparing them with the equivalent set of classical solutions.

We aim to determine the spacetime geometry and the semiclassical sources simultaneously. Given that the solutions cannot be found in closed form, we use analytical approximations and various numerical explorations to describe them. This combination allows us to present an adequate characterization of these solutions.

\subsection{The RSET and its usage in stellar physics}
\label{Sec:RSET}

Computing an exact expression for the RSET is far from straightforward. In conformally flat spacetimes with conformally invariant fields, this tensor is determined solely by the local geometry \cite{Frolov1987,Parker1993}, up to a collection of free parameters that depend on the fields under consideration.
In general static and spherically symmetric spacetimes, however, the exact RSET can only be computed numerically for the scalar  \cite{Anderson1995} and spin-$1/2$ \cite{Groves2002} fields, which hinders the task of finding solutions to \eqref{eq:semieinstein}. Despite technical difficulties, some numerical self-consistent solutions have been found \cite{Hochberg1997} using the analytical approximation to the exact RSET of a scalar field of Anderson et al. \cite{Anderson1995,Popov2003}. Introducing an elaborate form of the RSET allows to capture more of the physics of the system at the price of calculations becoming more intricate. Moreover, these involved expressions often carry along an increase in the degrees of freedom of the equations of motion, triggering the appearance of spurious solutions \cite{Simon1990,Parker1993}. 

As a consequence, it is standard to appeal to convenient analytical approximations where closed expressions of the RSET can be provided in particularly simple scenarios. This is the case of the Polyakov RSET \cite{Polyakov1981}, obtained through dimensional reduction to a $1+1$ manifold, and taking advantage of the conformal invariance of the reduced set of equations. After a point-splitting renormalization \cite{FullingDavies1977}, the resulting 2-dimensional RSET is then converted to a 4-dimensional quantity via a dimensional transformation that renders an independently conserved RSET in 4 dimensions. This approach only considers spherically symmetric fluctuations and ignores backscattering of the wave modes. However, it suffices to capture some of the most prominent features of vacuum states and has been used in numerous works (see e.g. \cite{Parentani1994,Ayal1997,Fabbri2005,Chakraborty2015,Ho2017,Carballo-Rubio2018a}). Coming from a calculation in a dimensionally reduced spacetime, the Polyakov RSET lacks knowledge about the $r=0$ point, and is indeed singular there.  Before attempting to construct regular, self-consistent solutions, this singularity has to be dealt with.

In a previous work \cite{Arrechea2020} we presented a regularization scheme for the Polyakov RSET. Following \cite{Parentani1994,Ayal1997} we constructed a Regularized Polyakov RSET devoid of the $r=0$ singularity, thus allowing for a self-consistent treatment of the field equations in vacuum. While sufficient for the analysis of the vacuum field equations, in this paper we will highlight some limitations of this regularization that motivate its generalization to adequately describe stellar configurations. Previous works \cite{Fabbri2006,Berthiere2017,Ho2018} already realized that static spacetimes in the Boulware vacuum state do not admit non-extremal horizons, and instead have them replaced by a wormhole neck that connects the asymptotically flat region to an internal null singularity. In \cite{Arrechea2020}, we obtained the complete set of solutions (which we refer to as semiclassical counterparts) to the spherically symmetric static vacuum field equations and proved that the set of solutions broadens when the Regularized Polyakov RSET is considered, admitting solutions with arbitrarily small wormhole necks. 
The counterparts that we obtained displace the horizon of a classical black hole to an asymptotic region ($r \to +\infty$) inside the wormhole neck, but at a finite proper distance from it. Moreover, this asymptotic horizon becomes singular (curvature scalars diverge there). When considering the RSET in the $s$-wave approximation \cite{Fabbri2006} this asymptotic null singularity becomes a proper naked singularity (i.e. it becomes timelike and uncovered by any horizon). This finding led us to conclude that the static vacuum solutions of semiclassical gravity are far from representing reasonable astrophysical objects by themselves (in this respect the situation is quite different than in classical general relativity; for a discussion of this issue see~\cite{Ensayo2021}). To obtain reasonable stellar configurations it is compulsory to add some classical matter component to the gravitational sources, and this is what we do in this paper.

We have also analyzed the nature of self-consistent solutions in the case of extremal horizons~\cite{Arrechea2021}. Given a background configuration with an extremal horizon one can show that the RSET associated with the Boulware vacuum diverges there \cite{Balbinot2007}. We have shown~\cite{Arrechea2021} that the self-consistent semiclassical solution makes this horizon singular (it accommodates a non-scalar curvature singularity).   
All in all we have accumulated strong evidence that semiclassical gravity does not allow for regular and asymptotically-flat geometries with static horizons. Then, the collapse of a star can either follow the standard Hawking evaporation paradigm (with non-static horizons and potential loss of information) or find a way to settle to an ultracompact static-equilibrium configuration without horizons~\cite{Ensayo2021}, precisely the configurations that are the subject of this work. 

Concerning semiclassical corrections to the Schwarzschild stellar interior solution, the amount of works is somewhat scarce. There exist calculations \cite{Hiscock1988} based on the Page-Brown-Ottewill approximation \cite{Page1982,Brown1985} that rely on the conformal invariance of the classical Schwarzschild stellar interior solution. Here, local semiclassical contributions amount to a perturbative correction over the classical spacetime. As the compactness of solutions approaches the Buchdahl limit, the RSET of a scalar field acquires negative energy densities at $r=0$. More recently, computations involving a non-local approximation to the RSET \cite{Satz2004} invalidate some of the points made in \cite{Hiscock1988} for the case of Newtonian stars, where non-local contributions are shown to dominate over local ones both inside and outside the stellar structure. Applications of the exact RSET (for fields of spin $0, 1/2$ and $1$) have been limited to, as far as we know, computing first-order corrections over the fixed Schwarzschild and Reissner-Nördstrom background spacetimes \cite{Jensen1989,Jensen1991, Anderson1995,Hiscock1997,Carlson2003}.  No calculation of the exact RSET exists for the matter region of stellar spacetimes, nor have backreaction effects been analyzed self-consistently. In the context of perturbation theory, semiclassical and effective quantum-gravitational corrections to the Schwarzschild stellar interior solution have been considered as well \cite{Campanelli1996,Calmet2019}. Recent works have addressed semiclassical corrections to stars with linear equations of state using the Polyakov approximation for the RSET \cite{Volkmer2019,Prasetyo2021}. 

A good candidate for backreaction studies would be the $s$-wave approximation used in \cite{Fabbri2006}, which is more refined than the Polyakov approximation as it does not neglects backscattering of the wave modes. The expressions found in \cite{Fabbri2005} are constructed following a mode decomposition which is consistent with wormhole and black-hole configurations. However, this decomposition is not adequate to analyze stellar-like configurations, where $r=0$ becomes a regular point of the spacetime. This is another reason why in this paper we stick to the Regularized Polyakov RSET (RP-RSET). Our approach is heuristic, aiming at understanding the limits of using the Polyakov approximation and its different well-motivated extensions.

Equipped with the Regularized Polyakov RSET, our aim is to obtain the complete set of static, spherically symmetric solutions to Einstein equations with a perfect fluid of constant density in the semiclassical theory (the classical counterparts were found by Lemaître \cite{Lemaitre1997}, with the exception of one solution \cite{Krasinski1997}). Constant-density solutions depict inhomogeneous and isotropic cosmologies. Among all the cosmological spacetimes analyzed, we focus on stellar spacetimes: those which have a surface that connects smoothly with the Schwarzschild vacuum solution. The Schwarzschild stellar interior solution belongs to this latter family. We will obtain self-consistent solutions for sub-Buchdahl (with compactness $C<8/9$) as well as super-Buchdahl configurations ($8/9<C<1$). There will be situations in which the semiclassical solutions here obtained are non-perturbative, in the sense that they do not have a classical counterpart in the $\hbar\to0$ limit.

This work is organized as follows. We will start in the next section by presenting a structured summary of the catalogue of solutions that we have found. A table will allow a clear comparison of the classical and semiclassical situations. We will also show a second table containing pictorial examples for each of the situations. Before presenting these tables, we will introduce an important aspect of stellar equilibrium configurations that we have denoted {\em criticality}. Criticality is related to the existence of constant masses introduced by hand and will serve as a classifying criterion. After that, the following sections will provide the technical details associated with each class of solutions, both classical and semiclassical. Section \ref{sec:classical} contains a review on the classical equations of stellar structure and their constant-density solutions. These solutions are already in the literature but we describe them here for easier comparison with the semiclassical case. In addition, our presentation of this section is original as it emphasizes the interplay between regular and non-regular solutions. Later, in section \ref{section:semiclassical}, we review the construction of the Regularized Polyakov RSET and write down the self-consistent semiclassical equations. In section \ref{section:semiclassicalsols} we turn to the analysis of the solutions to the self-consistent semiclassical field equations, the core of the paper, in which the notion of criticality is more subtle than in the classical case. We have nevertheless been able to characterize completely the space of solutions, obtaining numerical solutions of particular interest as well as analytical approximate expressions in certain regimes. Section \ref{section:discussion} will provide some conclusions and discuss aspects that could be studied in future investigations.

\section{Catalogue of solutions}\label{section:classification}

This section contains a summary of all the findings of this work contextualized and compared with the classical theory. We start by introducing the necessary preliminaries to present our classification scheme.

We consider the static and spherically symmetric line element
\begin{equation}\label{eq:metricrcoord}
ds^{2}=-e^{2\phi(r)}dt^{2}+\frac{1}{1-C(r)}dr^{2}+r^{2}d\Omega^{2}.
\end{equation}
Here, $d\Omega^{2}$ is the line element of the unit 2-sphere, $e^{2\phi(r)}$ represents the redshift function of the geometry, which is related to the redshift suffered by outgoing light rays. These become unable to escape to infinity when \mbox{$\phi\to-\infty$}, so the redshift function encodes how close the geometry is to having a horizon. The other function, $C(r)$, denotes the compactness of the geometry. It is equivalent to $2m(r)/r$, where $m(r)$ is the Misner-Sharp mass of the geometry \cite{Misner1964,Hernandez1966,Hayward1994}. Compactness represents the amount of mass contained within a spherical surface of radius $r$. In the classical vacuum, the metric \eqref{eq:metricrcoord} has the particular form $e^{2\phi}=1-C=1-2M/r$, and has a horizon at $r=2M$, with $M$ being a positive constant, the ADM (Arnowitt-Deser-Misner) mass. 

When a matter fluid is introduced in the form of some stress-energy tensor (SET), the relation $e^{2\phi}=1-C$ no longer holds. In this situation, a surface of unit compactness is not necessarily associated with a vanishing redshift function. For some of the geometries that we will discuss, it will be convenient to use a different (proper) radial coordinate $l$ defined through the relation
\begin{equation}\label{eq:cambiovar}
\frac{dr}{dl}=\pm\sqrt{1-C}.
\end{equation}
The coordinate $l$ can run along the entire real line, being particularly well adapted to study wormhole spacetimes, characterized by the existence of a minimal surface, and cosmological spacetimes, which can display multiple radial origins $r=0$. The resulting line element \eqref{eq:metricrcoord} then becomes
\begin{equation}\label{eq:metriclcoord}
ds^{2}=-e^{2\phi(l)}dt^{2}+dl^{2}+r(l)^{2}d\Omega^{2}.
\end{equation}

Let us now consider two definitions that will describe part of the solutions discussed in this paper:

* \emph{Strict stellar spacetime:} A regular geometry in which matter extends from $r(l_{0})=0$, representing the center of the structure, up to a finite radius $r(l_{\rm S})=R$. The geometry for $l>l_{\rm S}$ is the asymptotically flat Schwarzschild solution for the classical field equations or its semiclassical counterpart \cite{Arrechea2020} for the semiclassical equations. At the center $l=l_{0}$ (we will set $l_{0}=0$ in the following without loss of generality), the geometry must be regular, in particular having finite curvature scalars.

Computing the Kretschmann scalar $\mathcal{K}=R_{\mu\nu\rho\lambda}R^{\mu\nu\rho\lambda}$ yields
\begin{align}\label{eq:Kretschmann}
\mathcal{K}=
\frac{4}{r^{4}}\left\{ \left[-1+\left(r'\right)^{2}\right]^{2}\right.
&
+2\left[\left(r'\phi'\right)^{2}+\left(r''\right)^{2}\right]r^{2}\nonumber\\
&
+\left.\vphantom{\left[-1+\left(r'\right)^{2}\right]^{2}}\left[\left(\phi'\right)^{2}+\phi''\right]^{2}r^{4}\right\},
\end{align}
where $'$ denotes the derivative with respect to the $l$ coordinate. From Eq. \eqref{eq:Kretschmann} it follows that the regularity of curvature invariants at $l=0$ implies, for a strict stellar spacetime in which $r(0)=0$, that the metric functions must behave as
\begin{equation}\label{eq:regcond}
e^{2\phi(l)}=
\zeta+\lambda l^{2}+\mathcal{O}\left(l^{3}\right),~
r(l)=
 l+\gamma l^{3}+\mathcal{O}\left(l^{4}\right),
\end{equation}
where $\zeta>0$, $\lambda$, and $\gamma$ are constants fixed by solving the field equations for some SET. From Eqs. \eqref{eq:cambiovar} and \eqref{eq:regcond} we realize that strict stellar spacetimes must have $C(l\to0)\to0$. We will see below that, if there is no classical matter (the only source being the semiclassical vacuum) the above conditions cannot be fulfilled for any nonzero ADM mass $M$, while setting $M=0$ recovers Minkowski spacetime.

* \emph{$\epsilon$-strict stellar spacetime:} This is a possibly irregular spacetime (e.g. with diverging curvature invariants) but such that it does not show any signs of these possible irregularities if analyzed only for radii larger than some $r_{\epsilon}=r(l_\epsilon) \ll R, r_{\epsilon}>0$. By this we specifically mean that the pressure and compactness are finite for $l>l_\epsilon$, and that
\begin{equation}\label{eq:corecond}
|C(r_\epsilon)|< 2M_{\rm P} {r_\epsilon^2 \over l_{\rm P}^3 }= 2\rho_{\rm P} r_\epsilon^2.
\end{equation}
The radius $r_{\epsilon}$ represents an internal close-to-Planckian sphere and this last condition implies that, whatever happens inside this core, its effective mass (either positive or negative) does not exceed Planckian values. By construction, all strict stellar spacetimes are \emph{$\epsilon$-strict} spacetimes for arbitrary values of $\epsilon$ down to $\epsilon=0$.

\subsection{Classical equations of stellar equilibrium}

The SET of a perfect fluid is given by
\begin{equation}\label{eq:classSET}
T_{\mu\nu}=(\rho+p)u_{\mu}u_{\nu}+pg_{\mu\nu},
\end{equation}
where $\rho$ and $p$ are the energy density and the isotropic pressure of the fluid, measured by an observer comoving with the static fluid with $4$-velocity $u^{\mu}$. Possible contributions to the curvature coming from shear stress, fluid viscosity, or heat transfer are not included in this model. 
The $tt$ and $ll$ components of the Einstein equations resulting from considering the SET \eqref{eq:classSET} are
\begin{align}\label{eq:ttclasica}
-2r''r+1-(r')^{2}=~
&
8\pi r^{2} \rho,\\
\label{eq:rrclasica}
2r r' \phi'-1+(r')^{2}=~
&
8\pi r^{2}p.
\end{align}
In addition, covariant conservation of the SET provides the continuity equation
\begin{equation}\label{eq:consclasica}
p'=-(\rho+p)\phi'.
\end{equation}
If we interpret a relativistic star as a finite potential well, the continuity equation \eqref{eq:consclasica} guarantees that any decrease of the redshift function with decreasing $l$, or deepening of the potential, is compensated by a corresponding growth in the fluid pressure.

Equations (\ref{eq:ttclasica})-(\ref{eq:consclasica}) form a closed system of differential equations as long as we supply them with an equation of state that relates pressure and density. In the present work we will consider the equation of state
\begin{equation}\label{eq:constdens}
\rho=\text{const}.
\end{equation} 
This idealized incompressible fluid is insensitive to changes in pressure \cite{Misner1974}. This equation of state both allows for a simple treatment and uncovers interesting phenomena. For instance, as the energy density is independent from pressure, it allows for a better understanding of how the fluid arranges itself towards attaining equilibrium. In addition, the density profile \eqref{eq:constdens} saturates one of the hypotheses of Buchdahl's theorem \cite{Buchdahl1959,Urbano2018} stating that energy density must be non-increasing towards the surface. With this equation of state we can see in a clear form the appearance of the Buchdahl compactness bound.

We proceed by constructing the differential equation for the pressure known as the TOV (Tolman-Oppenheimer-Volkoff) equation, obtained by replacing Eq. \eqref{eq:consclasica} in Eq. \eqref{eq:rrclasica},
\begin{equation}\label{eq:TOVclas}
p'=-\frac{(\rho+p)\left[8\pi r^{2}p+1-(r')^{2}\right]}{2rr'}.
\end{equation}
This relation guarantees that pressure decreases monotonically outwards as long as $r'(l)>0$ and the numerator remains positive. Turning points for pressure can take place only if the numerator vanishes. This can occur either because the Misner-Sharp mass is negative [which implies $r'>1$ in virtue of \eqref{eq:cambiovar}], or because pressure reaches sufficiently negative values. These situations are realizable in the uniform density case and will be explored in section~\ref{sec:classical}.

\subsection{Criticality} \label{subsec:criticality}

Now we are going to introduce the notion of criticality in the context of the classical solutions of stellar equilibrium, which will be later transported to the semiclassical solutions. In general terms, the integration of equation \eqref{eq:ttclasica} with the change of variable \eqref{eq:cambiovar} leads to
\begin{equation}\label{eq:compclas}
r'=\pm\sqrt{1-\frac{8\pi r^{2} \rho}{3}-\frac{M_{0}}{r}}.
\end{equation}
In this equation there is an integration constant $M_{0}$, first noticed by Tolman and Volkoff \cite{Tolman1939, Volkoff1939}, that accounts for a constant mass in the spacetime.

By inspection of Eq. \eqref{eq:compclas} together with condition \eqref{eq:regcond} it becomes evident that having a nonzero $M_{0}$ produces a curvature singularity at the radial origin. Let us also highlight that, by replacing Eq. \eqref{eq:compclas} inside the TOV equation \eqref{eq:TOVclas}, we observe that the latter admits a complete analytical solution only in the $M_{0}=0$ case (progress towards obtaining analytical solutions for nonzero $M_{0}$ was made by Wyman \cite{Wyman1949}), requiring numerical integration otherwise. 

Endowing the spacetime with a constant mass, generating a singularity at $r=0$, implies that the solution acquires features of vacuum geometries. These are characterized by the mass being a constant parameter of the solution and not a quantity identified with some well-defined physical source. In that sense, $M_{0}$ can be either positive or negative. A positive $M_{0}$ indicates the presence of a positive, singular mass, endowing the solution with a singular horizon at some $r(l_{\rm div})>0$ where the pressure diverges. The final configuration resembles a black hole surrounded by matter forced to maintain hydrostatic equilibrium, causing the horizon to become singular. On the other hand, a negative $M_{0}$ introduces a naked singularity in the spacetime, as in the negative-mass Schwarzschild solution. This negative mass exerts a repulsive force that, in a sense, aids the fluid towards attaining equilibrium, but at the cost of introducing a singularity at $r=0$.

Analyzing how the total ADM mass relates to the matter content of the spacetime, we can find a correspondence between three notions of mass: the ADM mass, the mass coming from the fluid energy density $\rho$, and $M_{0}$, given by
\begin{equation}\label{eq:massrelation}
    M_{\text{ADM}}=M_{\text{cloud}}+M_{0}.
\end{equation}
Here, $M_{\text{cloud}}$ equals the outcome of the integral
\begin{equation}\label{eq:massint}
M_{\text{cloud}}=\int_{0}^{R}dr\,4\pi r^{2}\rho.
\end{equation}
When the ADM mass is equal to $M_{\text{cloud}}$ we are in the critical situation. 
Consider integrating the equations of stellar equilibrium from the surface of a star of radius $R$ and total mass $M$ inwards. Since $M_{\text{cloud}}$ is related to the energy density of the sphere of fluid, the value of $\rho$ that enforces $M_{0}=0$ in \eqref{eq:massrelation} is given by 
\begin{equation}\label{eq:critdens}
\rho=\rho_{\rm{c}\text{-clas}}=\frac{3C(R)}{8\pi R^{2}}
\end{equation}
and we will refer to this particular value as the critical density of the geometry. Any deviation from the critical value $\rho=\rho_{\rm{c}\text{-clas}}$ results in a non-critical solution with a nonzero $M_{0}$ that accounts for the respective excess or defect in mass. Particularly, an under-density (sub-critical case) translates into a positive $M_{0}$ to account for the missing mass in the right hand side of \eqref{eq:massint}, while an over-density (super-critical case) is balanced by a negative $M_{0}$. 

Non-critical constant-density solutions have been sparsely noticed in the literature. These were first analyzed by Oppenheimer and Volkoff \cite{Oppenheimer1939, Volkoff1939}, while further insight was provided by Wyman \cite{Wyman1949}. Since the equation for the compactness
in the classical equations \eqref{eq:rrclasica} is readily integrable, relation \eqref{eq:critdens} alone guarantees regularity in the compactness. In the semiclassical theory, however, the equation for the compactness is inextricably linked with that of the redshift function and it is difficult to discern whether negative energies, which have the potential to tame divergences in $p$, originate from semiclassical zero-point energies or from a super-critical unbalance. When integrating the semiclassical equations from the surface of a star of radius $R$ and mass $M$ inwards, it is not directly clear which density parameter should be used for the integration. One has to (numerically) explore different values of $\rho$ and discern the precise value that separates two types of behavior. This is the reason behind the need to properly understand both critical and non-critical configurations. 

Let us adopt the following definition: 

* \emph{Critical stellar spacetime:} As we have discussed, when integrating inwards from a radius $R$, with compactness $C(R)<1$ and density $\rho$, the classical equation for the compactness exhibits a qualitative change of behavior when going from $\rho<\rho_{\rm c}$ to $\rho>\rho_{\rm c}$, where $\rho_{\rm c}$ stands for a critical value of the density. In the classical case, this follows straightforwardly from Eq.~\eqref{eq:compclas}, as the integration constant $M_0$ in the latter equation changes sign. As we will show, we find equivalent changes in behavior in the semiclassical case. We will call a configuration \emph{critical} when it is precisely the separatrix between two different behaviors of the compactness, which in the classical case corresponds to a configuration with regular compactness and $M_0=0$. However, notice that our definition of criticality does not imply regularity. On the one hand, the pressure can be divergent in some critical solution. On the other hand, as we will show in the semiclassical case using the RP-RSET, some critical solutions lack a strictly regular compactness at the radial origin. All strict stellar spacetimes are critical stellar configurations, but the converse is not true. As we will show, around critical solutions the semiclassical equations uncover new forms of \emph{$\epsilon$-strict} stellar spacetimes which are absent in classical gravity. 

Finally, notice that criticality is a common property of stellar spacetimes and not just an artifact of considering a constant-density equation of state. The observations raised here are expected to apply to a broad class of equations of state even if, for them, a relation such as Eq.~\eqref{eq:compclas}, where $M_{0}$ appears explicitly, cannot be derived.

\subsection{The catalogue of solutions} \label{subsec:catalogue}

The purpose of this subsection is two-fold. First, it is aimed to serve as a distilled guide for the content and main results of the rest of the paper; second, it is devised as a map of the classical and semiclassical sets of stellar configurations with constant density. We encourage the reader to return to this catalogue at any point for guidance while reading the rest of the paper.

\begin{figure*}
    \centering
    \includegraphics[width=0.6\textwidth]{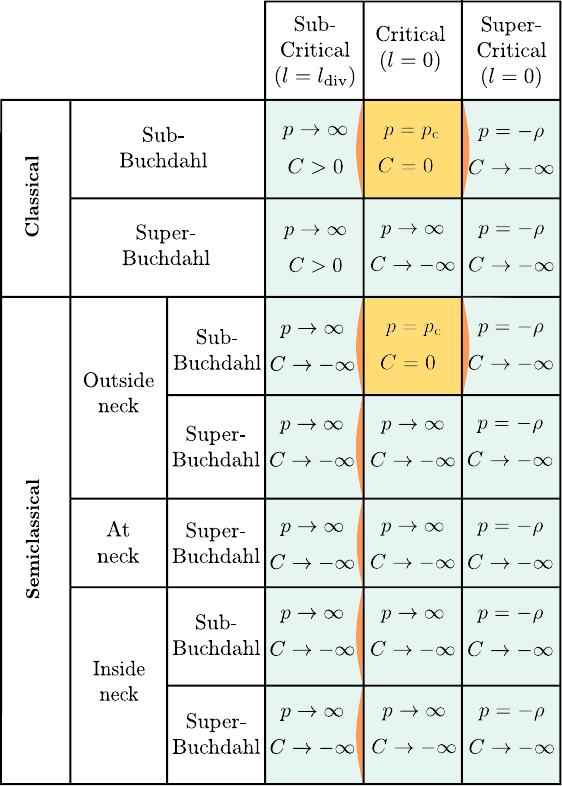}
    \caption{This table shows the complete set of classical and semiclassical stellar solutions of constant density. We distinguish whether the energy density $\rho$ takes values below, above, or at the critical value $\rho_{c}$; if the compactness is below or above the Buchdahl limit; and, for semiclassical stars, if their surface is located outside, inside, or at the neck itself. Each cell shows the behavior of pressure and compactness at the smallest value of $l$ in the domain of definition of the solution. Light-green (light grey) cells correspond to singular geometries. Yellow cells (grey) are strict stellar spacetimes. 
    The orange colour (black) percolating into the rightmost part of sub-critical cells and the leftmost part of super-critical cells represents \emph{$\epsilon$-strict} spacetimes. This family of spacetimes includes the subset of sub-critical solutions with small wormhole necks. See Figure \ref{fig:tablefigs} for the respective numerical solutions for each cell of the table.} \label{fig:table}
\end{figure*}

\begin{figure*}
    \centering
    \includegraphics[width=0.99\textwidth]{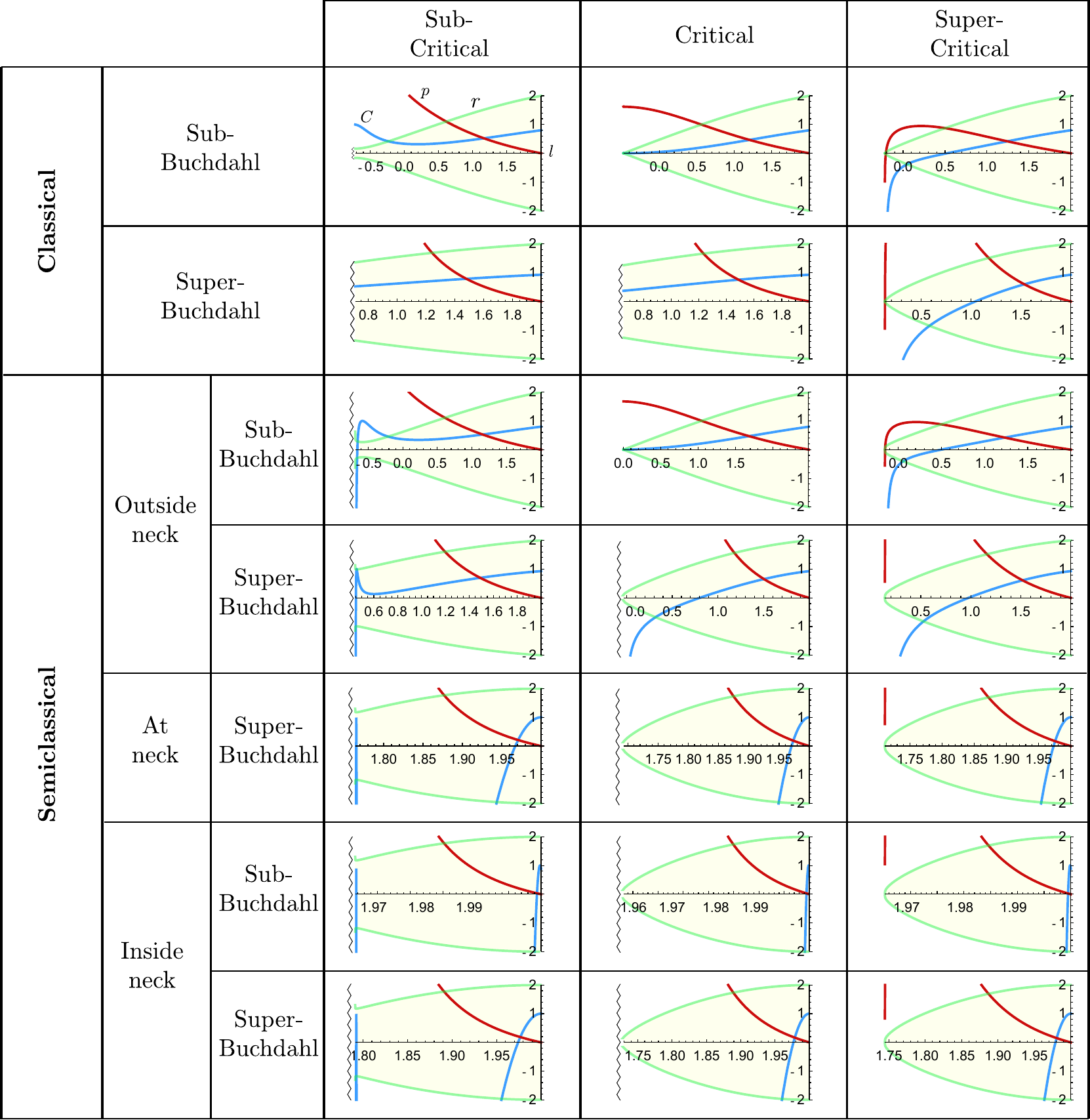}
    \caption{This table shows numerical integrations for each of the distinct regimes that we have found exploring classical and semiclassical stellar solutions of constant density. The criteria for classifying the solutions is the same followed in Fig. \ref{fig:table}. 
    The three regimes (sub-critical, critical and super-critical) appear represented for sub-Buchdahl and super-Buchdahl stars in both classical and semiclassical theories. In the semiclassical case, distinction is made on whether the star surface is located outside, inside, or at the neck of the vacuum wormhole geometry. In the semiclassical regime we show \emph{$\epsilon$-strict} solutions in the cases where no regular critical solutions exist.
    Each cell shows a numerical integration of a constant-density star. All but critical sub-Buchdahl stars, which are integrated from $l=0$ outwards, are  integrated from the surface $l_{\rm S}=R=2$ inwards until the centre of spherical symmetry or a singularity is reached. We stick to the following color criteria for the represented functions throughout the rest of the paper: the shape function $r(l)$ is represented in green, the pressure $p(l)$ in red, and the compactness function $C(l)$ in blue (colors online). The region where the classical fluid is present is filled in yellow for pictorical purposes. Spacetime singularities are depicted by a vertical zigzag line and correspond in every case with a divergent pressure. In super-Buchdahl super-critical plots the pressure grows inwards outside the plot window (with no divergences), to just come back inside the plot window when closer to the radial origin. Inside-the-neck sub-Buchdahl stars have $C(R)<8/9$ and their compactness function grows to $1$ inside the structure, generating a local maximum in the shape function $r(l)$ just below the surface. The semiclassical critical profiles from the fourth row downwards are pictorical representations of how the respective exact solutions would look like, rather than complete numerical integrations. This is due to the numerical instability of our numerical algorithm at the critical density. We have attached a Mathematica notebook that generates all the plots appearing in this table. The reader can access it in order to consult the values of $C(R)$, $\rho$ and $\alpha$ used for integrating the equations. See Fig. \ref{fig:table} for the asymptotic behaviors of the pressure and the compactness in each situation.}  \label{fig:tablefigs}
\end{figure*}

All the stellar solutions described in this paper are listed in the table of Figure \ref{fig:table}. In addition, Figure \ref{fig:tablefigs} shows illustrative numerical plots that highlight the overall features of the solutions described in the table in Fig. \ref{fig:table}. These figures are organized as follows:
\begin{itemize}
    \item First of all we distinguish between the classical and the semiclassical theory based on the Regularized Polyakov RSET. In both cases, we discriminate between sub-Buchdahl and super-Buchdahl stars, depending on whether their surface compactness is below or above the Buchdahl limit (given by the most compact strict stellar spacetime in each situation).
    In the classical case and for a star of constant density, this limit corresponds to $C(R)=8/9$.
    In the semiclassical theory there is no clear notion of Buchdahl limit due to the introduction of a preferred length scale $l_{\rm P}$. Now, the maximum $C(R)$ allowed by strict stellar spacetimes depends on the values of $R,~\rho$, and the particular regularization scheme adopted for the Polyakov RSET. In our semiclassical integrations, we do not impose any additional restriction on how large the values of the classical pressure can become as long as they are finite. We do this to make the discussion as close as possible to the analysis of classical configurations approaching the Buchdahl limit, in which the same logic is followed.
    \item Taking a stellar radius $R$ and a surface compactness $C(R)$ we can integrate the equations of equilibrium inwards for different values of $\rho$. By changing the parameter $\rho$ one realizes that there is a gross change in behavior for the compactness function when passing through a critical value $\rho_{\rm c}$. Attending to this value, we separate the different solutions as being sub-critical, critical, or super-critical.
    \item  For the semiclassical case, we distinguish three possibilities depending on where the star surface connects with the vacuum solution. Since the vacuum solution has a wormhole shape, the matter boundary can be located outside, inside, or at the neck itself. The value of $\rho_{\rm {c}}$ changes strongly depending on the region where the surface is located. Regardless, we find a similar distinction between critical and non-critical geometries.
    \item For each of the cells in the classification scheme from Fig. \ref{fig:table} (see Fig. \ref{fig:tablefigs} for the corresponding numerical solutions) we have added the asymptotic behavior of the pressure $p(l)$ and the compactness $C(l)$ at the smallest value of $l$ reached by each solution. This corresponds to $l=0$ for stars which extend to $r=0$, independently of whether the configuration is regular or singular there, or to some $l=l_{\text{div}}>0$ for stars with a singularity at $r(l_{\text{div}})>0$. 
    \item Finally, the cells corresponding to strict stellar spacetimes have yellow background and those in which we find \emph{$\epsilon$-strict} stellar configurations have orange background. 
\end{itemize}
We will describe the different regimes shown in Fig. {\ref{fig:table}} in the remaining of the section.

Let us start this summary from the sub-critical sub-Buchdahl corner of the classical solutions (Subsec. \ref{sec:C-subc-subB}). These configurations are irregular. When the density reaches the critical value $\rho_c$ for a given compactness the geometry becomes a strict stellar spacetime (Subsec. \ref{sec:C-c-subB}). Going into the super-critical regime (Subsec. \ref{sec:C-superc-subB}), the compactness $C(r)$ becomes irregular at the origin, diverging to $-\infty$. For a small window of densities just above the critical solution $\rho_{\rm{c}}$ we find \emph{$\epsilon$-strict} stellar configurations. 

Focusing now on the sub-critical super-Buchdahl cell we have again irregular solutions (Subsec. \ref{sec:C-subc-subB}).
The difference with the sub-Buchdahl case is that when reaching the critical density $\rho_c$, although the compactness function is well-behaved at the radial origin, the pressure diverges before reaching the origin (a divergence at the origin happens precisely in the Buchdahl limit; Subsec. \ref{subsec:C-c-superB}). Going further into the super-critical regime (Subsec. \ref{sec:C-superc-subB}) one is able to find solutions for which the pressure is regular until the origin, but that is at the cost of making a highly irregular compactness. \emph{$\epsilon$-strict} configurations are only found very close to criticality at the Buchdahl limit or below it.

Turning now to the semiclassical counterparts, the Schwarzschild vacuum geometry is drastically modified by semiclassical corrections in the Boulware vacuum state, becoming an asymmetric wormhole \cite{Arrechea2020} (see Subsec. \ref{Sec:VacuumSol} for a brief description of this geometry and Fig. \ref{fig:wormhole} for an illustrative numerical solution). This leaves three distinct regions (outside, inside, or at the wormhole neck itself) in which to match the vacuum geometry with the surface of a star. 

Let us start the route from the sub-critical, sub-Buchdahl, outside-the-neck configurations.
These are irregular configurations that display characteristics from vacuum solutions, i.e. they are singular wormhole-like configurations (Subsec. \ref{sec:S-subc-subB}). For the critical case we find strict stellar spacetimes (Subsec. \ref{sec:S-c-subB}) that amount to perturbative corrections of the classical regular stars. On the other side, super-critical configurations display naked singularities, with negative divergent compactness and Misner-Sharp mass at $r=0$ (Subsec. \ref{sec:S-superc-subB}). 

Passing to the sub-critical super-Buchdahl case, we again find wormhole-like configurations 
(Subsec. \ref{sec:S-subc-subB}). In the same manner, the super-critical regime exhibits naked singularities (Subsec. \ref{sec:S-superc-subB}). We find that the critical solution is one with a special profile of divergent pressure and compactness, which appears resilient to quantum corrections (Subsec. \ref{sec:S-c-superB}). An important difference between the classical and semiclassical super-Buchdahl critical configurations is that, in the latter case, close to criticality we find an ample window of \emph{$\epsilon$-strict} configurations. In Fig. \ref{fig:tablefigs}, for the cases where no regular critical configuration exists, we have shown an example of these \emph{$\epsilon$-strict} configurations instead.

Stars matched at the neck, analyzed in Subsec. \ref{subsec:at-neck}, show as well three distinct regimes (sub-critical, super-critical and critical) which depict asymptotic behaviors similar to those of the super-Buchdahl outside-the-neck case, depending on whether the density is above or below the critical value $\rho_{\rm c}$. Additionally, we find that the surface of the star displays different properties depending on the value of $\rho$. For sufficiently small $\rho$, the surface of the star corresponds to a minimal surface for the shape function $r$, or neck. By increasing $\rho$, this bouncing surface for the shape function gets pushed towards the interior of the star, disappearing eventually for $\rho\geq\rho_{\rm c}$.

The situation for stars inside the neck (Subsec. \ref{subsec:beyond-neck}) can be summarized saying that there are again three regimes, sub-critical, super-critical, and critical, with the same asymptotic behaviors seen for the super-Buchdahl, outside-the-neck case. The only caveat is that the critical density $\rho_{\rm c}$ increases as the surface of the star $R$ is moved away from the neck, becoming trans-Planckian not far from it (in proper distance, see Fig. \ref{fig:BeyondNeckDens} below). Therefore, reaching criticality for these configurations requires extremely dense classical fluids that compensate the negative masses generated by vacuum polarization \cite{Ho2018}.

\section{Classical solutions} \label{sec:classical}

Next we turn to the analysis of the set of solutions to the classical equations of stellar equilibrium for a perfect fluid of constant density. Throughout this section we will describe the solutions from the first two rows in Figs. \ref{fig:table} and \ref{fig:tablefigs}. We have found that it is more convenient to start the analyses by considering how the equations of equilibrium integrate outwards starting from a regular radial origin. 

\subsection{Solutions with a regular center}\label{sec:C-c-subB}

The first set of solutions we are going to describe can be seen as part of inhomogeneous cosmologies. In this context, they were analyzed by Lemaître \cite{Lemaitre1997} and later by Tolman \cite{Tolman1939}. Here we shall recall these analyses using our notation and perspective. Some of these solutions can be used to build interiors of stellar spacetimes. The suitable interiors retrieved in this first analysis are all critical, by construction, and result in sub-Buchdahl configurations (the only ones that are regular in the classical theory).

First, integrating Eq. \eqref{eq:compclas} returns
\begin{equation}\label{eq:rsolclas}
r=\frac{\sin{\left(A l\right)}}{A},
\end{equation}
with $A=\sqrt{8\pi \rho/3}$. The periodic character of the shape function $r(l)$ is consistent with a cosmological interpretation. The shape function extends between two zeroes which correspond to two poles of the inhomogeneous cosmologies.

With both $\rho$ and $r$ known, integration of the TOV equation \eqref{eq:TOVclas} is straightforward and yields
\begin{equation}\label{eq:psolclas}
p=\rho\left\{\frac{2}{3}\left[1-B_{0}\cos\left(A l \right)\right]^{-1}-1\right\}.
\end{equation}
Let us now extract the physical content of the above expression. The central pressure $p(0)$ is determined by the integration constant $B_{0}$. For any $B_{0}\neq1$ pressure is finite at $l=0$, so these kind of geometries (those with regular center), will be analyzed first. By varying the value of $p$ at the origin, we can divide cosmological solutions in the following three families, with their respective separatrices:

\begin{enumerate}
\item $p(0)>-\rho/3$. This guarantees that both the strong energy condition (SEC) and the null energy condition (NEC) hold at $l=0$. As Fig. \ref{fig:critclas} shows, the resulting cosmologies are regular everywhere and have a pressure that decreases between the two poles. Solutions with $p(0)>0$ reach a surface of zero pressure at $l_{\rm S}$ where the geometry can be matched with the Schwarzschild vacuum geometry, and thus resemble strict stellar spacetimes (Fig. \ref{fig:critclas}). On the other hand, solutions with $p(0)<0$ lack such surface and therefore resemble inhomogeneous cosmologies.
\begin{figure}
    \centering
    \includegraphics[width=\columnwidth]{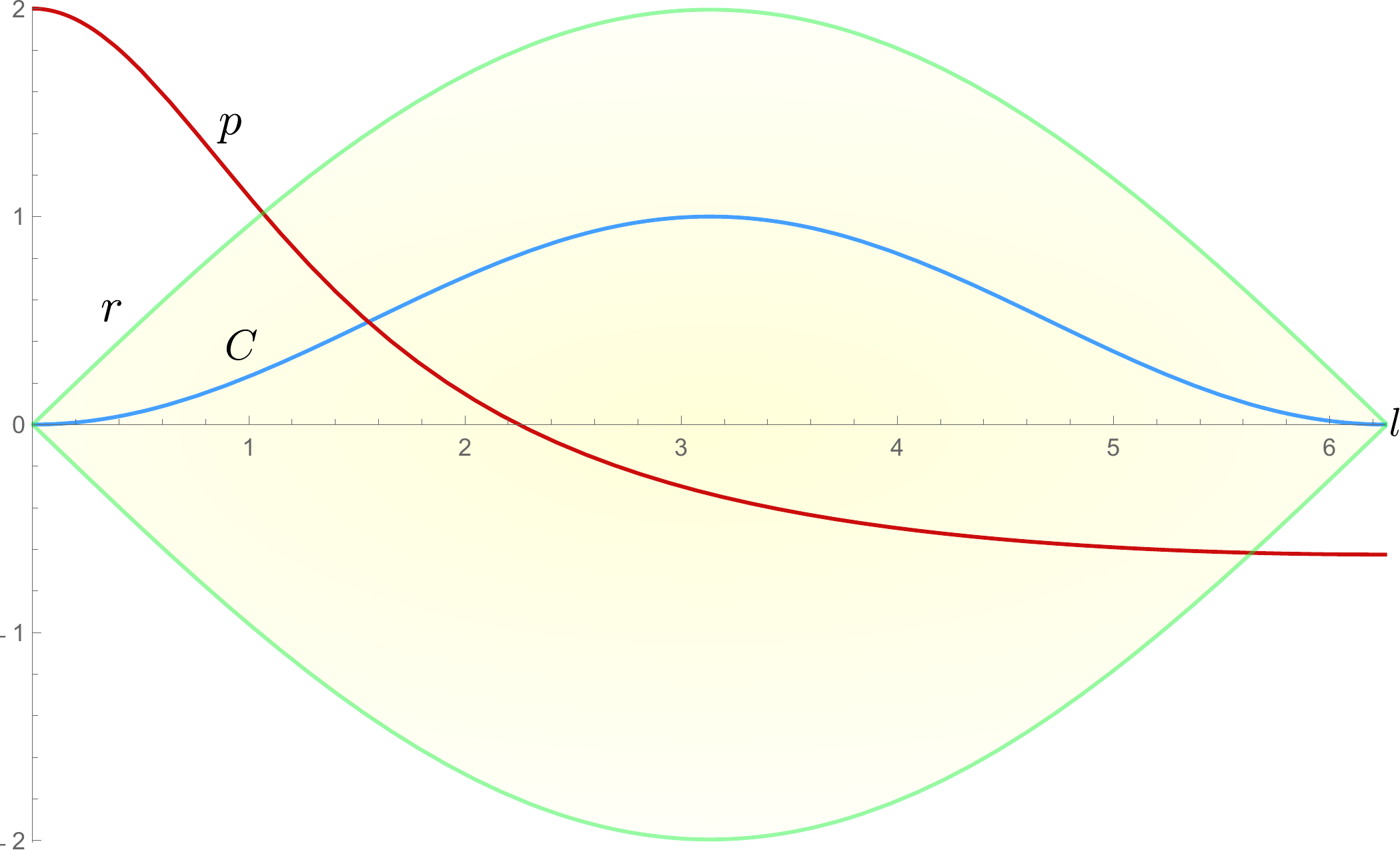}
    \caption{Plot of a positive pressure solution to the equations of structure. The above and below green lines represent the shape function $r(l)$ and the region in between both curves has been coloured for pictorial purposes. The blue line denotes the compactness function of the geometry $C(l)$, which reaches $1$ at the radial maximum and vanishes at the poles. The red curve is the pressure of the solution (in units of $\rho$) for a star with $\rho=0.03$ and $p(0)=2\rho$. Notice that the region of positive pressure corresponds to a relativistic star with $C(l_{\rm S})\simeq0.82$. This solution has an enormous density compared to that of astrophysical objects, but the physics of classical critical solutions is scale-invariant.}
    \label{fig:critclas}
\end{figure}
\item $-\rho<p(0)<-\rho/3$. In this case the SEC is violated at $l=0$ while the NEC holds. Solutions with \mbox{$p(0)\in-\left(\rho/3,2\rho/3\right)$} correspond to the mirror-reflected version of type $1$ profiles. Can one construct a regular star whose interior corresponds to this left-hand-side pole of the solution? In these interior solutions
pressure decreases inwards from its zero value at the star's surface, becoming all the way negative. This interior geometry can be matched with a patch of the Schwarzschild vacuum spacetime. However, in this case the shape function $r(l)$ is initially increasing (at the surface) towards the interior, so one cannot smoothly connect (without introducing a shell of matter) this interior with a patch of Schwarzschild that extends towards the asymptotically flat region; one could only connect this interior with a Schwarzschild patch covering $r<R$. Therefore, these solutions do not serve to construct regular stars. Decreasing the central pressure below $p(0)<-2\rho/3$
maintains the previous characteristics: pressure increases outwards and crosses zero at a finite radius. The difference with the previous situation is that if one now continued the internal solution beyond the surface of vanishing pressure, one would uncover a curvature singularity at
\begin{equation}\label{eq:presdiv}
l_{\text{div}}=\frac{\text{arccos}\left(1/B_{0}\right)}{A}.
\end{equation}
This singularity has an infinite positive pressure. The right-hand part of these geometries (that is, beyond the surface of zero pressure outwards) cannot be used to construct regular hydrostatic equilibrium configurations. As we will see shortly these solutions appear when integrating critical super-Buchdahl stellar configurations from the stellar surface inwards.

\item $p(0)<-\rho$. This guarantees SEC and NEC are violated. Pressure decreases from the radial origin outwards, eventually diverging towards $-\infty$ at \eqref{eq:presdiv}. Later we will briefly comment on these solutions, since they describe the interior patch of the gravastar model \cite{Mazur2004}.
\end{enumerate}

For the sake of completeness, let us comment briefly about the separatrices between families 1-3. The case between the first and second type of solutions corresponds to Einstein's static and homogeneous universe, where pressure is constant [by virtue of \eqref{eq:redsfunclas} and \eqref{eq:consclasica}] and equal to $-\rho/3$. The instability of this model has a long and interesting story (see for example \cite{BarceloVolovik2004}).

The remaining separatrix solution saturates the null energy condition with constant pressure equal to $-\rho$. Addition of Eqs. \eqref{eq:ttclasica} and \eqref{eq:rrclasica} leads to the relation
\begin{equation}
\frac{r''}{r'}=\phi',
\end{equation} 
which results in 
\begin{equation}
e^{2\phi}=\left[\cos\left(A l\right)\right]^{2},
\end{equation}
presenting a horizon at $l=\pi/2A$. This metric corresponds to de Sitter spacetime in static coordinates
\begin{equation}
ds^{2}=-\left[\cos\left(A l\right)\right]^{2}dt^{2}+dl^{2}+\frac{\left[\sin\left(A l\right)\right]^{2}}{A^{2}}d\Omega^{2},
\end{equation} 
revealing the existence of a cosmological horizon at $r(l_{\rm H})=A^{-1}$.

\subsection{Critical Buchdahl and super-Buchdahl solutions}\label{subsec:C-c-superB}

Note that taking $B_{0}=1$ in \eqref{eq:presdiv} makes the pressure diverge at $l=0$. 
The redshift function, obtained from integrating Eq.~\eqref{eq:rrclasica},
\begin{equation}\label{eq:redsfunclas}
e^{2\phi}=e^{2\phi_{0}}\left[B_{0}\cos\left(A l\right)-1\right]^{2},
\end{equation}
vanishes at $l=0$ for $B_{0}=1$. Here, $\phi_{0}$ is an irrelevant integration constant that amounts to a rescaling of $t$. For this particular $B_{0}$, the pressure \eqref{eq:psolclas} is found to diverge at the origin as
\begin{equation}\label{eq:presbuch}
p\simeq\frac{1}{2\pi l^{2}}.
\end{equation}

Since this solution does not have a regular center, we appeal to integrations from the star surface $l_{\rm S}$ to explore this and other similar cases. For that purpose, one needs to impose the following boundary conditions at the surface of the star:
\begin{align}\label{eq:boundcondclas}
p\left(l_{\rm S}\right)
&
=0,\quad r\left(l_{\rm S}\right)=R,  \\
\phi\left(l_{\rm S}\right)
&
=\frac{1}{2}\ln\left[1-C\left(l_{\rm S}\right)\right],\quad r'\left(l_{\rm S}\right)=\sqrt{1-C\left(l_{\rm S}\right)},\nonumber
\end{align}
with $C(l_{\rm S})=2M_{\text{ADM}}/R$. With these boundary conditions, solution \eqref{eq:presbuch} corresponds to a surface compactness
\begin{equation}\label{eq:Buchdahl}
C(l_{\rm S})=1-\left[r'\left(l_{\rm S}\right)\right]^{2}=8/9.
\end{equation}
This result denotes the maximum compactness of regular perfect fluid spheres in hydrostatic equilibrium, or Buchdahl limit \cite{Buchdahl1959}. Stellar configurations that have isotropic pressures, have an outwards non-increasing $\rho$, and whose exterior geometry is the Schwarzschild vacuum geometry are subject to the upper compactness bound \eqref{eq:Buchdahl}. More compact (super-Buchdahl) stars will have the surface of infinite pressure gradually moved from $r=0$ towards $r=R$. Beyond this curvature singularity we can find another geometric patch extending up to $r=0$, in which $p$ takes values below $-\rho$.
Matching these two solutions through a regularizing shell in the limit $C(l_{\rm S})\to1$ displays a gravastar geometry, a stellar model whose interior is supported by a cosmological constant \cite{Mazur2015}. This model has been proposed as a candidate for ultra-compact objects that relies on classical properties of the Schwarzschild interior solution in the ultra-compact limit \cite{Mazur2004}.

This ends our discussion concerning classical critical solutions, which will guide us in the classification of their semiclassical counterparts. In the following we turn to the analysis of non-critical configurations, which lack a regular center from the start. 

\subsection{Sub-critical solutions} \label{sec:C-subc-subB}

The analysis of solutions out of criticality is interesting because in the semiclassical case it is not directly clear how to associate a failure in criticality to the value of the mass at the origin. An understanding of the role played by non-criticality at the classical level will therefore allow us to distinguish between critical and non-critical solutions in the semiclassical case.

\begin{figure}
    \centering
    \includegraphics[width=\columnwidth]{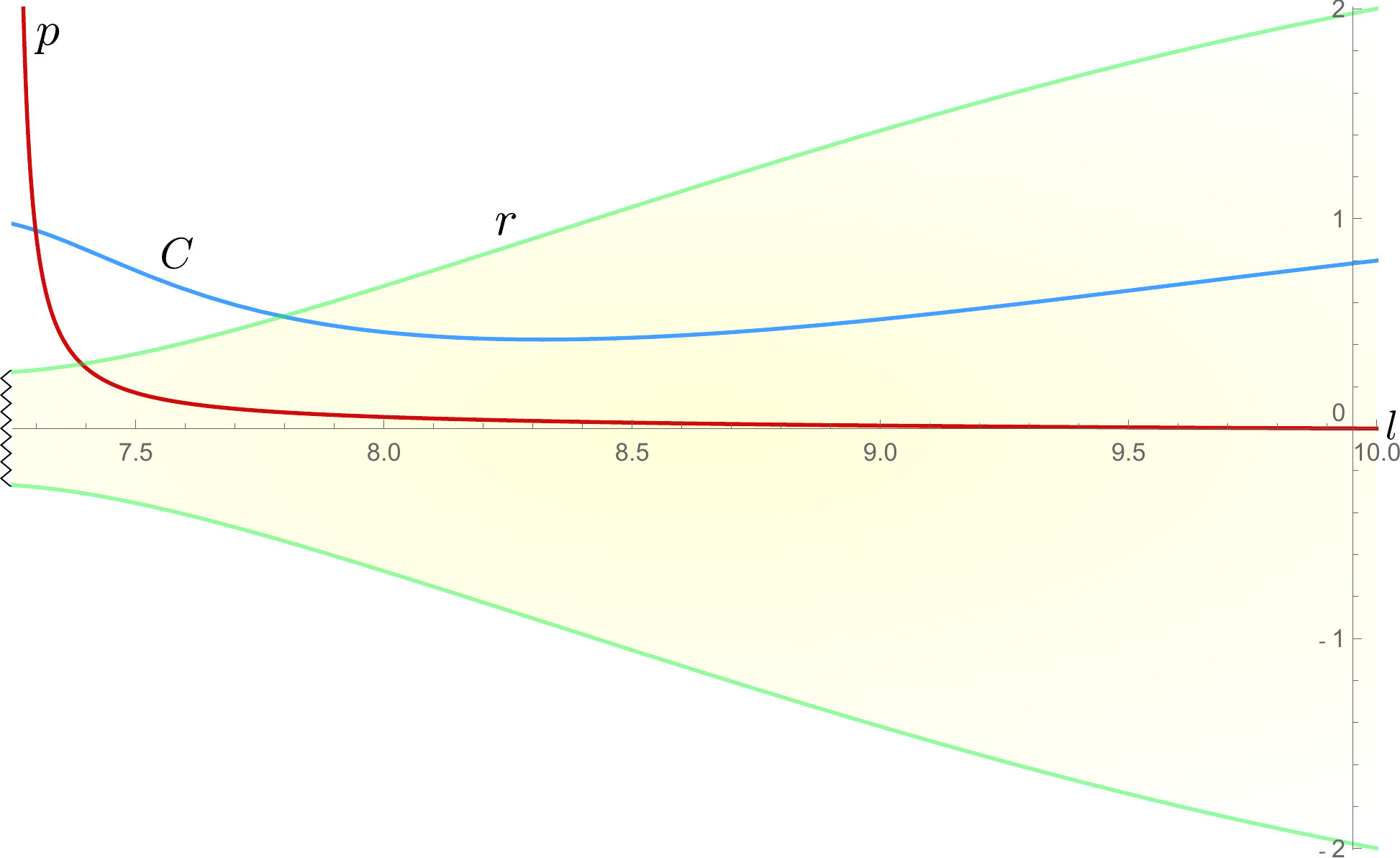}
    \caption{Plot of a sub-critical, sub-Buchdahl star with $R=2, C(R)=0.8$ and $\rho=0.84\rho_{\rm{c}\text{-clas}}$. Green lines represent the shape function $r(l)$, while red and blue lines denote the functions $p$ and  $C$, respectively. The presence of a positive constant mass $M_{0}\simeq0.07$ generates a (singular) event horizon at $l\simeq7.25$.}
\label{fig:subcritclas}
\end{figure}

Let us start by describing what is seen in the inwards integration of a sub-critical star. Reference to these solutions can be found in \cite{Volkoff1939, Tolman1987}. An example of a sub-critical sub-Buchdahl star is shown in Fig. \ref{fig:subcritclas}. By imposing $\rho<\rho_{\rm{c}\text{-clas}}$ the geometry acquires a positive constant mass $M_{0}$. The gravitational effect of this mass is perceived by the fluid, which responds to it with an increase in pressure. This increase happens more quickly than in the critical case as to compensate for the extra gravitational pull induced by $M_{0}$. As we deepen through the star, compactness passes through a turning point and starts increasing as the radius decreases.
Not far below this turning point, the pressure diverges and the geometry has a curvature singularity, as seen in Fig. \ref{fig:subcritclas}.

Let us derive the form of this curvature singularity by solving the continuity equation for the perfect fluid of constant density \eqref{eq:consclasica}
\begin{equation}\label{eq:solcont}
p=-\rho+\kappa e^{-\phi(l)},
\end{equation}
where $\kappa$ is a constant of integration with dimensions of inverse of length squared. This expression ensures that the pressure is infinite at any surface of zero redshift function, i.e. when $\phi(l_{\text{div}})\to-\infty$. Assuming that such surface exists, we approximate the TOV equation at leading order in the pressure as
\begin{equation}\label{eq:gradpressubcrit}
p'(l)=-\frac{4\pi r p^{2}+\mathcal{O}(p)}{\sqrt{\displaystyle 1-\frac{8\pi r^{2}\rho}{3}-\frac{M_{0}}{r}}}.
\end{equation}
Take into account that the pressure diverges while the denominator in \eqref{eq:gradpressubcrit} is still non-vanishing. In this regime, we can assume the following behavior for the pressure
\begin{equation}
p\simeq\frac{p_{+}}{\left(l-l_{\text{div}}\right)^{n}}, \qquad p_{+}>0,\qquad n>0.
\end{equation}
Replacing this ansatz in Eq.~\eqref{eq:gradpressubcrit} and solving for $p_+$ and $n$ we find
\begin{equation}
p\simeq\left(l-l_{\text{div}}\right)^{-1}\frac{\sqrt{\displaystyle 1-\frac{8\pi r_{\text{div}}^{2}\rho}{3}-\frac{M_{0}}{r_{\text{div}}}}}{4\pi r_{\text{div}}},
\end{equation}
where $r_{\text{div}}=r(l_{\text{div}})$. The pressure diverges positively at the surface $l=l_{\text{div}}$, whose location depends on the boundary conditions of the star and, consequently, on $M_{0}$. By decreasing $\rho$ (increasing $M_{0}$), this divergence approaches the surface of the star. Equivalently, the more super-Buchdahl the star is, the further the pressure divergence moves towards the surface of the star. Recall that, for the super-Buchdahl critical case, we know the position of the infinite pressure divergence in terms of boundary conditions. This explicit expression is lost in the sub-critical situation, since we lack a complete analytic solution. 

The resulting geometry resembles a black hole surrounded by matter forced to maintain equilibrium, causing a runaway in the pressure of the fluid. This divergence in the pressure takes place at the same position at which the redshift function, obtained from solving Eq.~\eqref{eq:solcont} in the $l\to l_{\text{div}}$, vanishes:
\begin{equation}
e^{2\phi}\simeq\left(\frac{l-l_{\text{div}}}{l_{\text{div}}}\right)^{2}.
\end{equation}
Since this geometry is not vacuum but filled with a perfect fluid, there is a curvature singularity at the horizon. This is foreseeable by recalling that horizons are incompatible with matter fluids in hydrostatic equilibrium. 

\subsection{Super-critical solutions}\label{sec:C-superc-subB}

Now, we turn to the analysis of super-critical configurations, where we distinguish between sub- and super-Buchdahl stars. 

\begin{figure}
    \centering
    \includegraphics[width=\columnwidth]{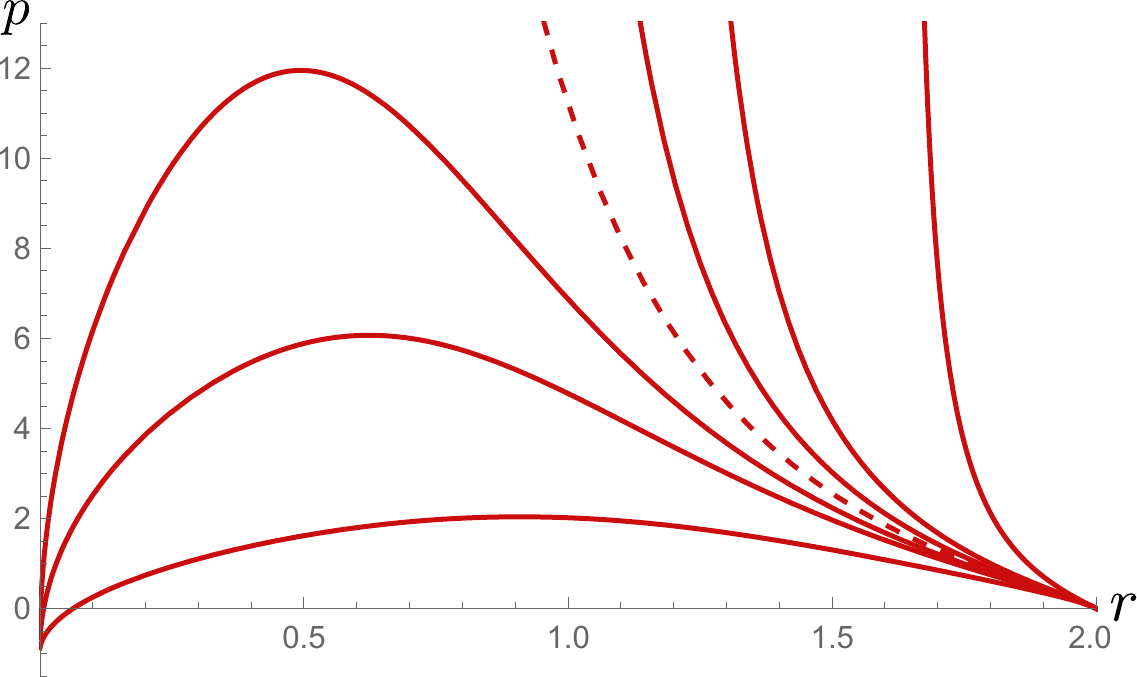}
    \caption{Plot of the pressure profile for a super-critical, super-Buchdahl star with $C(R)=0.96$. The curves denote the pressure $p(r)$ (in units of $\rho$) for the values of the energy density (from right to left) $\rho/\rho_{\rm{c}\text{-clas}}=1, 1.8, 2, 2.13, 2.26, 2.4$ and $3$. The dashed curve ($\rho/\rho_{\rm{c}\text{-clas}}\sim2.13$) corresponds to a separatrix for which pressure diverges at $r=0$. Note how the divergence in pressure of super-Buchdahl stars moves inwards as the density increases. 
    An increase of the negative mass $M_{0}$ finally regularizes the pressure, which tends to the value $p=-\rho$ at the origin.
    }
\label{fig:supercritclas}
\end{figure}

Recall (Subsec. \ref{sec:C-subc-subB} or Eq. \eqref{eq:compclas}) that in the sub-Buchdahl case, the effect of going super-critical (i.e. taking $\rho>\rho_{\rm{c}\text{-clas}}$) is to add a negative mass $M_{0}$ to the spacetime. The repulsive effect that this negative mass exerts on the fluid makes pressure reach a maximum value at some $r>0$. This can be viewed in the vanishing of the numerator of Eq. \eqref{eq:TOVclas} when $r'$ [as of Eq. \eqref{eq:compclas}] becomes large enough as to compensate for the positive $1+8\pi r^{2}p$ term. In this case, the perfect fluid extends up to $r=0$, where a naked curvature singularity resides. Figure \ref{fig:supercritclas} shows examples of pressure profiles for several super-critical configurations. Notice how the growth of the pressure is dampened as the solutions are made increasingly super-critical. Integrating \eqref{eq:compclas} in the $r\to0$ limit leads to the relation
\begin{equation}\label{eq:supercritr}
r\simeq\left[\frac{3 \sqrt{\abs{M_{0}}} }{2}l\right]^{2/3}.
\end{equation}
Therefore, there exists a neighbourhood of $r=0$ in which the geometry is well-approximated by the Schwarzschild vacuum solution with negative ADM mass.

The TOV equation \eqref{eq:TOVclas} can be integrated in terms of analytical functions by making the coordinate change \eqref{eq:cambiovar} and taking the limits $r\to0$ and $C\to-\infty$, which yields
\begin{equation}
p'\simeq\frac{(\rho+p)}{2r}.
\end{equation}
Integrating and replacing \eqref{eq:supercritr} we obtain
\begin{equation}\label{eq:supercritpres}
p\simeq-\rho+M_{0}^{-2}\left(\frac{l}{|M_{0}|}\right)^{1/3}.
\end{equation}
In the presence of a constant negative mass, the pressure acquires the equation of state of vacuum energy in the limit $r\to0$ as a consequence of the gravitational repulsion induced by $M_{0}$. Note that this finite value for the central pressure is reached with infinite derivative, which results in the redshift function being divergent in the $l\to 0$ limit as
\begin{equation}\label{eq:supercritredsclas}
e^{2\phi}\simeq\left(\frac{\abs{M_{0}}}{l}\right)^{2/3}.
\end{equation}

In some situations, semiclassical contributions can appear as a cloud of negative mass in the spacetime. The pressure-regularizing effect of this cloud is similar to that of super-criticality. We will revisit this discussion in the analysis of semiclassical solutions. Here, given a super-Buchdahl star, gradually increasing $\rho$ (decreasing $M_{0}$) displaces the pressure divergence towards the radial origin, eventually making pressure finite for densities above some $\rho=\rho_{\text{reg-p}}$. These aspects apply to more generic equations of state as well \cite{Smoller1997,Woszczyna2015,Bratek2019,Anastopoulos2020}. This value of the density constitutes an infinite-pressure separatrix between super-critical solutions singular and regular in pressure (see the dashed line in Fig. \ref{fig:supercritclas}). Hence, solutions with $\rho>\rho_{\text{reg-p}}$ will be regular in the pressure (although the pressure gradient diverges at $l=0$) but irregular in the compactness.
The value of $\rho_{\text{reg-p}}$ increases with the surface compactness of the star, eventually diverging towards $+\infty$ in the $C(l_{\rm S})\to1$ limit. The particular features of this separatrix solution are analyzed right below.

\subsubsection{Infinite pressure separatrix}\label{subsec:separatrix}

The separatrix solution lies between super-critical configurations irregular and regular in the pressure. Consider a super-critical solution extending to $r(0)=0$. Since $M_{0}<0$, the solution for the shape function $r(l)$ around the origin always obeys Eq.~\eqref{eq:supercritr}, and the pressure function can only follow two paths: either it goes to a constant value at $r=0$, which has to be exactly $-\rho$, in virtue of \eqref{eq:supercritpres}, or it diverges necessarily towards positive infinity at $r=0$. The separatrix solution corresponds to this last possibility. 

To derive the precise form of the divergence in pressure, we expand the TOV equation \eqref{eq:TOVclas} in the $l\to0$ limit under the assumption that $p\gg\rho$, yielding
\begin{equation}\label{eq:pressuresep}
p'\simeq-\frac{4\pi r p^{2}}{r'}.
\end{equation}
Now, assuming the following ansatz for the pressure 
\begin{equation}
p=\frac{p_+}{l^{n}},\quad n>0,
\end{equation}
where $p_+$ is a positive dimensionless constant, and replacing Eq.~\eqref{eq:supercritr} and this ansatz in Eq.~\eqref{eq:pressuresep}, we find 
\begin{equation}\label{eq:separatrixconst}
n=2,\quad p_+=1/3\pi.
\end{equation}
Therefore, the pressure diverges, in the $l\to0$ limit, with the same power of $l$ as in the separatrix \eqref{eq:presbuch} between sub-Buchdahl and super-Buchdahl configurations. However, the way the areal radius of spheres $r(l)$ approaches the origin $l=0$ differs in both cases. In terms of the shape function \eqref{eq:supercritr}, solution \eqref{eq:separatrixconst} takes the form
\begin{equation}\label{eq:separatrixpresclas}
p\simeq\frac{3 |M_{0}|}{4\pi r^{3}},
\end{equation}  
revealing a direct dependence in the constant mass $M_{0}$. A pictorial representation of this separatrix is shown in Fig. \ref{fig:supercritclas}. On the other hand, the separatrix described around Eq.~\eqref{eq:presbuch} satisfies
\begin{equation}\label{eq:separatrixpresclas2}
p\simeq\frac{1}{2\pi r^{2}},
\end{equation}
revealing that the leading behavior in the pressure is independent of $M_{0}$ since this separatrix corresponds to a critical configuration. 

The separatrix solution \eqref{eq:separatrixpresclas} was analyzed in \cite{Volkoff1939, Wyman1949} (see \cite{Raychaudhuri1951} for a compelling physical interpretation) and is particularly interesting because semiclassical corrections deform this solution into a separatrix for the compactness as well (i.e. a critical configuration). The relevance of this separatrix will be clear when analyzing the corresponding semiclassical situation.

\section{Semiclassical stellar equilibrium}\label{section:semiclassical}

In the following we are going to obtain the semiclassical self-consistent counterparts to the previous classical set of solutions. In doing so, first we need to address several aspects pertaining to semiclassical gravity. We introduce, as a source of curvature, the expectation value of the stress-energy tensor of a single massless, minimally coupled scalar field. Such stress-energy tensor demands a renormalization procedure in order to account for the genuine contribution of zero-point energy to curvature. Given the explained difficulties of handling the exact expressions for the RSET in $(3+1)$ dimensions, here we appeal to the Polyakov approximation. In this approximation, the RSET comes from uplifting to a $(3+1)$-manifold the RSET calculated over a $(1+1)$-spacetime. This RSET is renormalized following the covariant point-splitting scheme presented in \cite{FullingDavies1977}.
Taking advantage of conformal symmetry, a closed expression of the RSET can be provided in $(1+1)$ dimensions. Its components, in $(t,l)$ coordinates, take the form:
\begin{align}\label{eq:2dimRSET}
\langle\hat{T}_{tt}\rangle^{(2)}=
&
\frac{l_{\rm P}^{2}}{2}\left[\left(\phi'\right)^{2}+2\phi''\right]e^{2\phi}+\langle\text{STD}\rangle,\nonumber\\
\langle\hat{T}_{ll}\rangle^{(2)}=
&
-\frac{l_{\rm P}^{2}}{2}\left(\frac{\phi'}{r'}\right)^{2}+\langle\text{STD}\rangle,\nonumber\\
\langle\hat{T}_{tl}\rangle^{(2)}=
&
\langle\hat{T}_{lt}\rangle^{(2)}=0.
\end{align}
The terms $\langle\text{STD}\rangle$ are the state-dependent parts of the Polyakov RSET. We are taking the expectation value of the RSET to be in the Boulware vacuum state, so $\langle\text{STD}\rangle=0$ and asymptotic observers measure zero particle content. In this paper we will always use the Boulware vacuum as this is the natural vacuum state for genuinely static and asymptotically flat configurations. 

The $(3+1)$ Polyakov RSET is obtained from promoting to a $(3+1)$-spacetime the $(1+1)$-dimensional version of the RSET \eqref{eq:2dimRSET}
\begin{equation}\label{eq:4dimRSET}
\langle\hat{T}_{\mu\nu}\rangle^{(\rm{P})}=\frac{1}{4\pi r^{2}}\delta^{a}_{\mu}\delta^{b}_{\nu}\langle\hat{T}_{ab}\rangle^{(2)},
\end{equation}
where latin indices take $2$ values. Consequently, the Polyakov RSET contains no information about angular pressures: the $\theta\theta$ and $\varphi\varphi$ components are zero. Its simplicity (namely, the absence of higher-derivative terms) favors the search of self-consistent solutions to \eqref{eq:semieinstein}. Albeit its simple form, the Polyakov RSET properly encapsulates the most prominent features of vacuum states \cite{Fabbri2005}, such as non-local effects due to the quantum vacuum that can encompass horizon-sized regions \cite{Parentani1994, Chakraborty2015}. However, it does leave aside several physical contributions: it ignores backscattering of the field modes due to the gravitational potential and it neglects higher multipoles in the spherical harmonic expansion of the scalar field. The prefactor $1/r^2$ in the $(3+1)$ Polyakov RSET \eqref{eq:4dimRSET} is fixed by the requirement that it is covariantly conserved with no further modifications. However, this causes its components to become irregular at $r=0$, even for geometries that fulfill the regularity conditions \eqref{eq:regcond} imposed by the finitude of the Kretschmann invariant \eqref{eq:Kretschmann}. This cannot occur with an exact RSET, as the exact field modes must be regular in a regular geometry and so must be its associated RSET. The divergence at the radial origin of the $(3+1)$ Polyakov RSET simply points out its failure to approximate the exact RSET even qualitatively when approaching $r=0$. Thus, dealing with the singular character of the Polyakov RSET is compulsory for finding regular, self-consistent solutions. 

\subsection{Regularized Polyakov RSET} \label{sec:rset}

In the need to craft a RSET that is both regular at $r=0$ and self-consistently tractable, we proposed \cite{Arrechea2020} (following \cite{Parentani1994, Ayal1997}) the simplest regularization scheme one can think of: one based on introducing a cutoff to the RSET value at the radial origin. In this way we introduced a new Regularized Polyakov RSET (RP-RSET) where the $(t,l)$ components of the RP-RSET are defined as
\begin{equation}\label{eq:regularization}
\langle\hat{T}_{ab}\rangle^{(\rm{RP})}=\frac{r^{2}}{r^{2}+\alpha l_{\rm P}^{2}}\langle\hat{T}_{ab}\rangle^{(\rm{P})},
\end{equation}
were $\alpha>0$ plays the role of a regulator. Covariant conservation of the RSET now requires the introduction of non-zero angular components. All in all, our proposal for the components of the RP-RSET are:
\begin{align}\label{eq:RPRSET}
\langle\hat{T}_{tt}\rangle^{(\rm{RP})}=
&
\frac{l_{\rm P}^{2}}{8\pi \left(r^{2}+\alpha l_{\rm P}^{2}\right)}\left[\left(\phi'\right)^{2}+2\phi''\right]e^{2\phi},\nonumber\\
\langle\hat{T}_{ll}\rangle^{(\rm{RP})}=
&
-\frac{l_{\rm P}^{2}}{8\pi \left(r^{2}+\alpha l_{\rm P}^{2}\right)}\left(\frac{\phi'}{r'}\right)^{2},\nonumber\\
\langle\hat{T}_{\theta\theta}\rangle^{(\rm{RP})}=
&
\frac{\langle\hat{T}_{\varphi\varphi}\rangle^{(\rm{RP})}}{sin^{2}\varphi}=-\frac{\alpha\left(l_{\rm P}^{2}r\phi'\right)^{2}}{8\pi \left(r^{2}+\alpha l_{\rm P}^{2}\right)^{2}}.
\end{align}
The angular components are also well-behaved at $r=0$ (as the $tt$ and $ll$ components) and decay sufficiently fast with the radial distance as to ensure that the Polyakov RSET is recovered for $r\gg \sqrt{\alpha}l_{\rm P}$. We will always take values of the regulator $\alpha>1$. This is so because, although $\alpha>0$ is sufficient for regularity of the RP-RSET at $r=0$, the self-consistent semiclassical equations (namely, the dependence of $\langle\hat{T}_{tt}\rangle^{(\rm{RP})}$ on $\phi''$) move the singularity of the RSET from $r=0$ to $r=l_{\rm P}\sqrt{1-\alpha}$. Imposing $\alpha>1$ displaces the singularity out of the domain of the radial coordinate~\cite{Arrechea2020}. 

The regularization scheme that we have adopted is by no means unique, since there exists an infinite number of regularizing functions that ensure that the regularized RSET fulfills the desired properties. Let us note also that the non-conservation induced by the regularization \eqref{eq:regularization} can be compensated by the introduction of angular components as long as the whole construction remains static. In dynamical scenarios, however, it is an arduous task to find a regulating function that renders the RSET regular and covariantly conserved simultaneously. We believe this is why in some works the RSET is left non-conserved \cite{Parentani1994, Ayal1997}. For us, the choice of regulating function \eqref{eq:regularization} bows to a minimalist approach, in an attempt to modify the Polyakov RSET in the mildest possible way while serving our purposes. The choice of a better regulating function should ideally contain information about characteristics of the spacetime geometry close to the radial origin, and be capable of reproducing the physics predicted by more elaborate approximations to the RSET \cite{Hiscock1988}. It will be important to keep this in mind when extracting conclusions from our analysis of the semiclassical equations.

\subsection{Review of exterior vacuum solutions}
\label{Sec:VacuumSol}

\begin{figure}
    \centering
    \includegraphics[width=\columnwidth]{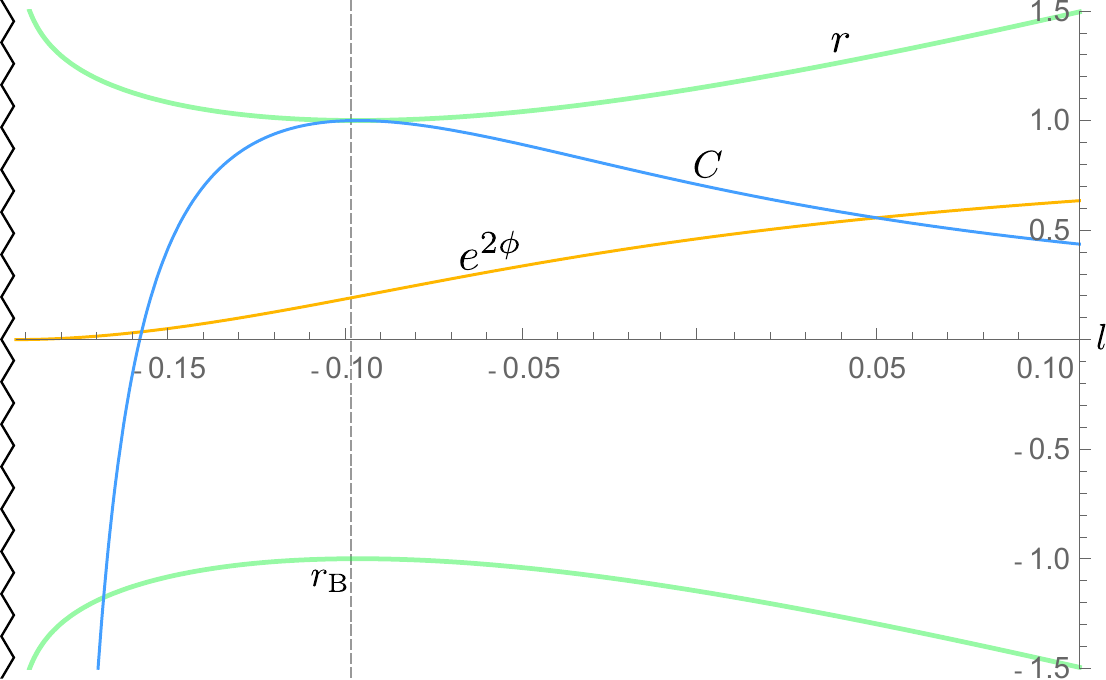}
    \caption{Numerical representation of the Schwarzschild vacuum geometry. The right hand side is the asymptotically flat region of the spacetime. In an inwards integration, a minimal surface or throat is encountered for the shape function $r(l)$ (green curve, in units of the neck radius), which connects to a null singularity at finite affine distance. This singularity has a negative infinite mass associated with it, which can be related to a runaway of vacuum polarization, as shown by the blue curve denoting the compactness $C(l)$ and the yellow curve representing the redshift function $e^{2\phi(l)}$. We have chosen $M=0.05$ and $\alpha=1.01$ to better highlight the characteristics of the solution, but the qualitative behavior of the geometry does not change for greater values of the ADM mass.}
    \label{fig:wormhole}
\end{figure}

Equipped with the RP-RSET, in previous works we obtained the complete set of semiclassical solutions for a single quantum massless scalar field in the absence of classical matter \cite{Arrechea2020} and also in the presence of a Coulombian electromagnetic field \cite{Arrechea2021}. See Fig. \ref{fig:wormhole} for a numerical example of the first case. Different patches of the vacuum solutions to the semiclassical equations will constitute the exterior geometry of semiclassical stars. Moreover, we find that lessons from the semiclassical counterpart of the Schwarzschild vacuum geometry are of relevance in order to discuss the semiclassical counterparts of the Schwarzschild stellar interior solutions.

One important characteristic of the set of semiclassical vacuum solutions is that, contrary to what happens for the classical vacuum, they are devoid of horizons of any kind. This result is foreseeable if one combines the following series of arguments. It is well known that the Boulware vacuum state is singular at the classical event horizon \cite{BirrellDavies1982}. This divergence is linked to the choice of mode decomposition of the field with respect to the Killing vector $\partial_{t}$, which becomes null at the event horizon. As a consequence, the Regularized Polyakov RSET is singular at the horizon of the Schwarzschild metric. By assuming that the metric \eqref{eq:metriclcoord} has a non-extremal horizon at some finite $l=l_{\rm H}$, so that $e^{2\phi}\propto(l-l_{\rm H})$, we observe that the semiclassical energy density
\begin{align}\label{eq:rhosemiL}
\rho_{\text{se}}
&
=-\langle\hat{T}^{t}_{t}\rangle^{(\rm{RP})}=\frac{l_{\rm P}^{2}}{8\pi (r^{2}+\alpha l_{\rm P}^{2})}\left[\left(\phi'\right)^{2}+\phi''\right]
\end{align}
diverges as
\begin{equation}\label{eq:rhodiv}
\rho_{\text{se}}\propto-\left(l-l_{\rm H}\right)^{-2}.
\end{equation}
Taking into account the backreaction of zero-point energies on the classical geometry makes the horizon disappear altogether as a consequence of the large vacuum polarization that builds up in its vicinity. The resulting semiclassical counterpart to the Schwarzschild vacuum geometry is an asymmetric wormhole, as depicted in Fig. \ref{fig:wormhole}. The classical horizon is replaced by a wormhole neck located slightly above the Schwarzschild radius. This neck connects an asymptotically flat region with a new singular asymptotic region whose singularity lies at a finite affine distance from the neck. Backreaction has turned the RSET regular everywhere except at this asymptotic singularity. 

As these wormhole solutions describe the exterior geometry of a semiclassical star, we immediately realize that the surface of the star (i.e. the position at which we start finding a non-vanishing classical matter contribution to the total SET) can in principle begin either outside the neck, innside the neck, or at the neck itself. We will analyze these situations in their respective sections. It is also interesting to recall~\cite{Arrechea2020} that once the vacuum geometry displays a, those solutions
become genuinely semiclassical as they do not have a classical limit.

Before passing to the analysis of the interior solutions, let us devote one moment to recall~\cite{Arrechea2021} what happens when using a Boulware vacuum RSET to modify an extremal horizon, instead of the non-extremal horizons relevant to the analyses on this paper. We have analyzed this issue in the context of the semiclassical counterparts of the Reissner-Nordstr\"om vacuum solutions. When analyzed in physical coordinates (those which are regular at the extremal horizon), the RSET components are divergent at the extremal horizon \cite{Balbinot2007}.
However, the corresponding semiclassical counterpart preserves the existence of a surface of zero red-shift, whose size and shape become modified in such a way that the geometry develops a non-scalar curvature singularity at the putative horizon. This result points out that the Boulware vacuum state has a strong incompatibility with horizons of any kind: it either destroys them, if non-extremal, or converts them into non-scalar curvature singularities, if extremal
~\cite{Arrechea2021}.

\subsection{Semiclassical equations of stellar interiors}
\label{section:semiclassicaleqs}

Let us now pass to the central part of the paper, the analysis of the internal stellar solutions under the hypothesis of having a classical matter component with a constant density $\rho$ in addition to the semiclassical contribution. 
The semiclassical field equations are obtained by plugging the RP-RSET components in Eq.~\eqref{eq:RPRSET} into Eq.~\eqref{eq:semieinstein},
\begin{align}\label{eq:ttsemi}
-2r''r+1-(r')^{2}=
&
8\pi r^{2} \rho
+\frac{r^{2}l_{\rm P}^{2}}{\left(r^{2}+\alpha l_{\rm P}^{2}\right)}\left[\left(\phi'\right)^{2}+2\phi''\right],\\
\label{eq:llsemi}
2r r' \phi'-1+(r')^{2}=
&
8\pi r^{2}p-\frac{r^{2}l_{\rm P}^{2}}{\left(r^{2}+\alpha l_{\rm P}^{2}\right)}\left(\phi'\right)^{2},
\end{align}
for the $tt$ and $ll$ components, respectively. The Regularized Polyakov RSET is independently conserved, and so is the classical SET. The system of equations is thus completed by the equation of conservation of the classical matter \eqref{eq:consclasica} and the equation of state of the uniform density fluid $\rho=\text{const}$.

We can construct the semiclassical version of the TOV equation~\cite{Carballo-Rubio2018a} by combining Eqs.~\eqref{eq:llsemi} and \eqref{eq:consclasica}
\begin{align}\label{eq:TOVsemi}
p'=
&
\left(\rho+p\right)r'\frac{\left(r^{2}+\alpha l_{\rm P}^{2}\right)}{l_{\rm P}^{2}r}\nonumber\\
&
\times\left(1\pm\sqrt{1+\left(\frac{l_{\rm P}}{r'}\right)^{2}\frac{8\pi r^{2}p+1-\left(r'\right)^{2}}{r^{2}+\alpha l_{\rm P}^{2}}}\right).
\end{align}
The semiclassical TOV equation is a quadratic polynomial for the gradient of the pressure. Therefore, two branches of solutions are present (already in vacuum) in the semiclassical theory,  given by the $\pm$ signs in Eq.~\eqref{eq:TOVsemi}. The $-$ sign or, as we shall call it, unconcealed branch, returns the classical TOV equation \eqref{eq:TOVclas} in the limit $l_{\rm P}\to0$ and can correspond, in many situations, to a quantum perturbation of the classical solution. On the other hand, the $+$ sign branch or concealed is intrinsically quantum and has no classical limit. This does not imply that the concealed branch is not physically relevant, since jumps between branches can occur when the radicand in \eqref{eq:TOVsemi} vanishes. For example, a branch jump takes place at the neck of the vacuum solutions (see Fig. \ref{fig:wormhole}), where the complete geometry is described by a combination of the unconcealed and concealed branches, resulting in a non-perturbative modification of the classical Schwarzschild solution. Therefore, the concealed branch is necessary to give a complete description of semiclassical solutions. Analytical solutions that describe ultra-compact horizonless stars have been found \cite{Carballo-Rubio2018a} for the concealed branch by solving Eq.~\eqref{eq:TOVsemi}.

For convenience in the upcoming analysis, field equations (\ref{eq:ttsemi}) and (\ref{eq:llsemi}) can be combined to construct a single differential equation that relates $\phi''(l)$ to the functions $\phi'(l),~\rho(l),~p(l)$ and $r(l)$. When expressed in terms of the Schwarzschild coordinates $(t,r,\theta,\phi)$, for which the metric takes the form \eqref{eq:metricrcoord}, this differential equation reveals convenient features that, in some situations, allow us to determine univocally the form of the solution. The change of variable from $l$ to $r$ amounts to the following replacements
\begin{align}
r'(l)\rightarrow\sqrt{1-C(r)},\quad \phi'(l)\rightarrow\psi(r)\sqrt{1-C(r)},
\end{align}
where $\psi(r)\equiv\phi'(r)$, the prime denoting now the derivative with respect to the coordinate $r$. The resulting differential equation is written as
\begin{align}\label{eq:eqpsi}
\psi'=
&
\mathcal{D}\left(\mathcal{A}_{0}+\mathcal{A}_{1}\psi+\mathcal{A}_{2}\psi^{2}+\mathcal{A}_{3}\psi^{3}\right),
\end{align}
where
\begin{align}
\mathcal{A}_{0}=
&
4\pi\left(\rho+3p\right),	\nonumber\\
\mathcal{A}_{1}=
&
4\pi r\left[3\left(\rho+p\right)+\frac{2l_{\rm P}^{2}}{r^{2}+\alpha l_{\rm P}^{2}}p\right]-\frac{2}{r},		\nonumber\\
\mathcal{A}_{2}=
&
8\pi	 r^{2}\left[\rho-p+\frac{l_{\rm P}^{2}(3p+\rho)}{2(r^{2}+\alpha l_{\rm P}^{2})}+\frac{l_{\rm P}^{2}r^{2}p}{(r^{2}+\alpha l_{\rm P}^{2})^{2}}\right] \nonumber\\
&
-\frac{2l_{\rm P}^{2}\left(r^{2}/2+\alpha l_{\rm P}^{2}\right)}{\left(r^{2}+\alpha l_{\rm P}^{2}\right)^{2}}-2,		\nonumber\\
\mathcal{A}_{3}=
&
\frac{l_{\rm P}^{2}r}{r^{2}+\alpha l_{\rm P}^{2}}\left\{4\pi r^{2}\left[\rho-p+\frac{2 l_{\rm P}^{2}r^{2}p}{(r^{2}+\alpha l_{\rm P}^{2})^{2}}\right]\right.		\nonumber\\
&
\left.\hspace{1.6cm}-\frac{\alpha l_{\rm P}^{4}}{(r^{2}+\alpha l_{\rm P}^{2})^{2}}-1\right\},\nonumber\\
\mathcal{D}=
&
\frac{r^{2}+\alpha l_{\rm P}^{2}}{(1+8\pi r^{2}p)\left[r^{2}+(\alpha-1)l_{\rm P}^{2}\right]}.
\end{align}
The right-hand side of \eqref{eq:eqpsi} is a third-order polynomial in $\psi$. Now, we can formally solve Eq.~\eqref{eq:consclasica} with the equation of state \eqref{eq:constdens} to yield Eq.~\eqref{eq:solcont}. In this way, replacing Eq.~\eqref{eq:solcont} into Eq.~\eqref{eq:eqpsi} will result in an integro-differential equation for the variable $\psi$ (although this is not the way we are going to solve the system of equations).

The convenience of this formulation comes from Eq.~\eqref{eq:eqpsi} being expressible as a first order differential equation for $\psi$ in several approximate situations. The clearest example being the vacuum case, obtained by taking $p$ and $\rho$ equal to zero in Eq.~\eqref{eq:eqpsi}. The resulting first-order differential equation has two exact analytical solutions
\begin{equation}\label{eq:exactsol}
\psi_{\pm}=-\frac{r^{2}+\alpha l_{\rm P}^{2}}{r l_{\rm P}^{2}}\left(1\pm\sqrt{1-\frac{l_{\rm P}^{2}}{r^{2}+\alpha l_{\rm P}^{2}}}\right),
\end{equation}
which allow us to restrict the region where the solution can take values, in virtue of the Picard-Lindelöf theorem. These vacuum exact solutions also happen to be exact in the general case. This becomes evident after rewriting \eqref{eq:eqpsi} in the form
\begin{equation}\label{eq:psifactsol}
\psi'=\mathcal{F}_{\text{Schw}}+\mathcal{G}\left(\rho,p,\psi,r\right)\left(\psi-\psi_{-}\right)\left(\psi-\psi_{+}\right),
\end{equation}
where $\mathcal{F}_{\text{Schw}}$ is the vacuum-portion of the right-hand side of Eq. \eqref{eq:eqpsi}, and
\begin{equation}
\mathcal{G}=\frac{4\pi l_{\rm P}^{2} r^{2}\left[\rho\left(1+r\psi\right)+p\left(3+r+\frac{2l_{\rm P}^{2}r}{r^{2}+\alpha l_{\rm P}^{2}}\right)\psi\right]}{\left[r^{2}+(\alpha-1)l_{\rm P}^{2}\right]\left(1+8\pi r^{2}p\right)}.
\end{equation}
Therefore, matter-dependent contributions vanish for $\psi=\psi_{\pm}$ in Eq. \eqref{eq:psifactsol}, leaving only the vacuum equation, for which they are exact solutions. Another situation where Picard-Lindelöf theorem can be applied requires assuming pressure to be much larger than the energy density. It will be useful afterwards to notice that, under the assumption $p\gg \rho$, all the $p$-dependence in Eq.~\eqref{eq:eqpsi} disappears, making it a first order differential equation for $\psi$.

\subsection{Semiclassical criticality}

The classical notion of criticality described in section \ref{subsec:criticality} is greatly affected by quantum corrections. 
The vacuum polarization of the scalar field generates a cloud of mass that coats the spacetime, extending to infinity. The semiclassical equivalent to relation \eqref{eq:massrelation} would now have $M_{\text{cloud}}$ defined as
\begin{equation}\label{eq:mcloud}
    M_{\text{cloud}}=\int_{0}^{\infty}dr\,4\pi r^{2}\left[\Theta(R-r)\rho+\rho_{\text{se}}\right],
\end{equation}
where $\Theta$ is the Heavyside step function and \mbox{$\rho_{\text{se}}=-\langle\hat{T}^{t}_{t}\rangle$} denotes the semiclassical energy density. In the vacuum portion of the spacetime, the only contribution to $M_{\text{cloud}}$ is semiclassical. It supplies a negative contribution in such a way that the Misner-Sharp mass grows from its asymptotic ADM value as we approach the surface of the object. Once inside the object, we have semiclassical as well as classical contributions to the density. As in the classical case, the Misner-Sharp mass can be ill-defined at the origin (recall that in the classical case this is exclusively related to the possible presence of an $M_{0}$ offset). The difference now is that the Misner-Sharp mass can fail to approach zero at the origin by different intertwined reasons. It might be that there is a mismatch between the internal mass and the classical density due to the presence of a nonzero $M_{0}$; it might also be that the semiclassical density \eqref{eq:rhosemiL} diverges at the origin; or it might be a combination of both.

As in the semiclassical case the equation for the compactness is intertwined with that of the pressure: given a star radius and compactness, we ignore a priori which value of $\rho$ we should use to find a regular compactness at the radial origin (i.e. a zero Misner-Sharp mass at the origin). As we will see, the situation is even more complicated, as in some important cases there does not exist a value of $\rho$ such that the compactness at the origin vanishes.
What we do find is that there always exists a value $\rho_{\rm c}$ of $\rho$ separating two rather different qualitative behaviors for the compactness. Therefore, in general terms we will say that a configuration is critical when its density is such that it represents a separatrix between these two regimes.

Having posed these difficulties and a definition of criticality, we now proceed to analyze the semiclassical set of solutions. First, we will study configurations with a regular origin, in the same spirit as we did in the classical analysis of cosmological solutions. Solutions with different sorts of irregularity will be analyzed in detail in the sections that follow.

\section{Semiclassical stellar-like solutions}
\label{section:semiclassicalsols}

The introduction of the RSET as an additional source of curvature makes exploring the space of stellar solutions of the semiclassical equations a more subtle task than in the classical theory. This difficulty can be attributed, in part, to the new length scale $l_{\rm P}$ that makes the physics of solutions sensitive to the overall size of the star. In this section we address every solutions belonging to the semiclassical sector of Figs. \ref{fig:table} and \ref{fig:tablefigs}.

\begin{figure}
    \centering
    \includegraphics[width=\columnwidth]{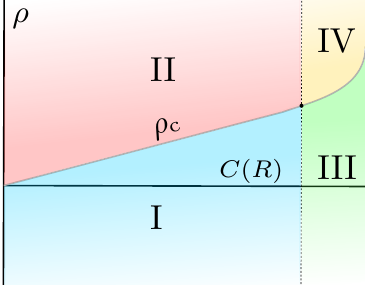}
    \caption{Pictorial representation of an $R\gg l_{\rm P}$ slice of the phase space of semiclassical constant-density stars. The vertical and horizontal axes represent the energy density and the surface compactness of stars. The curve $\rho_{\rm c}$ corresponds to a separatrix solution. The vertical dotted line here denotes the Buchdahl limit, in which the central black dot represents the most compact configuration strictly regular in both pressure and compactness, or Buchdahl solution. We distinguish four regions in the resulting figure: region I represents sub-critical sub-Buchdahl stars, region II is for super-critical sub-Buchdahl, region III is for sub-critical super-Buchdahl; and region IV represents super-critical super-Buchdahl stars. In the sub-Buchdahl semiplane (left-hand side of the vertical dotted line), the separatrix $\rho_{\rm c}$ corresponds to strict stellar spacetimes. For super-Buchdahl configurations this separatrix correspond to non-regular solutions, but in its neighbourhood we find quasi-regular configurations (i.e. \emph{$\epsilon$-strict} stellar configurations).}
    \label{fig:phasespace}
\end{figure}

Considering stars whose surface is outside the neck, Figure \ref{fig:phasespace} shows a pictorial representation of an $R\gg l_{\rm P}$ slice of the space of solutions. We distinguish four differentiated regions depending on whether the star is sub- or super-critical, and on whether its compactness surpasses Buchdahl limit or not. 
The central black dot represents the most compact configuration that is regular in both compactness and pressure.
We have observed that by increasing the density while decreasing the radius accordingly, this point can be moved towards higher values of the compactness. Here, we refer to this compactness bound as the Buchdahl bound for semiclassical stars sourced by the specific regularization of the Polyakov RSET that we are using. For this particular regularization, stars with large (stellar-like) radius and mass show a Buchdahl limit that corresponds to a perturbative correction over the classical compactness bound of $C(R)=8/9$. From now on, we distinguish between sub-Buchdahl or super-Buchdahl stars attending to this bound. This will be useful to divide the space of solutions in different regions, as in Fig. \ref{fig:phasespace}, although the reader should take into account that this definition has only an operational meaning, and cannot be directly identified as a Buchdahl limit in semiclassical gravity. In particular, defining this limit, which may exist or not, may require a better regularization of the Polyakov RSET. We are interested in probing whether it is possible to obtain regular or quasi-regular configurations largely surpassing the Buchdahl limit. Particularly, we will aim at the rightmost portion of the diagram \ref{fig:phasespace}, or ultra-compact limit, where semiclassical corrections meet the conditions to become comparable in magnitude to that of the classical SET, thus potentially inducing significant departures from the classical solutions. In what follows we will obtain the specific form of the solutions for all four regions, together with the form of the separatrix solutions $\rho_{\rm c}$.

We now turn to the analysis of integrations from the asymptotically flat region towards the center of the star. This treatment allows to better probe how the RSET acts in response to changes in the surface compactness and the classical density parameter used in the integrations. The boundary conditions required at the surface of the star follow from the classical ones \eqref{eq:boundcondclas}, with an extra condition for $\phi'$,
\begin{align}\label{eq:boundcond}
p\left(l_{\rm S}\right)
&
=0,\quad r\left(l_{\rm S}\right)=R,  \\
\phi\left(l_{\rm S}\right)
&
=\phi_{\rm S},\quad r'\left(l_{\rm S}\right)=\mp\sqrt{1-C(R)}.
\nonumber\\
\phi'\left(l_{\rm S}\right)
&
=\frac{R^{2}+\alpha l_{\rm P}^{2}}{R l_{\rm P}^{2}}\left[\sqrt{1+\frac{R^{2}}{R^{2}+\alpha l_{\rm P}^{2}}\frac{C(R)}{1-C(R)}}\pm1\right]\nonumber\\
&
~~\times\sqrt{1-C(R)},\nonumber
\end{align}
where $C(R)$ is the value of the compactness at the surface of the star.
Here, the $\pm$ signs select the side of the wormhole where the surface of the star is located. We choose the $+$ sign in $r'$ and the $-$ sign in $\phi'$ for stars whose surface lies outside the neck, and vice versa if the star surface is located inside the neck. Any other sign combination is not compatible with a stellar spacetime.

The above boundary conditions can be inserted in the semiclassical field equations to study how the RP-RSET behaves at the surface of stars in the $C(R)\to1$ limit. Computing the RSET over the classical background of the Schwarzschild interior solution causes that both the semiclassical energy density and pressure diverge at the surface of the star $R$ in the limit $C(R)\to 1$. This divergence appears both from the interior region of the star, where $\rho$ is constant and positive, and from the exterior, vacuum portion, where $\rho=0$. This is so because this limit corresponds to locating the surface on top of the event horizon, where the Boulware state is, by definition, singular.

In a self-consistent approach, on the contrary, the RSET backreacts on the metric and there is no horizon. Instead, we encounter a wormhole neck where the RSET components are finite and have no trace of divergences. Starting from Eqs. (\ref{eq:ttsemi},~\ref{eq:llsemi}), the RP-RSET components at the surface of a star in the $C(R)\to1$ limit are
\begin{align}\label{Eq:RSETvac}
p^{\text{r}}_{\text{se}}=
&
~\langle\hat{T}^{l}_{l}\rangle^{(\rm{RP})}=
-\frac{1}{8\pi R^{2}}+\order{\sqrt{1-C}},\nonumber\\
p^{\theta}_{\text{se}}=
&
~\langle\hat{T}^{\theta}_{\theta}\rangle^{(\rm{RP})}=
-\frac{\alpha l_{\rm P}^{2}}{8\pi R^{2}\left(R^{2}+\alpha l_{\rm P}^{2}\right)}+\order{\sqrt{1-C}},\nonumber\\
\rho_{\text{se}}=
&
-\langle\hat{T}^{t}_{t}\rangle^{(\rm{RP})}=
-\frac{1}{8\pi R^{2}}+\rho+\order{\sqrt{1-C}}.
\end{align}
Every component of the RP-RSET is finite and negative at the surface. In the ultracompact limit, the radial pressure and the energy density are able to compensate their $\order{l_{\rm P}^{2}}$ suppression, becoming comparable to the classical SET components. The finite jump in $\rho$ at the surface of the fluid sphere contributes positively to the semiclassical energy density. Consequently, the total energy density (the sum of classical and semiclassical contributions) will be positive at the surface given that 
\begin{equation}
\rho>\frac{1}{16\pi R^{2}}.   
\end{equation}
This result comes from a local analysis at the surface; the particular form of the RP-RSET at the bulk relies heavily on the classical pressure and density profiles. We expect more accurate approximations to the RSET to extend these negative semiclassical contributions to the interior of the star as well. For more realistic equations of state with vanishing energy density at the surface, the complete SET (the sum of the classical and quantum portions) violates all energy conditions at the surface of ultracompact stars.

Now, picture a numerical integration starting at the asymptotically flat region with a positive ADM mass. While in vacuum, compactness increases monotonically until the neck as in Fig. \ref{fig:wormhole}, and we can decide to locate the surface of the perfect fluid either outside or inside it. At the surface, compactness and radius are fixed, leaving the energy density $\rho$ as the only free parameter. With the aim of constraining the various possibilities embraced by the diagram in Fig. \ref{fig:phasespace}, we will first describe the behavior of stars situated at regions I and III in the diagram (sub-critical regime) and at regions II, IV (super-critical regime). Since stellar geometries should connect with the vacuum solution in the $\rho \to 0$ limit, we can always devise a star of any compactness that belongs to the sub-critical regime. Similarly, given a star with any $C(R)<1$, the super-critical regime is explored by increasing $\rho$ beyond the critical density. This classification is valid for stars located either outside or inside the neck, so we proceed by first investigating the former. Our results and acquired intuitions will extend to the study of the latter situation as well.

\subsection{Solutions with a regular center}
\label{sec:S-c-subB}

From the complete set of solutions, we want to extract first those solutions which are strict stellar configurations, i.e. which have a regular radial center. Recall that these configurations correspond to the critical sub-Buchdahl solutions in Figs. \ref{fig:table} and \ref{fig:tablefigs}. To obtain regular solutions to the semiclassical equations of structure, we proceed by performing numerical integrations from a regular origin. The following boundary conditions must be imposed at $r=0$ to integrate Eqs. \eqref{eq:consclasica}, \eqref{eq:ttsemi} and \eqref{eq:llsemi}:
\begin{align}\label{eq:regcondsemi}
r(0)=
&
~0,\quad \phi(0)=\phi_{\rm c},\quad p(0)=p_{\rm c}\nonumber\\
r'(0)=
&
~1,\quad \phi'(0)=0.
\end{align}
Integrations from a regular origin share many features with their classical counterparts. Given that the choice of $\phi_{\rm c}$ represents just a rescaling of time coordinate, the full space of solutions with regular origin is determined by the two-parameter set $(\rho, p_{\rm c})$.

\begin{figure*}
    \centering
    \includegraphics[width=0.99\textwidth]{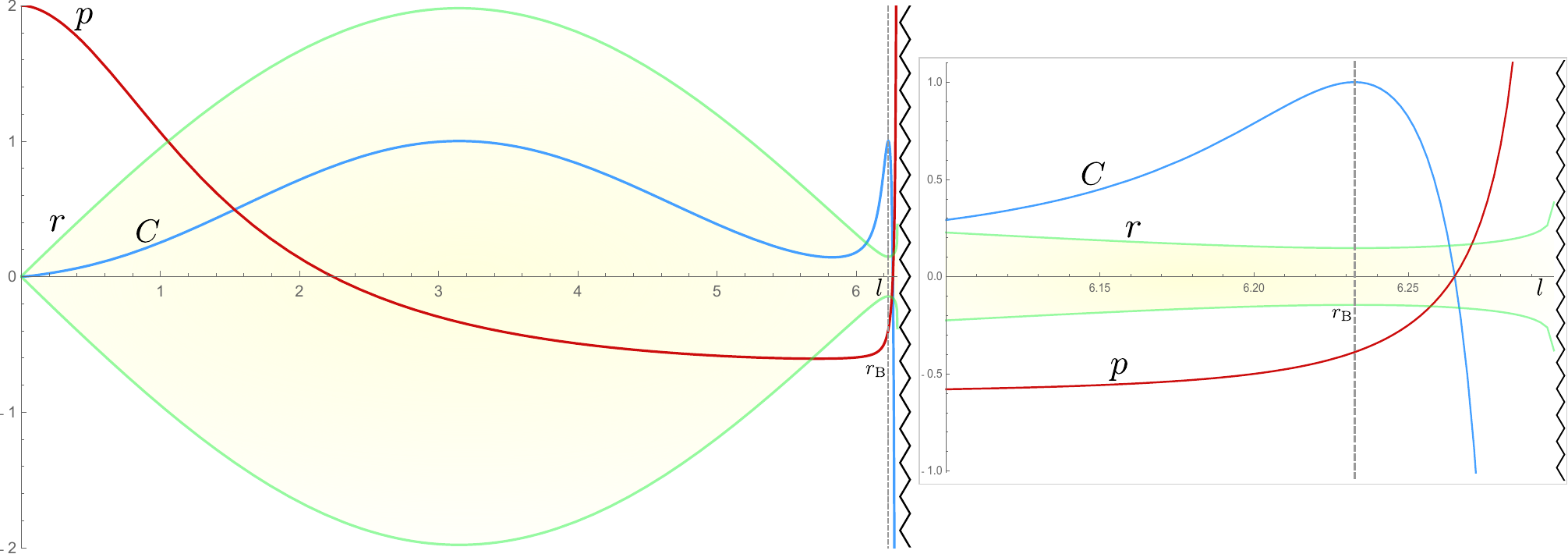}
    \caption{Plot of the semiclassical counterpart of figure \ref{fig:critclas} with $\rho=0.03$, $p_{\rm c}=2\rho$ and $\alpha-1=10^{-3}$. We have plotted the functions $r(l), C(l)$ and $p(l)$ (in units of $\rho$) and they appear in green, blue and red, respectively. The right pole of the geometry is shown in detail. Notice how the radial function has a minimal surface (vertical dashed line) at $l\sim 6.23 $ and the geometry connects to a singular region (vertical zigzag line) located at $r\to\infty$ but at finite proper distance from $l=0$.}
\label{fig:p2omegasemi}
\end{figure*}

\begin{figure}
    \centering
    \includegraphics[width=\columnwidth]{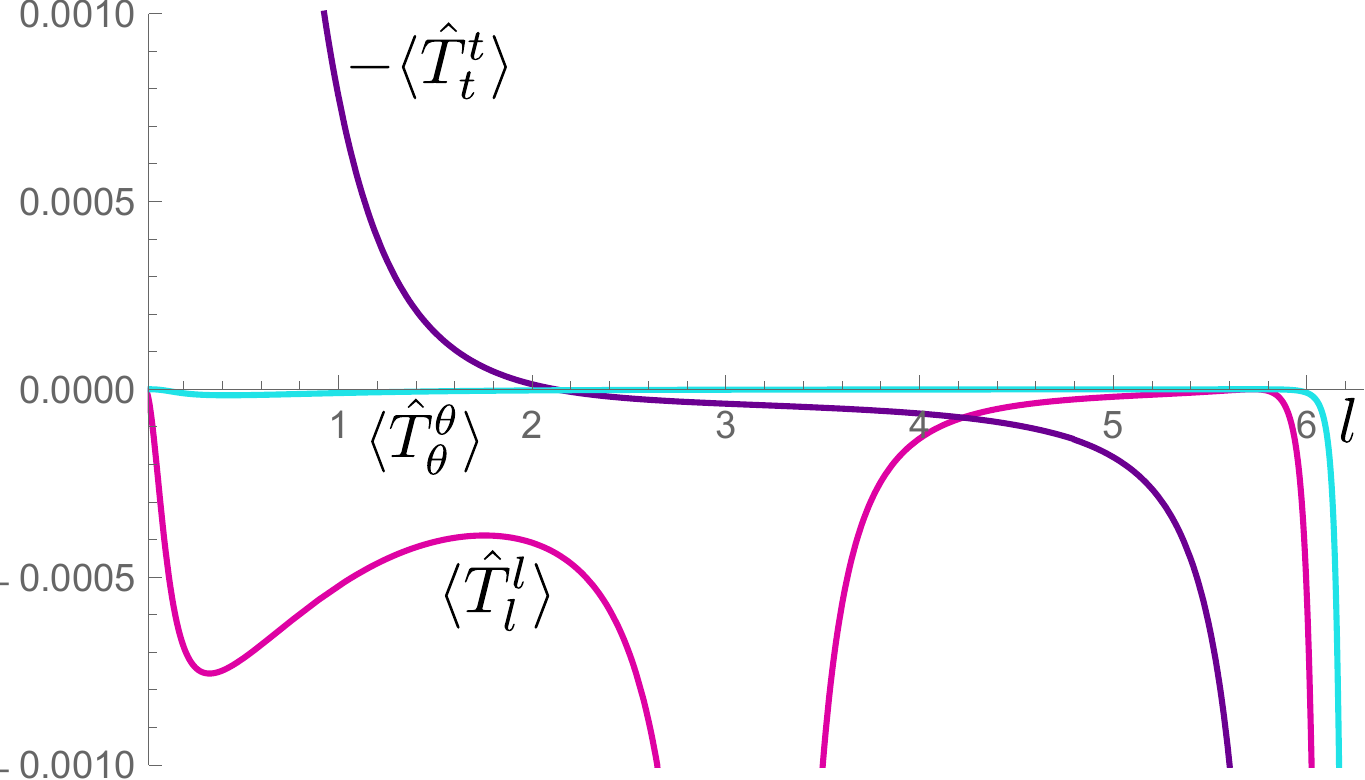}
    \caption{Plot of the RP-RSET components $-\langle\hat{T}^{t}_{t}\rangle$ (dark blue), $\langle\hat{T}^{l}_{l}\rangle$ (magenta) and $\langle\hat{T}^{\theta}_{\theta}\rangle$ (cyan) for the cosmological solution solution with $p_{\rm c}=2\rho$, $\rho=0.03$ and $\alpha-1=10^{-3}$.}
\label{fig:RSETp2omega}
\end{figure}

Depending on the relative values of $p_{\rm c}$ and $\rho$, three families of solutions are found, the separatrices between them corresponding, as in the classical case, to $p_{\rm c}/\rho=-1/3$ and $p_{\rm c}/\rho=-1$. In this section we focus on the semiclassical equivalent to the type 1 set of cosmological solutions, for which the NEC and SEC hold at $r=0$. A specific example has been plotted in Fig. \ref{fig:p2omegasemi} (see Fig. \ref{fig:RSETp2omega} for details on the RSET). Recall from section \ref{sec:C-c-subB} that the positive-pressure portion of type 1 solutions (i.e. those with $p_{\rm c},~\rho>0$) corresponds to  stellar spacetimes. This characteristic persists in the semiclassical theory, so we dedicate this section to exploring the semiclassical counterparts to these critical stellar spacetimes. Furthermore, extending the perfect fluid beyond the surface of zero pressure allows to find the semiclassical counterparts to the classical cosmological solutions. In this section, by counterparts, we are referring to the pair of classical and semiclassical solutions with the same $\rho$ and $p_{\rm c}$.

Firstly, we are going to describe the characteristics of the solutions that we have been able to find through numerical integrations. Unfortunately, this covers a quite limited range of initial conditions. This is so because of the numerical precision required to handle highly different scales. In the semiclassical approximation, the scale of semiclassical corrections is suppressed by $\l_{\rm P}$, and has to be resolved with the scale of typical compact objects, of the order of kilometers.
Thus, the results described in this section need to be extrapolated with care to stars of astrophysical size. An additional warning is that, as we will argue, some of the conclusions that one might extract from these solutions are opposite to those one might expect using more realistic astrophysical numbers and more refined approximations to RSET at the origin. 
With this caveat in mind, let us describe the characteristics of the numerical solutions and then what appropriate conclusions one can extract from them.

Figure \ref{fig:p2omegasemi} depicts the semiclassical counterpart to the classical cosmology from Fig.  \ref{fig:critclas}, with $\rho=0.03$ and $p_{\rm c}=2\rho$. Restricting ourselves to the positive pressure portion in Fig. \ref{fig:p2omegasemi}, we observe that the RSET contributes positively to the mass of the star on average (see the purple curve in Figure \ref{fig:RSETp2omega}). For a star that fulfills the regularity conditions \eqref{eq:regcond} and satisfies the SEC and NEC (the pressure is maximal at $r=0$), the semiclassical energy density \eqref{eq:rhosemiL} is positive at the center,
\begin{equation}\label{Eq:rhosemiorigin}
\rho_{\text{se}}=\frac{\lambda}{4\pi \alpha\zeta}>0,
\end{equation}
its magnitude being inversely proportional to the value of the regulator.  As a consequence of this we find that, as long as density remains within non-Planckian values, these semiclassical stars ---stars very small in astrophysical terms but still with non-Planckian classical densities--- are slightly less compact than their classical counterparts. 
All these stars are sub-Buchdahl and are more sub-Buchdahl than their classical counterparts. We have obtained the maximum compactness of strict regular spacetimes in terms of the density $\rho$. This curve always remains below the classical Buchdahl limit $C(R)=8/9$ for small densities (in the range of densities explored numerically). As $\rho$ increases, configurations that surpass the classical Buchdahl limit are obtained, but these remain outside the regime of validity of the semiclassical approximation as their density is trans-Planckian.

When these low-density stars are analyzed as integrations from the surface inwards, we find that, for counterparts of the same $R$ and $C(R)$, semiclassical critical stars happen to be less dense than classical critical stars, the remaining mass being supplied by the RSET so that relation \eqref{eq:massrelation} is fulfilled. This under-density then results in the classical fluid perceiving an amount of mass greater than the one generated by its own classical energy density and pressures, needing to reach central classical pressures greater than in the classical case to retain equilibrium. 

This result is counterintuitive with respect to initial expectations that one may have regarding semiclassical effects. Reasonably, we would have expected the total semiclassical energetic contribution to a star to be negative. In fact, the value of the semiclassical contributions to the local density when crossing the surface of the star is negative. This negativity increases as $C(R)$ approaches $1$, but decreases as the classical energy density is raised. At this point, we have two issues at stake. On the one hand, for sub-Buchdahl stars, the negativity of the semiclassical contribution is very small (it is suppressed by $l_{\rm P}$ and it is not amplified by the surface of the star being close to its gravitational radius). On the other hand, there is a strong dependence on the behavior of the RSET at the origin. In our approach, the regulating scheme for the RSET comes as a cutoff to the total magnitude of the RSET at the origin. Setting the value of $\alpha$ so that the RP-RSET is very suppressed, this suppression applies to the entire interior, diminishing also the RP-RSET at the surface of the star. In fact, in the limit $\alpha \to \infty$ one eliminates completely any semiclassical contribution, thus recovering the classical solutions.
On the other extreme, if we take $\alpha \sim 1$, then the RP-RSET at the origin of sub-Buchdahl regular stars is not suppresed by $l_{\rm P}$ and can lead to very large and positive semiclassical densities [as in Eq. \eqref{Eq:rhosemiorigin}]. Then, in all the numerical solutions, the central positive contribution to the semiclassical energy widely outstrips the mild negative energies at the surface, if any. To avoid these problems, one would need to consider sufficiently large stars so that there exists room to fix the regulator in a way that only affects the core of the star without affecting the surface. In addition, ideally, one would like to design a regulator bringing the RP-RSET close to an exact RSET. This better behaved RSET would be sensitive to the local characteristics of the geometry at the origin and so able to properly capture the physics close to the radial origin. For instance, for regular sub-Buchdahl configurations one expects the RSET to be also small at the origin as neither large curvatures nor horizons are present through the configuration. Notice that from these
arguments alone it is not straightforward to say anything about the Buchdahl limit itself.

In any case, the analysis reported here is valuable in clearly illustrating the limitations and strengths of the Polyakov and Regularized Polyakov RSETs. It is reasonable to expect that the RP-RSET should be a trustworthy approximation when the physics is driven by non-local effects generated at values of the radius close to where a horizon would have been classically located. On the contrary, it should not provide a reliable approximation when the physics is driven by the values of the RSET at the origin. This motivates our definition of \emph{$\epsilon$-strict} spacetimes as the solutions of relevance for extracting robust conclusions, since the behavior of any solution close to the origin is necessarily impacted by the choice of regulator. As the regular solutions described in this section are a subset of the \emph{$\epsilon$-strict} spacetimes, it cannot be assumed that these provide a typical description of the properties of this larger set of solutions. Nevertheless, the existence of a set of non-regular but \emph{$\epsilon$-strict} solutions provides further motivation to analyze alternative regularizations of the Polyakov RSET, exploiting the available freedom discussed mentioned in Sec.~\ref{sec:rset}, which may be sufficient to regularize these solutions as well.

\subsubsection{Cosmological solutions}

For completeness, as we did in the classical case, let us mention some particularities of the cosmological solutions, independently of whether they can be used as regular stellar interiors or not. Coming back to Fig. \ref{fig:p2omegasemi}, we observe that the resulting ``cosmology" never reaches its would-be right pole. This difference with respect to the classical cosmology from Fig. \ref{fig:critclas} comes from the aforementioned semiclassical contribution to $M_{\text{cloud}}$. In an outwards integration starting at the origin, the semiclassical energy density giving rise to such contribution begins as positive and changes sign eventually. In Fig. \ref{fig:RSETp2omega} we observe that the semiclassical energy density grows as the origin $r=0$ is approached, so that its weight at short distances is very significant. The overall effect of this mass cloud is to prevent the cosmology from being regular at its right pole. As this region is approached, the solution shows a minimal surface or neck that connects to an asymptotic, negative mass singularity, in the same fashion as in the vacuum solution, but now in presence of perfect fluid with divergent pressure. As an additional comment note that, although pressure has a second zero close to the neck, this surface does not connect with the Schwarzschild vacuum geometry in a way that resembles a stellar spacetime.

Now, decreasing $p_{\rm c}$ below zero results in configurations qualitatively similar to Fig. \ref{fig:p2omegasemi}, but with pressure everywhere negative in between the center and the neck.
Taking $p_{\rm c}=-\rho/3$ results in an Einstein static universe, which receives no semiclassical corrections whatsoever: the RP-RSET is identically zero. Going below this separatrix for $p$ changes the sign of the pressure gradient outside the radial origin, so that $p$ increases outwards. For $-2\rho/3\lesssim p_{\rm c}<-\rho/3$ the obtained cosmologies show no neck. Instead, a second $r=0$ is reached in a singular manner. This is so because the contribution to the Misner-Sharp mass that comes from the RP-RSET is now negative overall. In consequence, the solution tends to the semiclassical counterpart of the Schwarzschild geometry with negative asymptotic mass as the second $r=0$ surface is approached.

Taking $p_{\rm c}\lesssim-2\rho/3$ causes the neck to reappear, leading, once again, to an asymptotic singularity at radial infinity. This singularity moves towards smaller $l$ as $p_{\rm c}$ decreases. When the NEC is saturated, the divergence has engulfed the radial maximum and the shape function increases monotonically from $r=0$ outwards. Henceforth, all configurations show a negative-pressure divergence at $r\to+\infty$. Note that, owing to the curvature singularity at infinite $r$, these profiles cannot resemble the interior portion of gravastar solutions anymore since their shape functions do not match continuously with those of the positive-pressure portion of super-Buchdahl stars. 

In summary, the semiclassical counterparts to these cosmological spacetimes have acquired features from configurations with non-regular compactness profiles as far as the behavior of the putative right-hand-side pole is concerned.
This is due to the imbalance in mass that originates from quantum corrections as encapsulated in the Regularized Polyakov RSET being used. Thus, solutions with non-regular compactness solutions are important in the study of cosmologies with one regular center. In the next sections we derive the properties of solutions with irregular (non-critical) compactness in detail, using the notion of semiclassical criticality to catalogue them.

\subsection{Sub-critical configurations}
\label{sec:S-subc-subB}

We begin by considering a star with compactness well below the Buchdahl limit and density well below $\rho_{\rm c}$ (we are referring to the region I from Fig. \ref{fig:phasespace}). Taking $\rho=0$ we recover the vacuum solution, which has a wormhole neck at some radius $r_{\rm B}\gtrsim2M_{\text{ADM}}$ (the suffix $\rm B$ stands for a bouncing surface of the shape function). By matching the vacuum solution with the surface of a constant-density configuration with small, positive $\rho$ at some $R>r_{\rm B}$, the interior geometry resembles the vacuum solution (in the sense that it develops a wormhole neck in the interior) but with a perfect fluid added to it. Recall that, as we saw for  the cosmological solution from Fig. \ref{fig:p2omegasemi}, wormhole necks can appear in the presence of matter. In this section we prove this statement and obtain analytical approximations to this wormhole geometry in certain regimes: around the neck and in the singular asymptotic region deep inside the neck. The effect of increasing $\rho$ is to approach the critical solution $\rho_{\rm c}$ in the space of solutions from Fig. \ref{fig:phasespace}, pushing the wormhole neck [i.e. a surface where $C(r_{\rm B}=1$)] to smaller values of $r$ until it disappears for some $\rho_{\rm{c}}$.
Here, all solutions showing a wormhole neck will be called sub-critical. From a critical value of the density upwards (super-critical regime), we find that the geometries do not longer have a neck, having their shape functions extended until $r=0$. The separatrix solution sits, obviously, between both regimes.

\begin{figure*}
\centering
\includegraphics[width=0.8\linewidth]{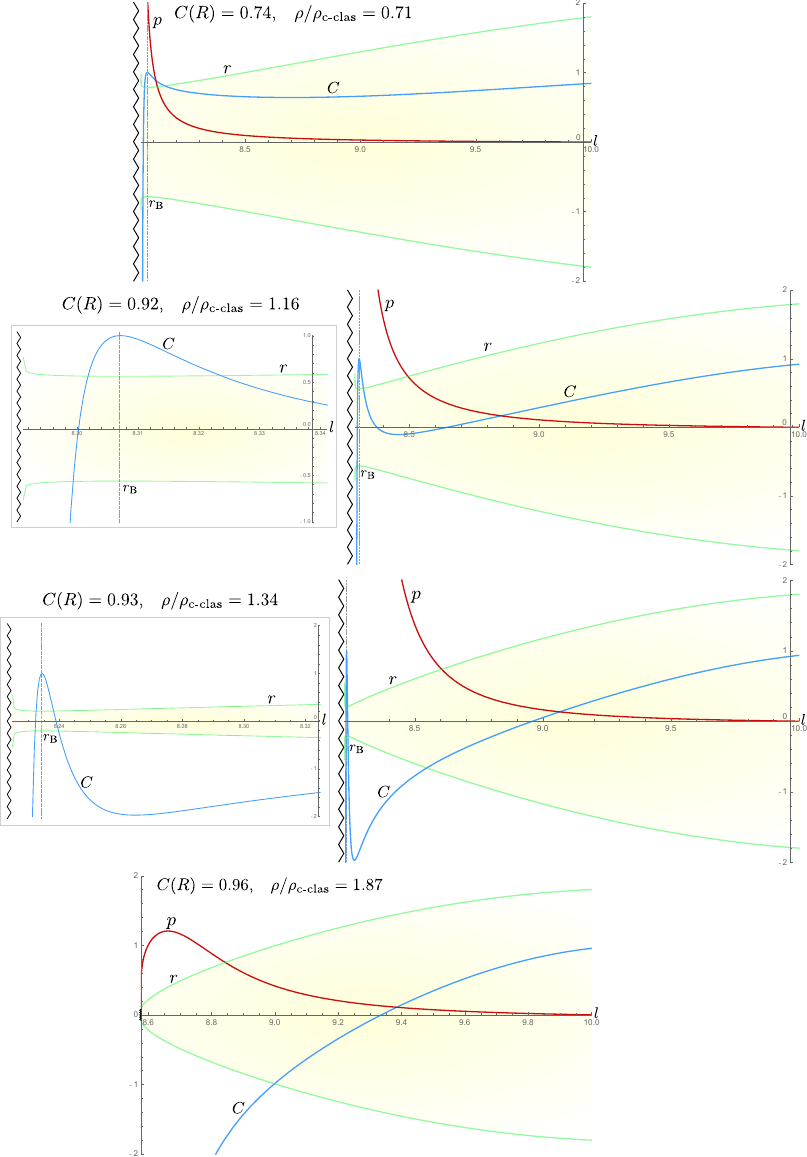} 
\caption{Semiclassical stars integrated from the surface. The green and blue curves denote $r(l)$ and $C(l)$, and the red curve represents the function $p(l)$. All integrations correspond to stars with $R=1.8$ and $\alpha-1=10^{-3}$. Their surface compactness and their $\rho/\rho_{\rm{c}\text{-clas}}$ quotients are, approximately and from top to bottom: $(0.84,0.71)$, $(0.92, 1.16)$, $(0.93, 1.34)$ and $(0.96, 1.87)$. The second and third panels show a zoomed plot of the near-neck region, highlighting the neck (vertical dashed line) and the singularity (zigzag line). Increasing $\rho$ generates a well of negative mass. This negative mass slows down the increase in pressure, causing a shrinkage of the wormhole neck. Eventually, the neck disappears leaving a naked singularity at $r=0$. In between sub-critical and super-critical configurations there is an infinite pressure separatrix solution. As the wormhole neck can be as small as desired by adjusting $\rho$, a mild deformation of the geometry at the core would suffice to make the whole construction regular. See Fig. \ref{fig:RSETnoncritsemi} for the RP-RSET components from each solution.}
\label{fig:noncritsemi}
\end{figure*}

\begin{figure*} 
  \centering
\includegraphics[width=\linewidth]{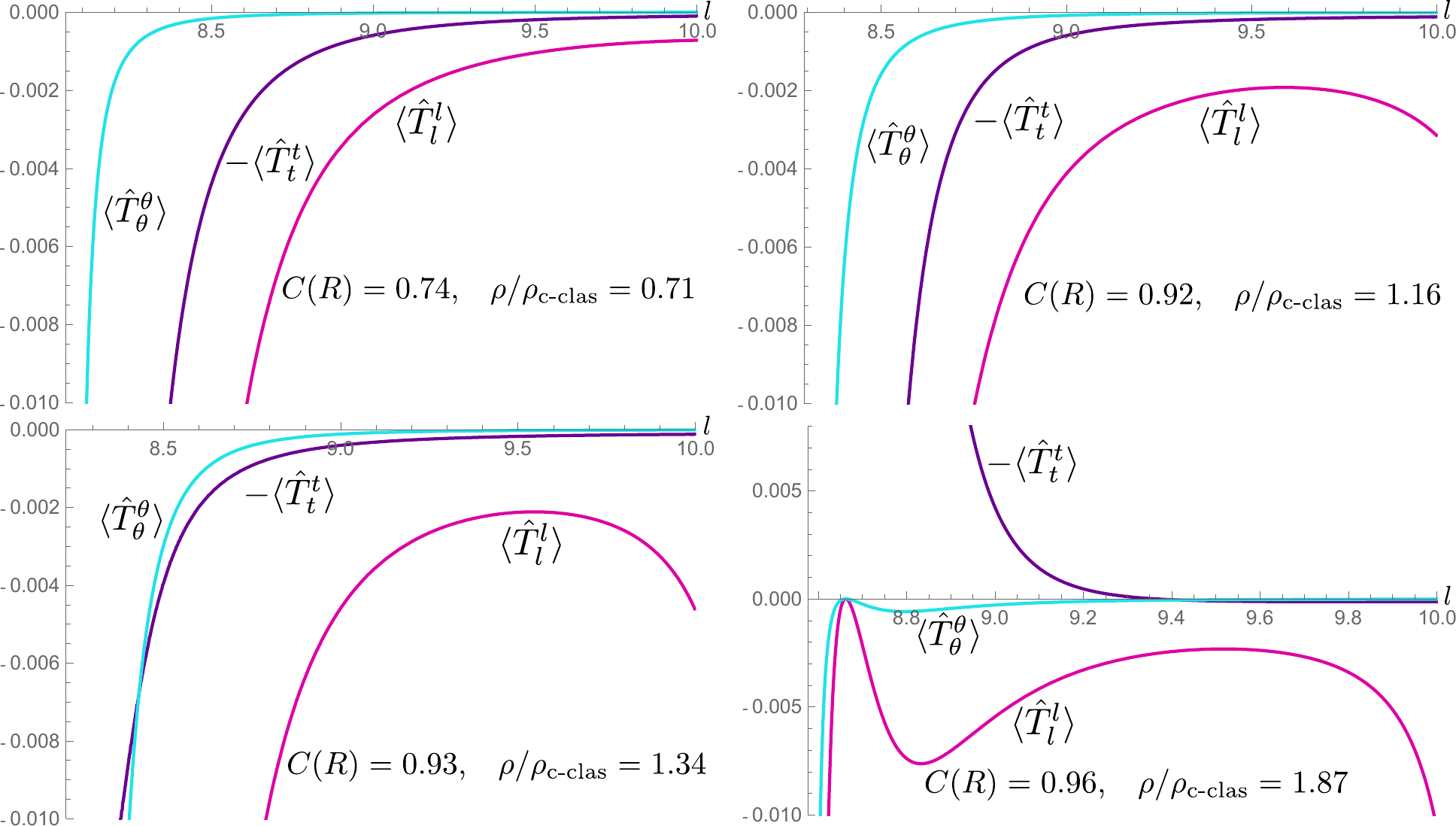}
\caption{RSET components $-\langle\hat{T}^{t}_{t}\rangle$ (dark blue), $\langle\hat{T}^{l}_{l}\rangle$ (magenta) and $\langle\hat{T}^{\theta}_{\theta}\rangle$ (cyan) for various stars integrated from the surface. All the integrations correspond to stars with $R=1.8$ and $\alpha-1=10^{-3}$.  They correspond to the solutions appearing in Fig. \ref{fig:noncritsemi}, whose surface compactness and $\rho/\rho_{\rm c \text{-clas}}$ quotients are, approximately: $(0.84,0.71)$ (top left), $(0.92, 1.16)$ (top right), $(0.93, 1.34)$ (bottom left), $(0.96, 1.87)$ (bottom right). Note the abrupt change in the sign of the semiclassical energy density in the transition from the sub-critical to the super-critical regime.}
\label{fig:RSETnoncritsemi}
\end{figure*}

\begin{figure}
\centering
\includegraphics[width=\columnwidth]{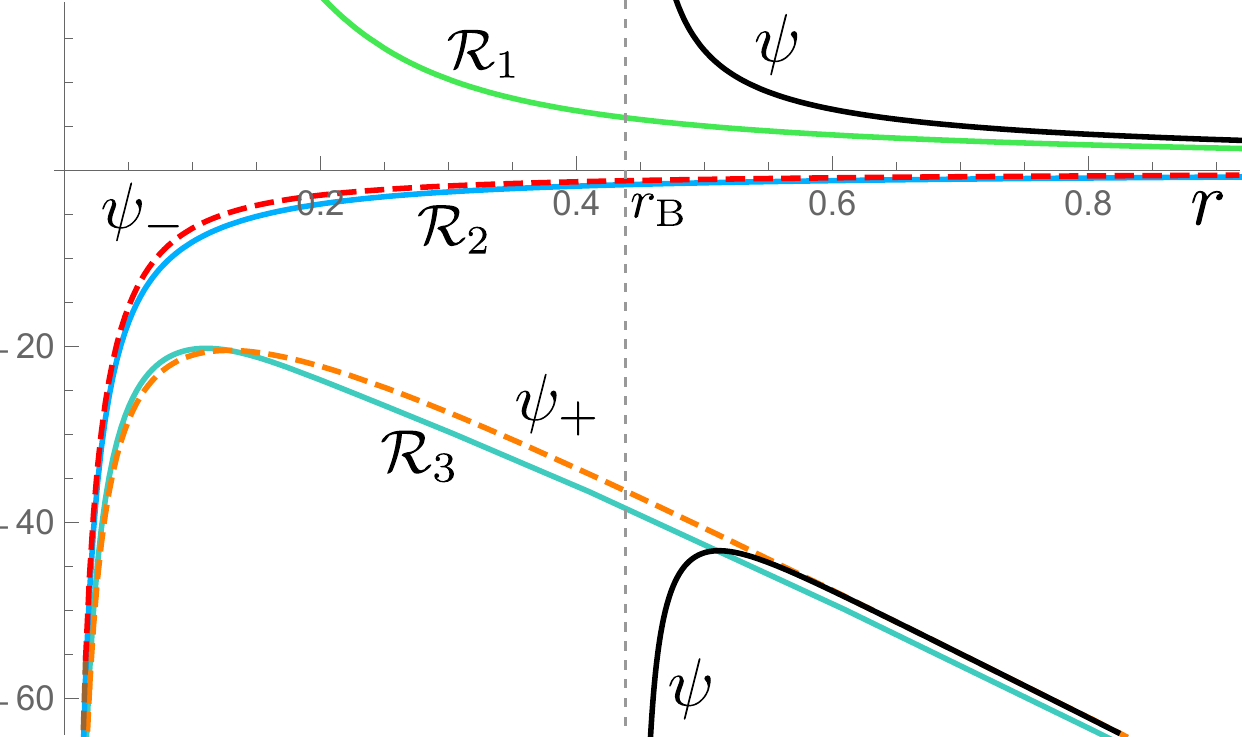}
\caption{Numerical plot of the roots $\mathcal{R}_{1},\mathcal{R}_{2}$ and $\mathcal{R}_{3}$ (green, blue and turquoise curves, respectively) together with the exact solutions $\psi_{\pm}$ (orange and red dashed curves) and an exact numerical solution in black (the neck radius $r_{\rm B}$ is represented by a vertical dashed line). The numerical solution corresponds to a sub-critical super-Buchdahl star with $R=2, C(R)=0.95$ and $\rho/\rho_{\rm c \text{-clas}}\simeq1.67$ with its neck at $l\simeq0.45$ (vertical dashed line). These values have been chosen to aid visualization. The upper portion of the exact solution lives in the unconcealed branch, whereas the bottom portion lives in the concealed branch. The concealed part of the exact solution gets confined between $\mathcal{R}_{3}$ and $\psi_{+}$, converting towards the vacuum solution asymptotically. }
\label{fig:roots}
\end{figure}

Let us consider in more detail the form of configurations belonging to the sub-critical regime and whose surface is located outside the neck. The first three panels in Fig.~\ref{fig:noncritsemi} describe configurations of this kind (see Fig.~\ref{fig:RSETnoncritsemi} for details on the RSET components). In virtue of Eq. \eqref{eq:TOVsemi}, pressure grows monotonically inwards as long as the squared root term is greater than unity. Similarly to the classical sub-critical case, the compactness function $C$, which decreases as we move away from the surface inwards, encounters a minimum value somewhere in the bulk of the configuration, triggering a runaway in the pressure. Restricting ourselves to the regime where the expression for the pressure in Eq.~\eqref{eq:solcont} can be well approximated by
\begin{equation}\label{eq:presapprox}
    p\simeq\kappa e^{-\int\psi dr},
\end{equation}
we find that Eq. \eqref{eq:eqpsi} is approximated by a first-order differential equation of the form
\begin{equation}\label{eq:eqpsipresion}
\psi'=\mathcal{H}\left(\psi-\mathcal{R}_{1}\right)\left(\psi-\mathcal{R}_{2}\right)\left(\psi-\mathcal{R}_{3}\right),
\end{equation}
where 
\begin{equation}
\mathcal{H}=-\frac{l_{\rm P}^{2}r}{2\left[r^{2}+l_{\rm P}^{2}\left(\alpha-1\right)\right]}\left[1-\frac{2 l_{\rm P}^{2}r^{2}}{\left(r^{2}+\alpha l_{\rm P}^{2}\right)^{2}}\right]
\end{equation}
and $\left\{\mathcal{R}_{i}\right\}_{i=1}^{3}$ are three roots with involved and lengthy expressions that depend on $r$, $\alpha$ and $l_{\rm P}$. Their approximate asymptotic forms  for $r\gg \sqrt{\alpha} l_{\rm P}$ are
\begin{align}
\mathcal R_{1,2}\simeq
&
\frac{3\pm\sqrt{33}}{4r},\quad\mathcal{R}_{3}\simeq
-\frac{2r}{l_{\rm P}^{2}}.
\end{align}
These roots appear plotted in Fig. \ref{fig:roots} alongside $\psi_{\pm}$ as defined in Eq.~\eqref{eq:exactsol}, and an exact numerical solution belonging to the sub-critical regime. While $\mathcal{R}_{1},~\mathcal{R}_{2}$ are monotonic, $\mathcal{R}_{3}$ reaches a maximum value precisely where the $\psi_{+}$ exact solution intersects $\mathcal{R}_{3}$. This observation will guide us in what follows since, as long as Eq. \eqref{eq:eqpsi} is well-approximated by a first-order differential equation, the shape of the solution $\psi$ is determined by $\psi_{\pm}$ and $\left\{\mathcal{R}_{i}\right\}_{i=1}^{3}$.

The approximate expression \eqref{eq:presapprox} implies that $\phi$ diverges towards negative values, for which its derivative $\psi$ needs to diverge towards $+\infty$ at some radius $r_{\rm B}$. Therefore, the right-hand side in Eq. \eqref{eq:eqpsipresion} can be approximated to cubic order in $\psi$. By solving this approximate equation and expanding the solution in the limit $r\to r_{\rm B}$ we find
\begin{equation}\label{eq:psirb}
    \psi\simeq\pm\sqrt{\frac{k_{0}}{4(r-r_{\rm B})}},
\end{equation}
with 
\begin{align}
k_{0}
&
=\frac{2\left[r_{\rm B}^{2}+(\alpha-1)l_{\rm P}^{2}\right]\left(r_{\rm B}^{2}+\alpha l_{\rm P}^{2}\right)^{2}}{r_{\rm B}l_{\rm P}^{2}\left[(r_{\rm B}^{2}+\alpha l_{\rm P}^{2})^{2}-2r_{\rm B}^{2}l_{\rm P}^{2}\right]}>0.
\end{align}
Expression \eqref{eq:psirb} shows that the modifications induced by the RSET change the rate at which a surface of zero redshift is approached. Classically, the Schwarzschild horizon is approached as $\psi\propto (r-r_{\rm H})^{-1}$. Due to the increase in order of the $\psi$ terms in Eq.~\eqref{eq:eqpsi} coming from semiclassical corrections, the classical Schwarzschild horizon is no longer part of the solution. Instead, we find that $\psi$ grows more slowly in the semiclassical theory, the precise form of Eq.~\eqref{eq:psirb} being integrable across the surface $r=r_{\rm B}$. The latter represents an asymmetric wormhole neck, where the shape function $r$ reaches a minimum value. Integrating Eq.~\eqref{eq:psirb} and returning to the $l$ coordinate, which is regular through the neck, the approximate behavior of the metric functions obtained is
\begin{equation}\label{eq:neckmetric}
r\simeq \frac{k_{1}}{4}\left(l-l_{\rm B}\right)^{2}+r_{\rm B},\quad \phi\simeq\frac{\sqrt{k_{0}k_{1}}}{2}\left(l-l_{\rm B}\right)+\phi_{\rm B},
\end{equation}
where $l_{\rm B}$ and $\phi_{\rm B}$ are the values of the proper coordinate and the exponent of the redshift function at the neck, and
\begin{equation}
k_{1}
=\frac{4\left(r_{\rm B}^{2}+\alpha l_{\rm P}^{2}\right)}{r_{\rm B}^{2}l_{\rm P}^{2}k_{0}}>0.
\end{equation}
Replacing these expressions in Eq. \eqref{eq:presapprox}, we see that the pressure 
\begin{equation}\label{eq:neckpres}
p\simeq p_{\rm B}\left[1-\frac{\sqrt{k_{0}k_{1}}}{2}\left(l-l_{\rm B}\right)\right]
\end{equation}
is finite and positive through the neck as well. Therefore, locally around the neck, the geometry resembles that of the vacuum solution from Fig. \ref{fig:wormhole}, but covered by a perfect fluid of constant density with pressures that exceed the value of the density [note that $\left(p_{\rm B}\gg \rho\right)$ by consistency with \eqref{eq:presapprox}].

Inside the neck, the solution jumps from the unconcealed to the concealed branch, where vacuum polarization grows unbounded. Following similar arguments to those in \cite{Arrechea2020} for the vacuum solution, we can determine the form of the metric in the new asymptotic region. In particular, noticing that $\psi$ takes the $-$ sign of \eqref{eq:psirb} at the interior (concealed) side of the neck, and that $r_{\rm B}>0$, $\psi$ always takes values below the three roots and the exact solutions that appear represented in Fig. \ref{fig:roots}. By consistency of Eq. \eqref{eq:eqpsipresion}, $\psi$ grows with $r$ until the most negative root, $\mathcal{R}_{3}$, is crossed. Beyond this point $\psi$ decreases linearly with $r$, taking values between the exact solution $\psi_{+}$ and the root $\mathcal{R}_{3}$. The former cannot be crossed in virtue of the Picard-Lindelöf theorem, and the latter cannot be encountered for a second time for self-consistency of \eqref{eq:eqpsipresion}. Thus, in the $r\to\infty$ limit, $\psi$ decreases linearly with $r$ (at leading order) and essentially corresponds to the vacuum solution.

To derive the asymptotic form of the metric deep inside the wormhole neck (in radial distance), we assume $\psi$ deviates slightly from the exact solution as $\psi\simeq\psi_{+}+\beta(r)$. Replacing this expression in Eq. \eqref{eq:eqpsipresion} and neglecting terms beyond linear order in $\beta$, we obtain
\begin{equation}
\beta'\simeq -\frac{2r}{l_{\rm P}^{2}}\beta+\order{\beta^{2}}.
\end{equation}
Integrating yields
\begin{equation}\label{eq:beta}
\beta\simeq - e^{-2r^{2}/l_{\rm P}^{2}}\beta_{0},
\end{equation}
where $\beta_{0}$ is a positive constant of integration of dimensions of inverse of length [the sign in Eq.~\eqref{eq:beta} is chosen so that the solution $\psi$ approaches $\psi_{+}$ from below]. Now, we further integrate $\psi$ to derive the compactness function and the asymptotic form of the metric. Written in Schwarzschild coordinates, it takes the approximate form
\begin{align}\label{eq:metricsing}
ds^{2}\simeq 
&
e^{-2r^{2}/l_{\rm P}^{2}}\left(\frac{r}{l_{\rm P}}\right)^{1-4\alpha}\left\{-a_{0}\left(1-\frac{l_{\rm P}^{2}}{8 r^{2}}\right)dt^{2}\vphantom{\left(\frac{r}{l_{\rm P}}\right)^{\frac{21-20\alpha}{8}}}\right.\nonumber\\
&
\left.+b_{0}\left(\frac{r}{l_{\rm P}}\right)^{2}\left[1-\frac{\left(9-32\alpha\right)l_{\rm P}^{2}}{r^{2}}\right]dr^{2}\right\}+r^{2}d\Omega^{2}.
\end{align}
Here, $a_{0}$ and $b_{0}$ are dimensionless integration constants. In view of the above expression, the metric has a null singularity at radial infinity, which is located at finite affine distance from the neck for all geodesic paths. In the asymptotic region, the pressure of the fluid diverges exponentially towards positive infinity. The compactness function diverges towards negative infinity exponentially as well, due to the presence of an infinite cloud of negative mass which is being generated by the vacuum energy of the scalar field. 

We observe that the characteristics of sub-critical solutions are identical to those of the vacuum solution i.e. an asymmetric wormhole with an interior null singularity at infinite $r$, but filled with an isotropic fluid of constant density and divergent pressures. Despite the classical SET being singular, the dominant contribution to the divergence in curvature invariants comes from semiclassical contributions, and differences between vacuum and matter geometries appear at subleading order in the approximate metric \eqref{eq:metricsing}. The uppermost panel in Fig. \ref{fig:noncritsemi} contains an example of a sub-Buchdahl, sub-critical star (see top left panel in Fig. \ref{fig:RSETnoncritsemi} for details on the RSET). 

Semiclassical stellar solutions can be interpreted as a mixture of competing classical and quantum contributions. Taking $\rho=0$ gives all predominance to the vacuum sector, while increasing $\rho$ endows the geometry with classical-like properties. On the other hand, as the compactness at the surface of the star $C(R)$ is increased (while keeping $\rho<\rho_{\rm{c}}$ at all times), the wormhole neck follows a trajectory similar to the infinite positive pressure divergence from the classical theory: it moves outwards as $C(R)$ approaches the Buchdahl limit. At this stage, keeping $C(R)$ fixed and giving predominance to the classical fluid (increasing $\rho$) effectively pushes the wormhole neck towards smaller radii. As a consequence of increasing $\rho$, a greater amount of the contribution to $M_{\text{cloud}}$ in Eq.~\eqref{eq:mcloud} is coming from the classical source rather than the semiclassical vacuum polarization.

The second panel in Fig. \ref{fig:noncritsemi} (top-right panel in Fig. \ref{fig:RSETnoncritsemi} for the RSET) exemplifies a super-Buchdahl, sub-critical star where $\rho$ has been chosen so that the neck is pushed inwards appreciably. Given a sub-critical super-Buchdahl configuration and increasing $\rho$ moves the position of the wormhole neck inwards. This is accomplished at the expense of generating a nucleus of negative mass whose repulsive force smears the growth in pressure.  The increase in $\rho$ makes the classical fluid contribution prevail, causing compactness to become negative, but not as negative as to compensate the growth in pressure, resulting in a wormhole. 

\subsubsection{Relevance of \emph{$\epsilon$-strict} stellar spacetimes and validity of the Polyakov approximation}
The third panel in Fig. \ref{fig:noncritsemi} (see bottom left panel in Fig. \ref{fig:RSETnoncritsemi} for details on the RSET) shows a geometry with its wormhole neck very close to the radial origin. This neck has a Planckian radius, and lies in the regime where the physics of the solution is subject to the particular regulator scheme adopted for the RP-RSET. Hence, the regime around where the wormhole neck is reached lies outside the domain of reliability of the Polyakov approximation.
Notice that these configurations have the compactness function bouncing from negative numbers to $C(r_{\rm B})=1$ at the neck. Hence, by moving the $\rho$ parameter, the compactness of these solutions can be made as small as desired arbitrarily close to $r=0$. In this precise sense, there is a family among all sub-critical semiclassical solutions that describes \emph{$\epsilon$-strict} spacetimes, as for these solutions the compactness can be made to obey the bound \eqref{eq:corecond} in a sphere of radius $r_{\epsilon}$ by taking a suitable $\rho<\rho_{\rm c}$. Obtaining a  strict stellar spacetime from configurations of this sort would amount to regularize their nucleus. As \emph{$\epsilon$-strict} spacetimes are absent in the classical space of super-Buchdahl solutions, semiclassical constant-density spheres of high compactness are one step closer to being regular than classical ones, precisely due to the way quantum corrections operate within these structures.

The existence of \emph{$\epsilon$-strict} spacetimes is in a way related to the failure of the Polyakov approximation to properly account for the contributions of vacuum polarization in presence of matter fluid spheres which extend all the way to $r=0$ [equivalently, to distances where the spacetime metric in Eq.~\eqref{eq:metricrcoord} cannot be dimensionally reduced to its non-angular sector accurately]. Were the spacetime geometry sourced by a RSET adequate for computing backreaction effects over regular stellar spacetimes, the resulting configurations might have been regular from the start. We are demanding from the RSET more than just yielding finite components at $r=0$, as we also look for a RSET that captures more accurately the physics at the nucleus of compact relativistic stars (i.e. the expected violation of energy conditions that the RP-RSET seems unable to reproduce at the core of regular stellar spacetimes that approach the Buchdahl limit \eqref{Eq:rhosemiorigin} but more precise, local approximations account for \cite{Hiscock1988}). The redshift function of classical Buchdahl stars vanishes exactly at $r=0$, as seen in Eq.~\eqref{eq:redsfunclas}. Thus, the result by Hiscock \cite{Hiscock1988} indicates that the RSET acquires a negative energy density when nearing a surface of zero redshift. As the Polyakov RSET is oblivious to the overall value of the redshift function [only their derivatives enter the field equations (\ref{eq:ttsemi},~\ref{eq:llsemi})], this characteristic is not being well-captured by this approximation.

The shrinkage of the neck as the density increases goes on until we encounter a separatrix solution with distinct features (see Subsec. \ref{sec:S-c-superB} below for details and Fig. \ref{fig:noncritsemi} for a series of configurations that approach this separatrix). For this solution, pressure
and compactness diverge towards positive and negative infinity, respectively, at $r=0$. This is a separatrix solution between two distinct behaviors in the pressure and in the compactness. Hence, attending to our definition of criticality from Subsec. \ref{subsec:criticality}, this solution corresponds to a critical (and singular) configuration.
Beyond this critical density $\rho_{\rm{c}}$, solutions have no neck and their shape function extends to $r=0$, but in a singular manner. These super-critical configurations are the ones analyzed in the next subsection.

\subsection{Super-critical stars}
\label{sec:S-superc-subB}

Returning to the phase space from Fig. \ref{fig:phasespace}, sub-critical solutions are situated between the pure vacuum solution, with $\rho=0$, and solutions which have regular pressure everywhere. Increasing the density allows to observe a transition between the former and the latter, the separatrix between both being $\rho=\rho_{\rm{c}}$. For stars well below the Buchdahl limit, everything indicates that the lowest value of the density that makes the neck vanish ensures the regularity of the structure. These solutions correspond to the configurations obtained integrating outwards from a regular radial origin (see Subsec. \ref{sec:S-c-subB}).
We find that this critical solution stops being regular beyond certain value of the compactness $C(R)$. This can be deduced from the fact that we have not been able to obtain solutions starting from a regular origin that end up corresponding to super-Buchdahl stars (excluding those with trans-Planckian $\rho$).

Picture now a super-critical star, for which the solution extends up to $r=0$. An example of this configuration appears in the bottom panel of Fig.~\ref{fig:noncritsemi} (see bottom right panel in Fig.~\ref{fig:RSETnoncritsemi} for the corresponding RSET components). Sufficiently close to the radial origin, the geometry can be approximated by that of the semiclassical Schwarzschild counterpart with negative mass ADM mass. By evaluating Eq. \eqref{eq:llsemi} in the $r\to0$ limit assuming a finite pressure at the origin, we obtain
\begin{equation}
\phi'\simeq
\frac{-\alpha+\sqrt{\alpha(\alpha-1)}}{r}r',\label{eq:supercritphi}.
\end{equation}
Notice that, in Schwarzschild coordinates, this corresponds to the exact solution $\psi_{-}$, which lives in the unconcealed branch and connects smoothly with the classical solution in the $l_{\rm P}\to0$ limit. Replacing Eq.~\eqref{eq:supercritphi} in Eq.~\eqref{eq:ttsemi}, we obtain the following relation for the shape function,
\begin{equation}\label{eq:supercritcomp}
r'\simeq
\left(\frac{|\tilde{M}|}{r}\right)^{(1+\alpha)\left(\sqrt{\frac{\alpha}{\alpha-1}}-1\right)}.
\end{equation}
Here, $\tilde{M}$ is a constant of integration related to the deviations of $\rho$ from $\rho_{\rm{c}}$. Integrating \eqref{eq:supercritcomp} returns the following asymptotic form of the radial function
\begin{equation}\label{eq:supercritrsemi}
r(l)\simeq\left(|\tilde{M}|^{-1+\sqrt{\frac{\alpha}{\alpha-1}}}l\right)^{\left[\alpha\left(\sqrt{\frac{\alpha}{\alpha-1}}-1\right)+\sqrt{\frac{\alpha}{\alpha-1}}\right]^{-1}},
\end{equation}
where, in the limit of big $\alpha$, or when the RP-RSET is fully suppressed, we recover the classical behavior \eqref{eq:supercritr}. The redshift function indeed diverges towards positive infinity in the limit $l \to 0$, 
\begin{equation}\label{eq:supercritreds}
e^{2\phi}\simeq\left(\frac{|\tilde{M}|}{l}\right)^{\frac{2}{1+2\sqrt{\frac{\alpha}{\alpha-1}}}},
\end{equation}
and the classical fluid acquires the equation of state of vacuum energy at the radial origin
\begin{equation}\label{eq:supercritpressemi}
p\simeq-\rho+\tilde{M}^{-2}\left(\frac{l}{|\tilde{M}|}\right)^{\frac{1}{1+2\sqrt{\frac{\alpha}{\alpha-1}}}}.
\end{equation} 
The finite value of the central pressure is approached with infinite gradient, as in the classical expression \eqref{eq:supercritpres}. The divergence of the pressure gradient is stronger than the classical one since the exponent of Eq.~\eqref{eq:supercritpressemi} vanishes in the limit $\alpha \to 1$. Vacuum polarization gets stimulated by the presence of this central negative mass, strengthening the super-critical singularity with respect to the classical situation.

We return now to the bottom picture in Fig. \ref{fig:noncritsemi}, which shows an example of a super-Buchdahl, super-critical star. The RP-RSET (bottom right panel in Fig. \ref{fig:RSETnoncritsemi}) shows drastic differences with the sub-critical case. Namely, $\rho_{\text{se}}$ changes sign with respect to its negative contribution at the surface, diverging towards positive infinity at $r=0$. The semiclassical pressures diverge towards negative infinity after having encountered a maximum. 
\subsection{Semiclassical infinite pressure separatrix}
\label{sec:S-c-superB}

The semiclassical separatrix between sub-critical and super-critical configurations is reminiscent of the classical separatrix in several aspects that we will detail in what follows. Let us work under the assumption that the separatrix solution has infinite pressure at the radial origin by similarity with the classical case in section \ref{subsec:separatrix}. First, we go back to Eq. \eqref{eq:eqpsi}, expand the right-hand side in powers of $r$, and neglect terms subleading in the pressure, as of \eqref{eq:presapprox}. The coefficients in Eq. \eqref{eq:eqpsi} become
\begin{align}\label{eq:eqpsifact}
    A_{0}\simeq
    &
    ~12\pi p,\nonumber\\
    A_{1}\simeq
    &
    ~4\pi r\left(3p+\frac{2p}{\alpha}\right)-\frac{2}{r},\nonumber\\
    A_{2}\simeq
    &
    -8\pi r^{2}p\left[1-\frac{3}{2\alpha}-\mathcal{O}\left(r^{2}/l_{\rm P}^{2}\right)\right]-\frac{2}{\alpha}-2,\nonumber\\
    A_{3}\simeq
    &
    -\frac{r}{\alpha}\left\{4\pi r^{2}p\left[1-\mathcal{O}\left(r^{2}/l_{\rm P}^{2}\right)\right]+\frac{1}{\alpha}+1\right\},\nonumber\\
    \mathcal{D}\simeq
    &
    ~\frac{\alpha}{(1+8\pi r^{2}p)(\alpha-1)}.
\end{align}
We arrange these coefficients in a particularly illustrative form, yielding
\begin{align}\label{Eq:Semi}
    \psi'\simeq \left\{\vphantom{\frac{r(\alpha+1)}{\alpha}\psi^{3}}\right. 
    &
    \left.4\pi p\left[3\alpha+\left(2+3\alpha\right)r\psi+\left(3-2\alpha\right)r^{2}\psi^{2}-r^{3}\psi^{3}\right]\right.\nonumber\\
    &
    \left.-\frac{2\alpha}{r}\psi-2(1+\alpha)\psi^{2}-\frac{r(\alpha+1)}{\alpha}\psi^{3}\right\}\times\nonumber\\
    &
   \frac{1}{(\alpha-1)(1+8\pi r^{2} p)}.
\end{align}
This expression is describing a competition between vacuum and matter contributions. By dropping the terms proportional to the pressure in Eq. \eqref{Eq:Semi} we obtain the solutions to the equation in vacuum \cite{Arrechea2020}
\begin{equation}\label{eq:psiseparatrix}
    \psi=-\frac{\alpha\pm\sqrt{\alpha(\alpha-1)}}{r},\qquad \psi=0,
\end{equation}
where only the $-$ sign returns the Schwarzschild solution in the classical limit (taking \mbox{$\alpha\to\infty$}, an infinitely suppressed RP-RSET). Note that Eq.~\eqref{eq:psiseparatrix} is equivalent to Eq.~\eqref{eq:supercritphi} for the super-critical case, but expressed in Schwarzschild coordinates. 

In the regime of approximation described by Eq. \eqref{eq:presapprox}, the pressure is proportional to the integral of $\psi$. Let us assume the ansatz
\begin{equation}\label{eq:separatrixpsi}
\psi=\frac{\eta}{r}.
\end{equation}
For this ansatz, the pressure becomes, in virtue of \eqref{eq:presapprox}
\begin{equation}\label{eq:separatrixpres}
p\simeq \kappa \left(\frac{r_{0}}{r}\right)^{\eta},
\end{equation}
where $r_{0}$ is an integration constant with dimension of length and $\eta$ needs to take positive values, since $\eta<0$ is not compatible with the infinite pressure assumption. 
Inserting Eqs.~\eqref{eq:separatrixpsi} and \eqref{eq:separatrixpres} in Eq.~\eqref{Eq:Semi}, we obtain
\begin{align}\label{Eq:SemiSimp}
    \psi'\simeq
    \left\{\vphantom{\frac{r(\alpha+1)}{\alpha}\psi^{3}}\right.
&
\left.4\pi \kappa\left(\frac{r_{0}}{r}\right)^{\eta}\left[3\alpha+\left(2+3\alpha\right)\eta+\left(3-2\alpha\right)r^{2}\eta^{2}-\eta^{3}\right]\right.\nonumber\\
   &
    \left.-\frac{(1+\alpha)\eta}{r^{2}}\left(2\alpha+2\eta+\eta^{2}\right)\right\}\nonumber\\
&
\times\frac{1}{(\alpha-1)\left[1+8\pi r^{2} \kappa\left(\frac{r_{0}}{r}\right)^{\eta}\right]}.
\end{align}
The value of $\eta$ determines which source, classical or quantum, provides the dominant contribution to the divergence in $\psi'$. For $\eta<2$, the vacuum terms carry the dominant divergence. For $\eta=2$, the terms in the first and second line all contribute at the same order, whereas for $\eta>2$, terms proportional to the pressure dominate both the numerator and the denominator in Eq.~\eqref{Eq:SemiSimp}. Let us explore these possibilities.

Replacing the derivative of Eq.~\eqref{eq:separatrixpsi} in Eq.~\eqref{Eq:SemiSimp} and taking $\eta=2$ (which equates vacuum and matter contributions) yields the following relation between integration constants,
\begin{equation}
\kappa=\frac{1+\alpha}{2\pi \alpha r_{0}^{2}}.
\end{equation}  
Replacing this behavior in the radial Einstein equation \eqref{eq:llsemi} (in Schwarzschild coordinates) we obtain
\begin{equation}
C\simeq\frac{4(1+\alpha)(-1+r_{0}^{2})}{(4+5\alpha)r_{0}^{2}}+\mathcal{O}\left(r^{2}\right),
\end{equation}
from where only the value $r_{0}=1$ returns a vanishing compactness at the radial origin. The solution
\begin{equation}
p=\frac{1+\alpha}{2\pi \alpha r^{2}},
\end{equation}
is reminiscent of the classical separatrix between critical sub- and super-Buchdahl configurations and connects smoothly with the classical Buchdahl solution \eqref{eq:separatrixpresclas2} in the $\alpha\to\infty$ limit. The semiclassical counterpart to that separatrix retains its critical character, in the sense that $C(l\to0)=0$, while the rate of growth of the pressure increases as $\alpha$ is decreased. Hence, semiclassical corrections contribution towards strengthening the divergence of the pressure in this separatrix. 

Once the Buchdahl limit is surpassed the separatrix solution takes a different form. By taking $\eta>2$ in Eq.~\eqref{Eq:SemiSimp}, we are assuming that pressure-dependent terms carry the leading-order divergences in the expansion. Therefore, Eq. \eqref{Eq:SemiSimp} can be reduced to
\begin{equation}
-\frac{\eta}{r^{2}}\simeq\frac{r^{-\eta}\left[3\alpha+\left(2+3\alpha\right)\eta+\left(3-2\alpha\right)r^{2}\eta^{2}-\eta^{3}\right]}{2(\alpha-1)r^{2-\eta}},
\end{equation}
from where the only positive solution is  $\eta=3$. We have a pressure profile of the form
\begin{equation}\label{eq:separatrixpres2}
p\simeq \frac{\tilde{\kappa}}{r^{3}}.
\end{equation}
where $\tilde{\kappa}$ is a positive integration constant of dimension length. Replaced in the equation for the compactness we find
\begin{equation}
C\simeq-\frac{8\pi \alpha \tilde{\kappa}}{(9+7\alpha)r}.
\end{equation}
The separatrix between sub-critical and super-critical configurations has an infinite compactness at the origin. This divergence in the compactness is weaker than the curvature singularity from super-critical configurations, which fits right in the separatrix between sub- and super-critical profiles in the super-Buchdahl case. Since the differential equation for the compactness \eqref{eq:ttsemi} is not integrable in terms of analytical functions, we do not know the specific form of the constant $\tilde{\kappa}$. Nevertheless, we expect it should present the correct classical limit.

Separatrices are only perturbatively deformed by semiclassical corrections. Solutions belonging to the sub-critical regime are wormhole geometries, whereas super-critical solutions are naked singularities. The separatrix solutions \eqref{eq:separatrixpres} and \eqref{eq:separatrixpres2} are modified perturbatively by regulator-dependent corrections. This is reminiscent, in a sense, to what happens in the vacuum situation, where the separatrix between wormhole geometries and naked singularities at $r=0$ is precisely Minkowski spacetime, for which vacuum polarization is exactly zero \cite{Arrechea2020}. The infinite pressure separatrices here obtained apparently exhibit a similar stability with respect to quantum corrections.

\subsection{Outside-the-neck stars and pressure regularization}
The analyzed behaviors for both sub- and super-critical stars (i.e. top and bottom regions of the phase space in Fig. \ref{fig:phasespace}) only depend on whether the value of $\rho$ is below or above $\rho_{\rm{c}}$, and are thus universal for stars either outside or inside the neck. Turning back to the diagram in Fig. \ref{fig:phasespace}, which qualitatively describes stars with their surfaces outside the neck, we now draw attention to the separatrix solution between regions III and IV (or the super-Buchdahl half-plane). 
Recall that numerical integrations for these regimes (sub-critical, critical and super-critical) appear represented in the fourth row of Fig.~\ref{fig:tablefigs}.
Beginning with a sub-critical configuration and integrating from the surface, we can estimate numerically from surface integrations (within some expected numerical uncertainty) the value of the density that sits between the less dense super-critical solution and the most dense sub-critical solution. 
To the limit of our numerical precision, this density value coincides with $\rho_{\rm c}$. Notice that with our definition of criticality, this coincidence between the pressure separatrix and the critical solution does not happen in the classical case: by increasing the parameter $\rho$ we first find $\rho_{\rm c}$, i.e. a change in behavior of $C$, and later on for $\rho_{\text{reg-p}}>\rho_{\rm{c}}$ we find the first solution for which pressure becomes finite at the origin. So, in what follows, we will make use of $\rho_{\text{reg-p}}$ to refer to the classical separatrix in pressure and $\rho_{\rm{c}}$ to denote the (semiclassical) critical solution, which is a separatrix in pressure as well.

\begin{figure}
    \centering
    \includegraphics[width=\columnwidth]{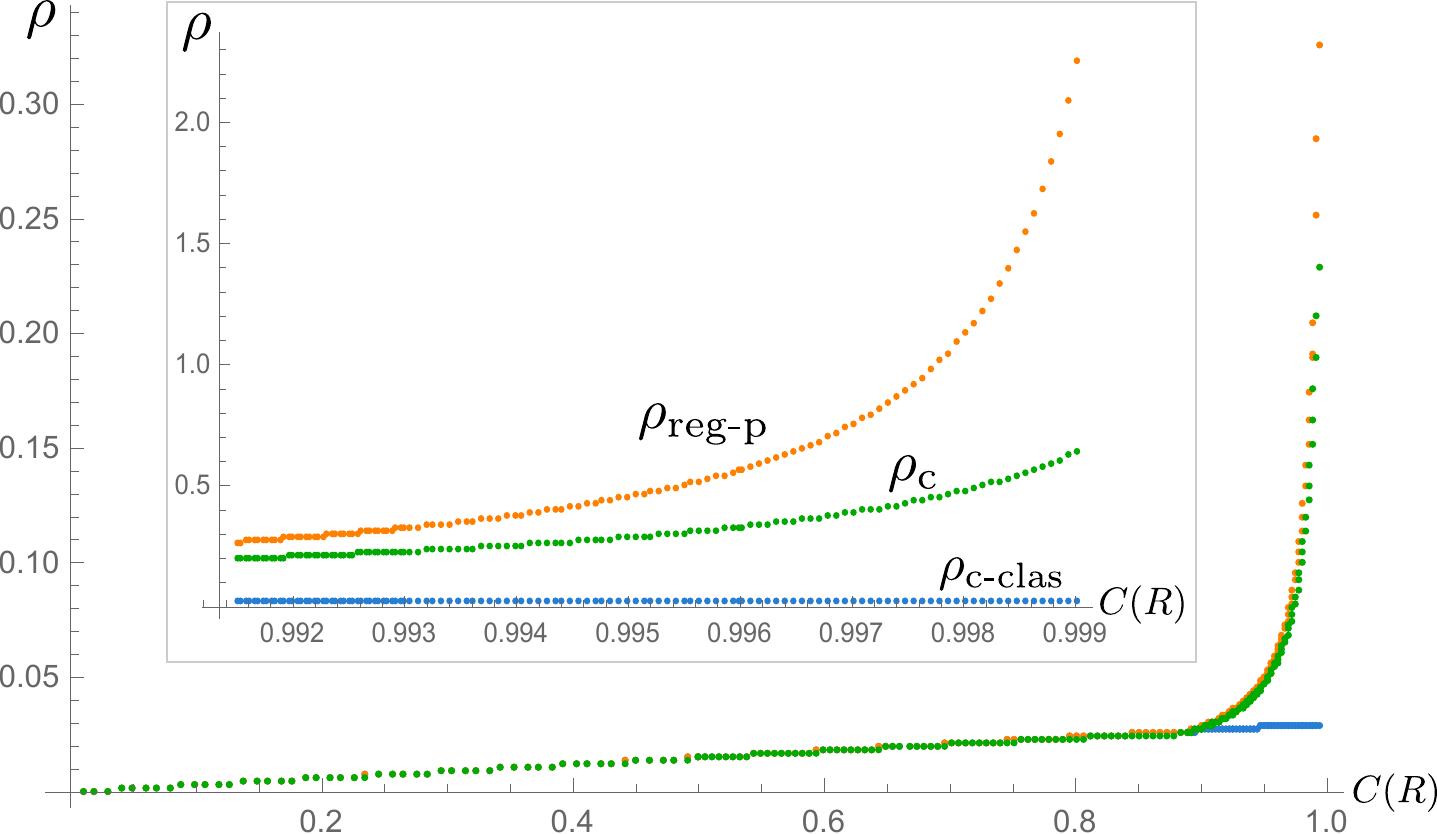}
    \caption{Plot of $\rho_{\text{reg-p}}$ in terms of the compactness for classical (orange) stars and of $\rho_{\rm{c}}$ for semiclassical (green) stars with $R=2$ and $C(R)\in\left(0,1\right)$. The blue line corresponds to the classical critical density \eqref{eq:critdens}. The orange curve diverges in the $C(R)\to1$ limit, whereas the green curve reaches a finite value, in this case $\rho_{\rm{c}}\left(C(R)\to1\right)\simeq1.366.$}
\label{fig:DensityComparison}
\end{figure}

\begin{figure}
    \centering
    \includegraphics[width=\columnwidth]{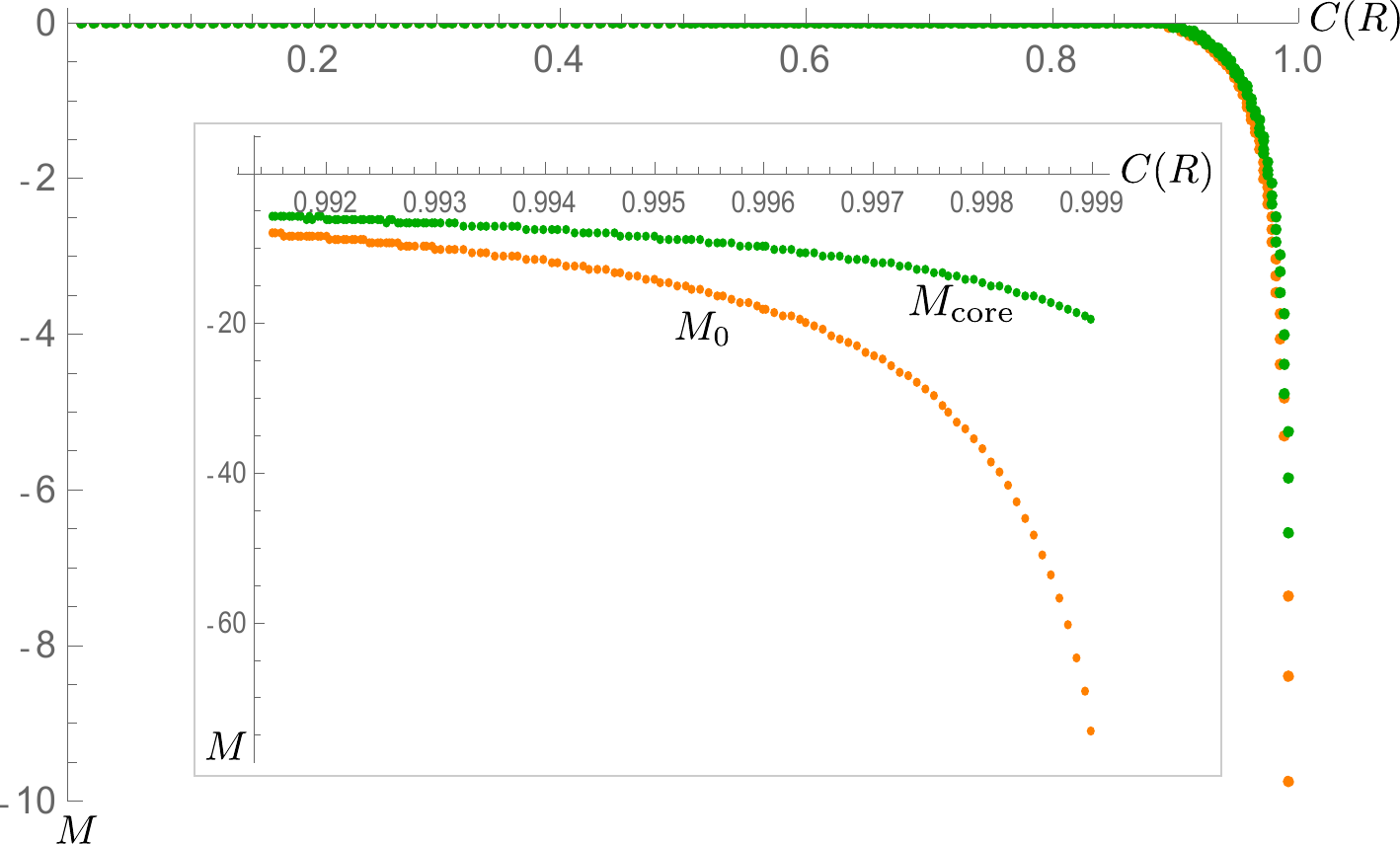}
    \caption{Plot of the Misner-Sharp mass at a central radius $r_{\text{core}}=\order{l_{\rm P}}$ in terms of the surface compactness for classical (orange) and semiclassical (green) stars with $R=2$. Notice how the orange curve diverges in the $C(R)\to1$ limit, as infinite negative masses are required to regularize the pressure in that limit. In the semiclassical case, since the surface of $C(R)=1$ is a wormhole neck. As pressure at the neck is finite [see Eq. \eqref{eq:neckpres}] the required negative mass is finite, in this case $M_{\text{core}}\left(C(R)\to1\right)\simeq-41.20.$ }
\label{fig:MassComparison}
\end{figure}

Let us numerically explore the behavior of the quantity $\rho_{\rm{c}}$ in different situations. For semiclassical stars with the same radius and compactness as their classical counterparts, $\rho_{\rm{c}}$ is appreciably smaller than the corresponding classical value $\rho_{\text{reg-p}}$. Figure \ref{fig:DensityComparison} shows a comparison between these two densities for stars of various $C(R)$, together with the line $\rho_{\rm{c}-\text{clas}}$. Remarkably, we find that $\rho_{\rm{c}}$ is finite in the limit $C(R)\to1$. In turn, the negative masses needed to halt the growth of the pressure are less negative for semiclassical stars with $C(R)\to1$, when compared to the classical case. See Fig.~\ref{fig:MassComparison} for a detailed plot of the approximate Misner-Sharp mass needed to regularize the pressure in each situation.

\begin{figure}
    \centering
    \includegraphics[width=\columnwidth]{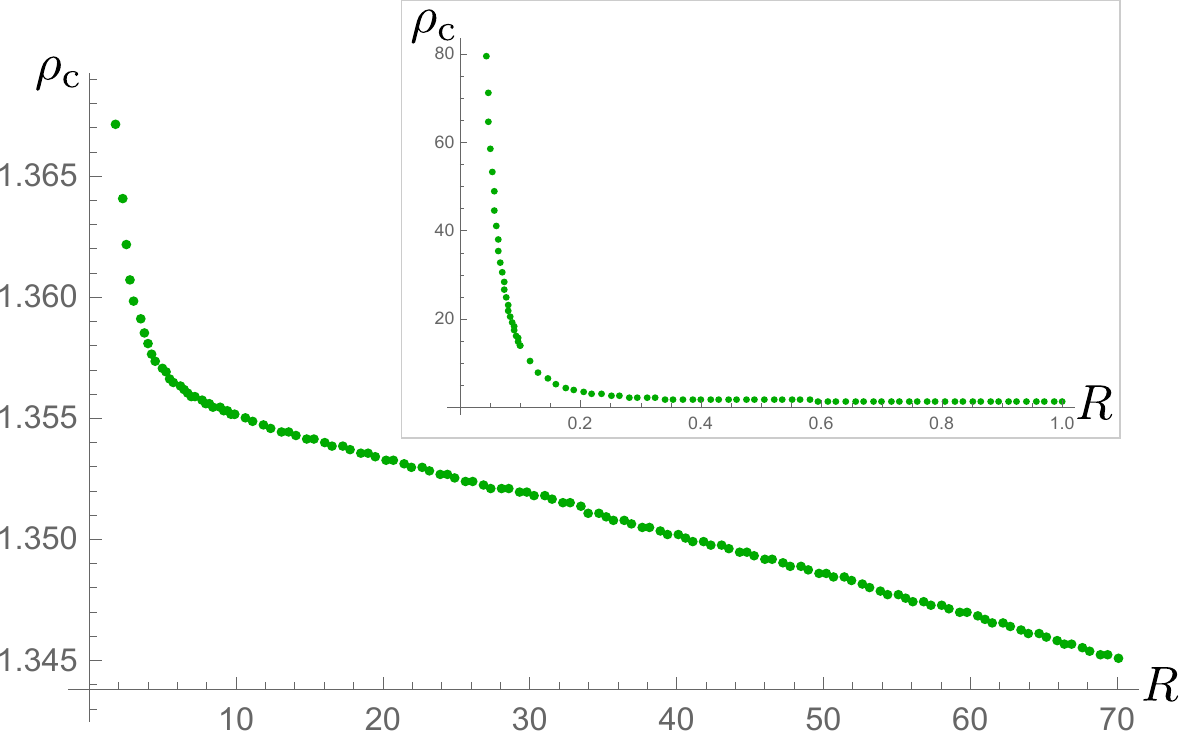}
    \caption{Plot of $\rho_{\rm{c}}$ for semiclassical stars of various radii and compactnes $C(R)=1-10^{-10}$. As $R$ shrinks, the separatrix density diverges, whereas for larger radii  (in Planck units) it decreases linearly.}
\label{fig:DensityRadius}
\end{figure}

The cause of this discrepancy between classical and semiclassical stars in the $C(R)\to1$ limit comes from the differences between their respective vacuum solutions. In the case of a classical star the surface where $C(R)=1$ is the horizon, resulting in infinite surface pressures in the $C(R)\to1$ limit, which can only be compensated by an infinite amount of negative mass at the origin. In the semiclassical case, however, the $C(R)\to1$ limit corresponds to taking the surface of the star towards the neck, where pressure is indeed finite, in virtue of Eq.~\eqref{eq:neckpres}. In consequence, a finite increase in $\rho$ regularizes the pressure profile of the configuration. As observed in Fig.~\ref{fig:DensityRadius}, the parameter $r_{\rm c}$ decreases linearly as the radius of the star is increased while keeping $C(R)$ fixed.

\begin{figure}
    \centering
    \includegraphics[width=\columnwidth]{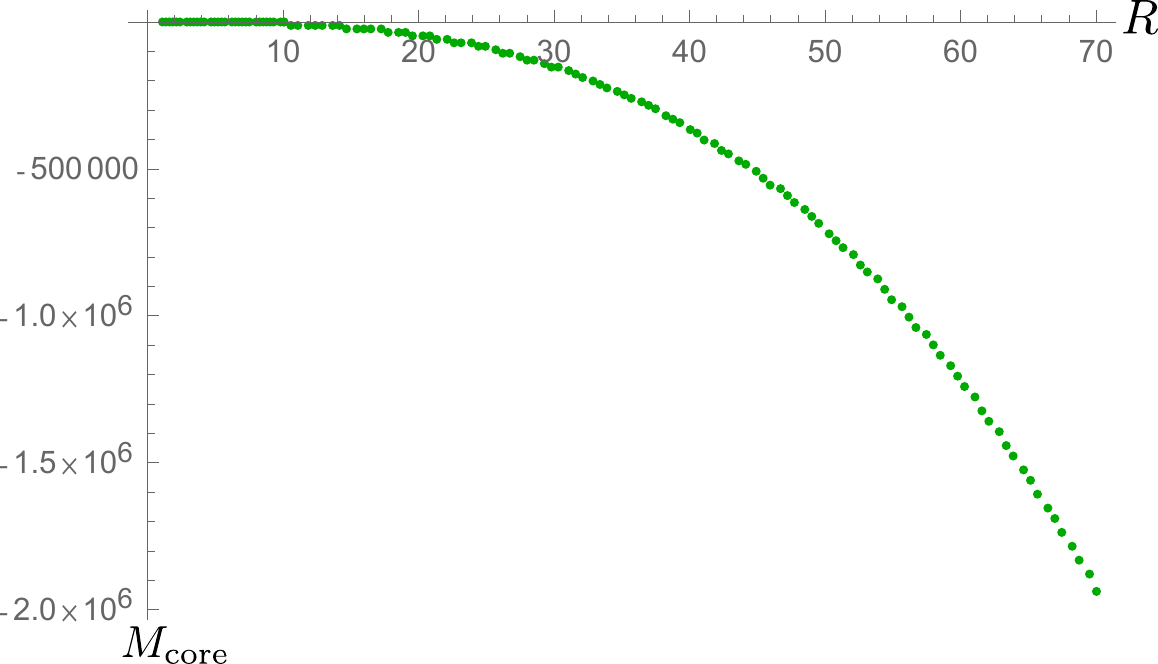}
    \caption{Plot of the central negative mass $M_{\text{core}}$ for semiclassical super-critical stars of various radii and surface compactness $C(R)=1-10^{-10}$. This estimated central mass is obtained by stopping the integration at a security radius far from the region where the Misner-Sharp mass diverges. Notice how the central mass has to be many orders of magnitude higher than the total mass of the star $M\simeq R/2$ for big stars. For small stars have their pressure regularized by negative central masses comparable to their total mass, indicating that the physics of Planckian stars may be different from that of astrophysical bodies.}
\label{fig:MassRadius}
\end{figure}

Finally, Fig.~\ref{fig:MassRadius} shows that the negative mass core induced by increasing $\rho$ grows much faster than the rate at which the total mass $M=R C(R)/2$ increases with $R$ while keeping $C(R)$ fixed. This core can be estimated obtaining the value of the Misner-Sharp mass at a security radius where the Misner-Sharp mass has not yet entered into a runaway regime. The values of the density required to strictly regularize the pressure of ultra-compact stars are therefore many orders of magnitude greater than the total Misner-Sharp mass associated with those stars. However, these densities are finite in the $C(R)\to1$ limit thanks to the energy-condition violating contributions of the RSET at the surface of ultracompact stars [recall Eq. \eqref{Eq:RSETvac}] 

Figures (\ref{fig:DensityComparison}-\ref{fig:MassRadius}) have been obtained by taking $\alpha-1=10^{-6}$. We have observed that increasing the value of $\alpha$ has the effect of making the solutions more alike to their classical counterparts. Consequently, it seems reasonable to assume that $\rho_{\rm{c}}$ approaches its classical value $\rho_{\text{reg-p}}$ in the limit $\alpha\to\infty$ as well.

\subsection{Inside-the-neck stars and pressure regularization}
\label{subsec:beyond-neck}
Up to now our analysis has focused on stars located outside the neck of the wormhole. In this section we turn to locating the surface of the star inside the neck, which in particular implies that these solutions have no well-defined classical limit. In this case, there is again a strong interplay between contributions coming from the vacuum and classical matter that results in sub-critical, critical and super-critical regimes. There exists also a distinction depending on whether the surface is located close to the neck (super-Buchdahl, [$C(R)\to1$]), or far from the neck (sub-Buchdahl, [$C(R)\ll1$]). The last two rows in Fig.~\ref{fig:tablefigs} display numerical plots for all these cases.

\begin{figure}
\centering
\includegraphics[width=\columnwidth]{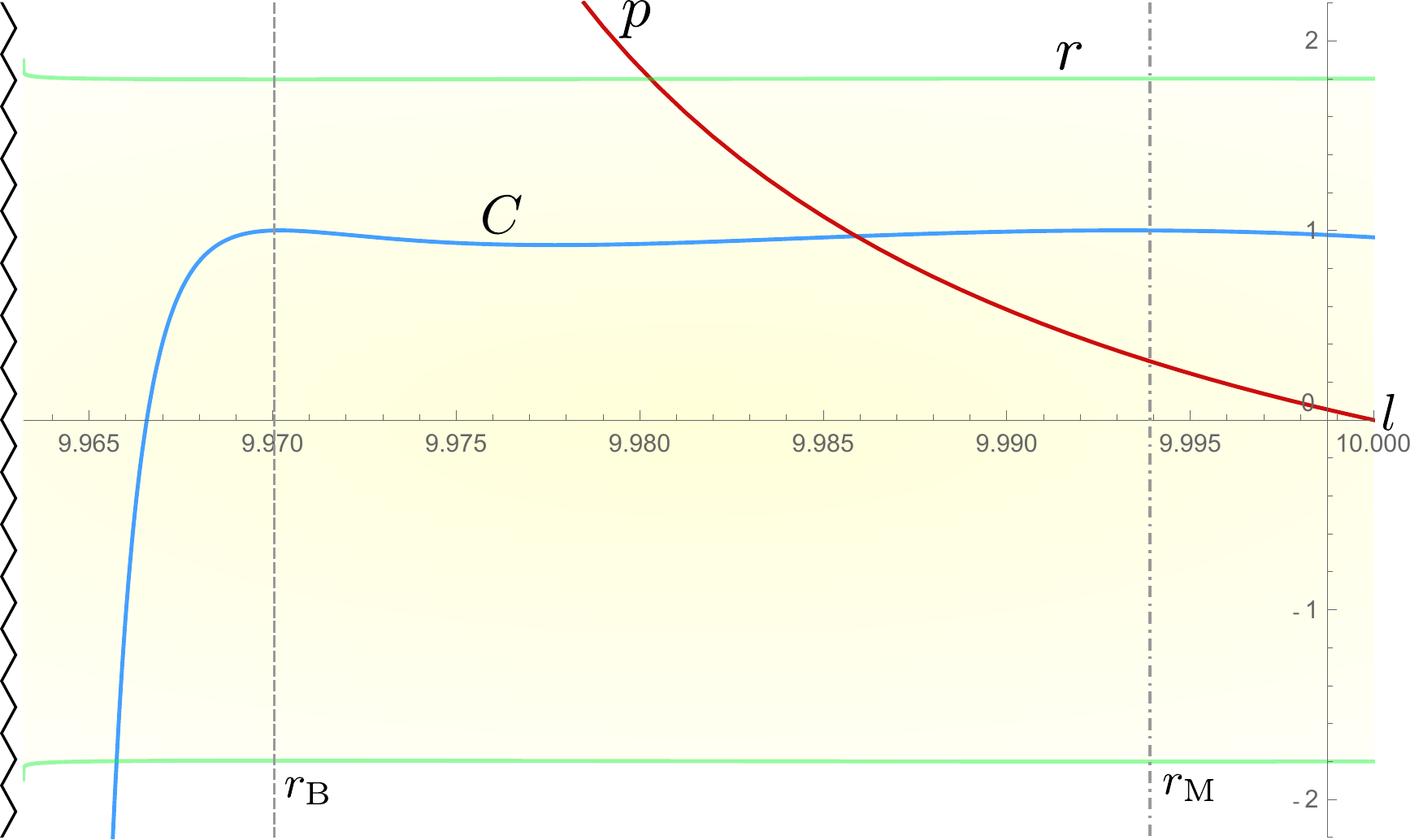}
\caption{Plot of a sub-critical star located inside the neck. The blue green and blue curves represent the shape function $r(l)$ and the compactness $C(l)$, respectively. The red curve is the pressure $p(l)$. The dashed and dot-dashed vertical lines represents the radial minimum $r_{\rm B}$ and maximum $r_{\rm M}$, respectively. The singularity is represented by a zigzag line. The parameters chosen are $R=1.8$, $C(R)=0.96$, $\rho/\rho_{\rm{c}\text{-clas}}=43.4$ and $\alpha-1=10^{-3}$.}
\label{fig:tubito}
\end{figure}

\begin{figure}
\centering
\includegraphics[width=\columnwidth]{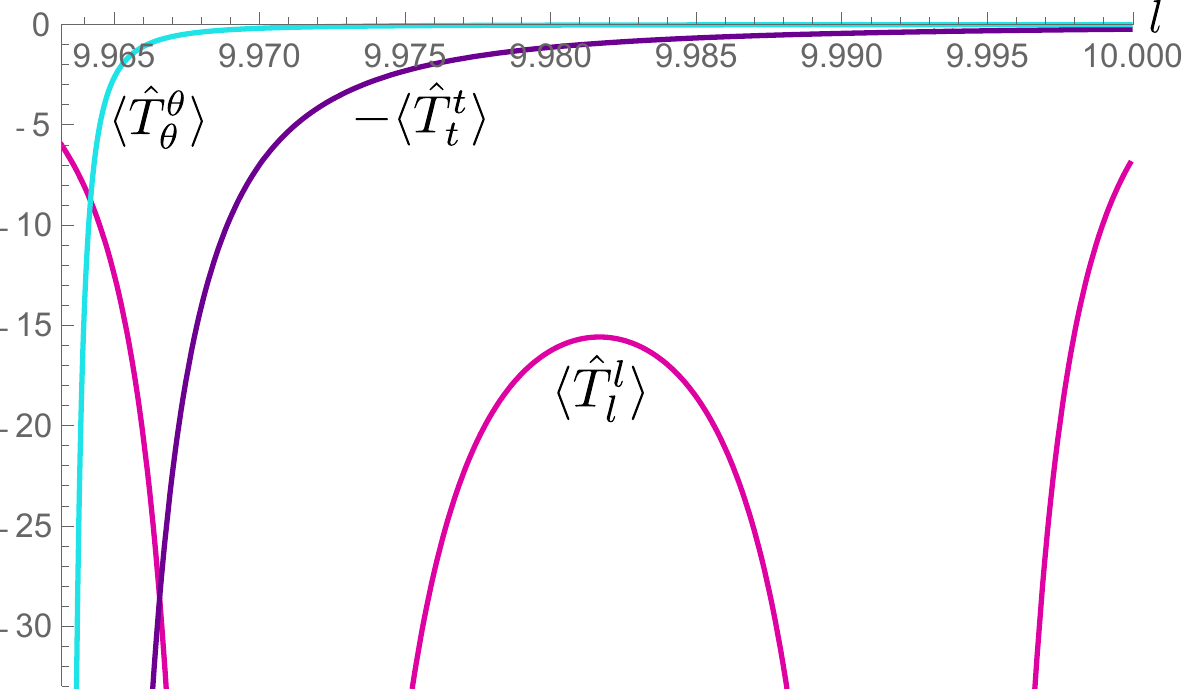}
\caption{RSET components $-\langle\hat{T}^{t}_{t}\rangle$ (dark blue), $\langle\hat{T}^{l}_{l}\rangle$ (magenta) and $\langle\hat{T}^{\theta}_{\theta}\rangle$ (cyan) for a sub-critical star located beyond the neck. The parameters of the integration are $R=1.8$, $C(R)=0.96$, $\rho/\rho_{\rm{c}\text{-clas}}=43.4$ and $\alpha-1=10^{-3}$.}
\label{fig:tubitoRSET}
\end{figure}

In the first of these scenarios the surface of the super-Buchdahl star is located very close to the neck but inside it. If the energy density is sufficiently big, a radial maximum takes place just below the surface of the star, inverting the tendency of the radial coordinate to increase as we deepen through the neck. To illustrate this, we work in Schwarzschild coordinates and consider a local analysis of Eq. \eqref{eq:eqpsi} around the surface of a star located inside the neck $r_{\rm B}$ but very close to it, so that the solution \eqref{eq:psirb} remains a valid approximate solution. This is guaranteed as long as \begin{equation}
r-r_{\rm B}\lesssim\frac{l_{\rm P}^{2}}{r_{\rm B}}.
\end{equation}
Now, expanding Eq. \eqref{eq:eqpsi} at leading order in $\psi$ while taking $p=0$ and $\rho$ positive and constant, the solution is
\begin{align}\label{eq:psimaximum}
\psi\simeq
&
- \frac{1}{l_{\rm P}}\left\{-\frac{l_{\rm P}^{2}\alpha\left(r^{2}-r_{\rm B}^{2}\right)}{\left(r^{2}+\alpha l_{\rm P}^{2}\right)\left(r_{\rm B}^{2}+\alpha l_{\rm P}^{2}\right)}-\alpha\ln\left(\frac{r^{2}+\alpha l_{\rm P}^{2}}{r_{\rm B}^{2}+\alpha l_{\rm P}^{2}}\right)\right.\nonumber\\
&
\left.+\left(1+\alpha\right)\ln\left[\frac{r^{2}+l_{\rm P}^{2}(\alpha-1)}{r_{\rm B}^{2}+\alpha l_{\rm P}^{2}(\alpha-1)}\right]-4\pi \rho \left(r^{2}-R^{2}\right)\right.\nonumber\\
&
\left.+4\pi \rho l_{\rm P}^{2}(\alpha-1)\ln\left[\frac{r^{2}+l_{\rm P}^{2}(\alpha-1)}{R^{2}+l_{\rm P}^{2}(\alpha-1)}\right]\right\}^{-1/2}.
\end{align}
For a positive (and sufficiently large) $\rho$, the term proportional to $(r^{2}-R^{2})$ is the dominant contribution to \eqref{eq:psimaximum}, which compensates the positive logarithmic terms from vacuum contributions (recall that, initially, $r$ increases as we move away from the surface towards the interior of the star). The interior of the squared root in \eqref{eq:psimaximum} vanishes at some radius $r_{\rm M}$ inside the star, generating a radial maximum and taking the solution back to the concealed branch. Once $\rho$ is large enough as to generate this radial maximum, we encounter again three different scenarios depending on whether $\rho$ is above or below its critical value. If $\rho<\rho_{\rm{c}}$, a second radial minimum or neck takes place after the first maximum (this is the situation depicted in Figs. \ref{fig:tubito} and \ref{fig:tubitoRSET}). The metric functions around this second neck have the form \eqref{eq:neckmetric} and connect with a null singularity. Further increments of $\rho$ displace this second neck towards smaller values of $r$ and eventually makes solutions super-critical if $\rho>\rho_{\rm c}$, showing finite pressures everywhere. Examples for each of these cases can be found in the last row from Fig. \ref{fig:tablefigs}.

\begin{figure}
    \centering
    \includegraphics[width=\columnwidth]{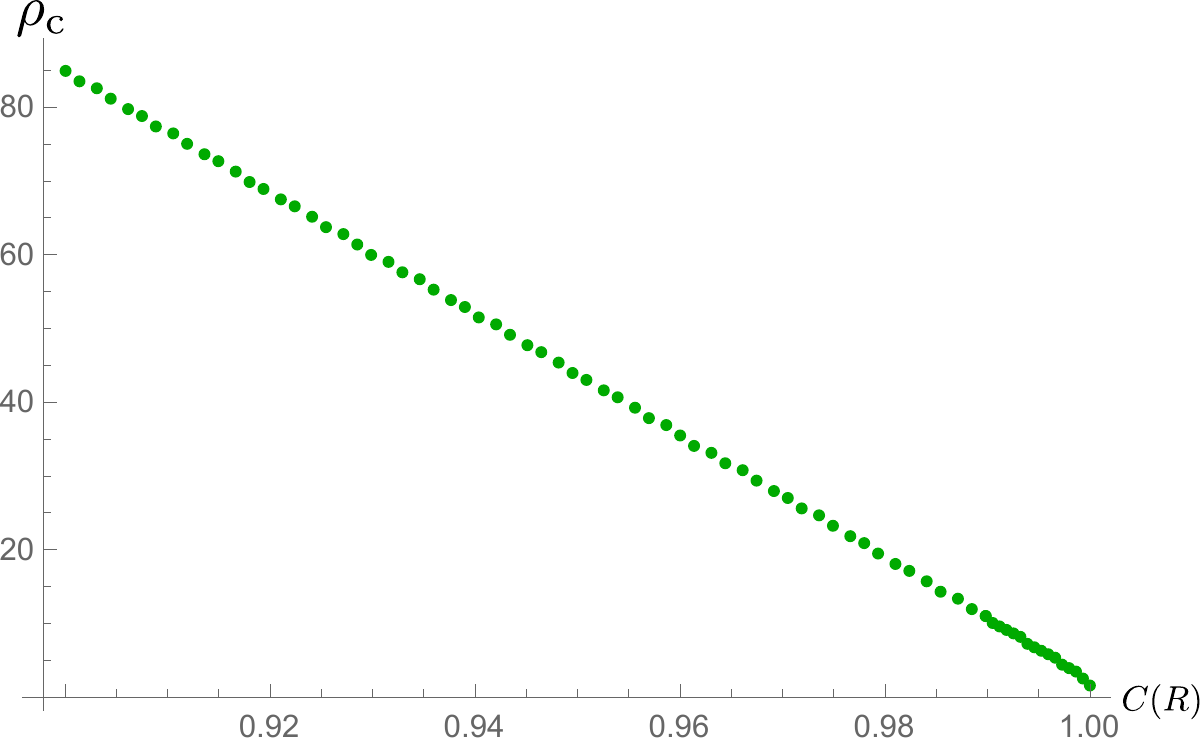}
    \caption{Plot of $\rho_{\rm{c}}$ in terms of the compactness for a star located inside the neck. The density required to regularize the structure grows linearly as compactness diminishes, reaching over Planckian densities very quickly.}
\label{fig:BeyondNeckDens}
\end{figure}

The second possibility is to consider matter located sufficiently deep inside the neck (in radial distance), the negative mass generated by the scalar field becomes comparable to that of the classical source. The three plots in the sixth line of Fig. \ref{fig:tablefigs} show the respective sub-critical, critical and super-critical regimes. Here we can observe that, unless the density of the fluid is increased sufficiently, the geometry will adopt the form \eqref{eq:metricsing} without reaching a radial maximum. Such geometries are completely dominated by vacuum polarization. In this regime, we cannot appeal to the local analysis of \eqref{eq:psimaximum}, and we are forced to solve numerically the complete equations. Figure \ref{fig:BeyondNeckDens} shows the value of the energy density $\rho_{\rm{c}}$ for stars of various surface compactness. These values of the compactness are directly linked to how far inside the neck we are locating the surface of the fluid (see the compactness curve in Fig. \ref{fig:wormhole}). The energy density $\rho_{\rm{c}}$ is found to grow linearly as compactness decreases (as we move the surface of the star far from the neck).

Super-critical configurations of this kind (inside-the-neck stars with a radial maximum) were analyzed in \cite{Ho2018}. 
The authors used the unregularized Polyakov RSET and thus the solutions obtained always displayed a singularity at $r=l_{\rm P}$. For this reason, the authors understood as regular stars those whose pressure remains finite up to a sphere slightly outside the Planck radius. However, due to this singlarity, these analyses are unable to distinguish whether the pressure regularization is attained at the expense of producing a super-critical configuration with a non-regular $C$, as it happens in the classical case. 
Here, by introducing a cutoff to the Polyakov RSET, we have found that all constant-density stars with their surface inside the neck are singular. Infinite pressure separatrices are still expected to be present, but the densities required to reach them are trans-Planckian.

\subsection{At-the-neck stars and pressure regularization}
\label{subsec:at-neck}
Finally, we want to wrap up this discussion by examining the particular case of fluid spheres whose surface is located at the neck of the vacuum wormhole geometry. The fifth line in Fig. \ref{fig:tablefigs} shows the critical and non-critical regimes for at-the-neck stars. As this wormhole neck $r_{\rm B}$ corresponds to $C(r_{\rm B})=1$, this scenario has no classical counterpart. In other words, it involves a non-perturbative departure from the classical situation. Furthermore, it allows to illustrate the aforementioned interplay between quantum and classical contributions to the spacetime geometry.

Combining Eqs. \eqref{eq:ttsemi} and \eqref{eq:llsemi} yields a differential equation of the form
\begin{align}\label{Eq:StarAtNeck}
    r''=\mathcal{E}\left(r^{6}\mathcal{B}_{0}+l_{\rm P}^{2}r^{4}\mathcal{B}_{1}+l_{\rm P}^{4}r^{2}\mathcal{B}_{2}+l_{\rm P}^{6}\mathcal{B}_{3}\right),
\end{align}
where 
\begin{align}
    \mathcal{B}_{0}=
    &
    ~4\pi l_{\rm P}^{2}\left(p-\rho\right)+\left(r'\right)^{2},\nonumber\\
    \mathcal{B}_{1}=
    &
    ~1-\mathcal{H}+8\pi l_{\rm P}^{2}\left[\alpha\left(p-\rho\right)-p\right]+\left(3\alpha-2\right)\left(r'\right)^{2},\nonumber\\
    \mathcal{B}_{2}=
    &
    ~\left(1-2\alpha\right)\mathcal{H}+2\alpha l_{\rm P}^{2}\left[1+2\pi \alpha l_{\rm P}^{2}\left(p-\rho\right)\right]\nonumber\\
    &
    +\alpha\left(3\alpha-2\right)\left(r'\right)^2,\nonumber\\
    \mathcal{B}_{3}=
    &
    ~\alpha\left(\alpha+1\right)\left[1-\mathcal{H}+\left(\alpha-1\right)\left(r'\right)^{2}\right],\nonumber\\
    \mathcal{E}=
    &
    ~\left\{l_{\rm P}^{2}r\left[r^{2}+\left(\alpha-1\right)l_{\rm P}^{2}\right]\left(r^{2}+\alpha l_{\rm P}^{2}\right)\right\}^{-1},
\end{align}
and
\begin{align}
    \mathcal{H}=
    &
    \left\{1+8\pi r^{2} p
    +\left[r^{2}+\left(\alpha-1\right)l_{\rm P}^{2}\right]\left(r'/l_{\rm P}\right)^{2}\right\}^{1/2}\nonumber\\
    &
    \times r' \sqrt{r^{2}+\alpha l_{\rm P}^{2}}.
\end{align}

Let us assume again that the semiclassical field equations are integrated inwards from an asymptotically flat region with positive ADM mass until the neck $r_{\rm B}$, where we decide to locate the surface of radius $R$ of a perfect fluid of constant and positive classical density $\rho$. Continuity of the metric at the neck demands the shape function and the pressure, which are the only unknown functions appearing in Eq.~\eqref{Eq:StarAtNeck}, obey expansions of the form 
\begin{align}\label{Eq:ExpansionsNeck}
    r(l)=
    &
    ~R+r_{1}\left(l-l_{\rm S}\right)^{2}+r_{2}\left(l-l_{\rm S}\right)^{3}+\mathcal{O}\left[(l-l_{\rm S})^{4}\right],\nonumber\\
    p(l)=
    &
    ~p_{1}\left(l-l_{\rm S}\right)+\mathcal{O}\left[(l-l_{\rm S})^{2}\right],
\end{align}
where $r_{1},~r_{2}$ and $p_{1}$ are arbitrary constants. 

Replacing expressions \eqref{Eq:ExpansionsNeck} in Eqs. \eqref{Eq:StarAtNeck} and \eqref{eq:TOVsemi}, we obtain the following values for the first coefficients in the expansion
\begin{align}
    r_{1}=
    &
    ~\frac{\left(R^{2}+\alpha l_{\rm P}^{2}\right)\left(1-4\pi R^{2}\rho\right)+\alpha l_{\rm P}^{4}}{2R\left(R^{2}+\alpha l_{\rm P}^{2}\right)\left[R^{2}+\left(\alpha-1\right)l_{\rm P}^{2}\right]},\nonumber\\
    p_{1}=
    &
    -\frac{\sqrt{R^{2}+\alpha l_{\rm P}^{2}}}{l_{\rm P}R}\rho,
\end{align}
where $p_{1}<0$ for any positive $\rho$, indicating that pressure always grows in the interior region of the star. The coefficient $r_{1}$, however, vanishes for the density value
\begin{equation}\label{Eq:RhoNeck}
    \rho_{\text{neck}}=\frac{1}{4\pi R^{2}}\left[1+\frac{\alpha l_{\rm P}^{4}}{\left(R^{2}+\alpha l_{\rm P}^{2}\right)^{2}}\right],
\end{equation}
hence becoming positive if $\rho>\rho_{\text{neck}}$ and negative if \mbox{$\rho<\rho_{\text{neck}}$}. The density $\rho_{\text{neck}}$ marks the boundary between two distinct geometries: For $\rho<\rho_{\text{neck}}$, the geometry is qualitatively similar to the vacuum solution depicted in Fig. \ref{fig:wormhole} (below the surface, $r(l)$ increases as $l$ decreases), whereas for $\rho>\rho_{\text{neck}}$ the geometry is such that, just below the surface, the shape function diminishes with $l$, resembling the first stages of a stellar configuration. Note that, if $\rho<\rho_{\rm{c}}$, a neck will nevertheless appear inside the region filled with matter, endowing the solution with a sub-critical character.

The particular case where $\rho=\rho_{\text{neck}}$ is characterized by having $r_{1}=0$. The next-order coefficient in the expansion of the shape function, evaluated for $\rho=\rho_{\text{neck}}$, yields
\begin{equation}
    r_{2}=-\frac{\left[\left(R^{2}+\alpha l_{\rm P}^{2}\right)^{2}-2R^{2} l_{\rm P}^{2}\right]\left[\left(R^{2}+\alpha l_{\rm P}^{2}\right)^{2}+\alpha l_{\rm P}^{4}\right]}{6 R^{2} l_{\rm P}\left[R^{2}+\left(\alpha-1\right)l_{\rm P}^{2}\right]\left(R^{2}+\alpha l_{\rm P}^{2}\right)^{5/2}},
\end{equation}
which is a negative quantity. Therefore, this solution also has a shape function that decreases just below the surface of the star. It corresponds to a sub-critical configuration since, as long as $R\gg l_{\rm P}$, it is guaranteed that \eqref{Eq:RhoNeck} is smaller than $\rho_{\rm c}$ (we infer this by extrapolating the tendency observed in Fig. \ref{fig:DensityRadius} to stars of large radii).

At-the-neck stars clearly show that the predominance of vacuum effects (in the Polyakov approximation) drives the solution towards the formation of a wormhole neck, an ``opening" of the spacetime geometry, which eventually leads to an asymptotic singularity. The predominance of classical matter, on the other hand, contributes towards ``closing" the geometry and forming a fluid sphere. The interplay between these two effects is what eventually gives rise to \emph{$\epsilon$-strict} stellar spacetimes. These correspond to nearly ``closed" configurations in which vacuum polarization effects end up dominating at the core of the star. In order for the configuration to reach $r=0$ in a regular manner it must be sourced by an RSET that properly accounts for vacuum polarization at the core of compact stars.

\section{Further discussion and conclusions}\label{section:discussion}

We have used a Regularized Polyakov RSET as a toy model to incorporate semiclassical corrections into the equations of stellar equilibrium. We expect this RSET to qualitatively capture the semiclassical effects caused by a geometry that is near horizon formation, that is, describing ultracompact configurations. On the other hand, the regularization of the Polyakov RSET introduces ambiguities that have an impact in the understanding of the solutions of the semiclassical equations of stellar equilibrium.
\begin{figure*}
    \includegraphics[width=0.99\textwidth]{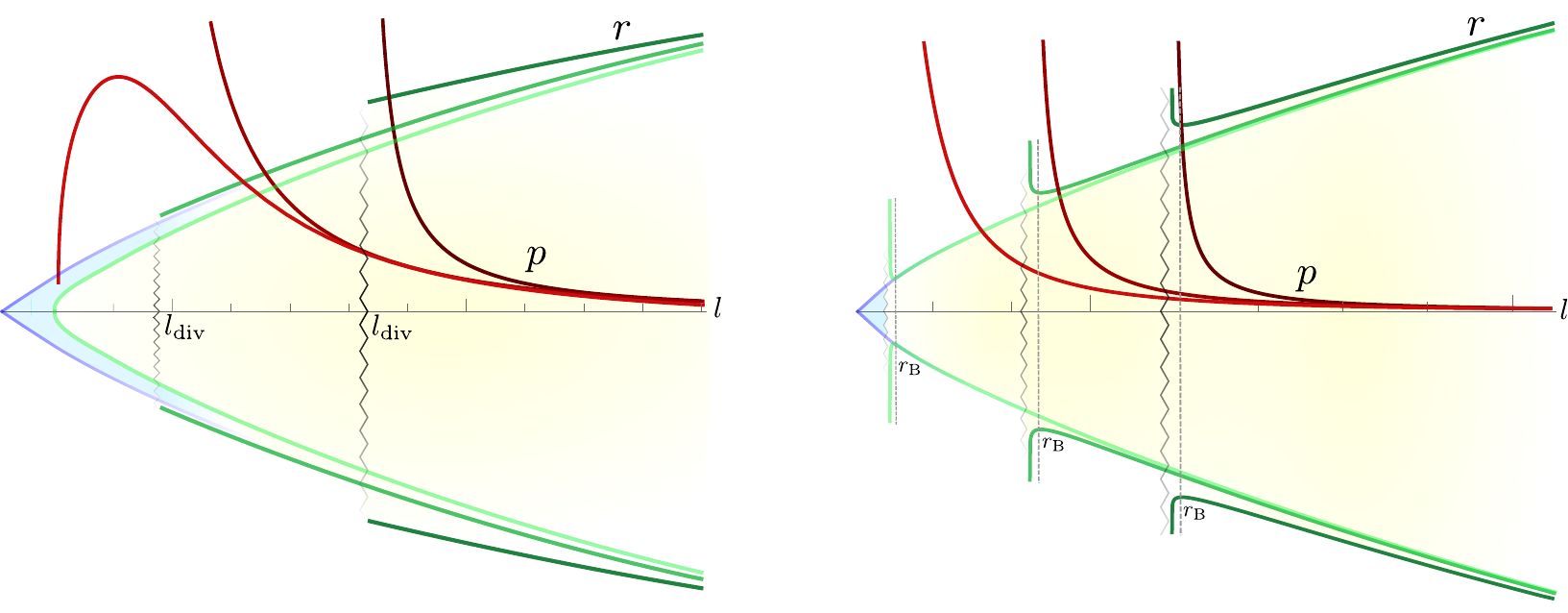}
    \caption{Left panel: series of classical super-Buchdal stars approaching a finite pressure solution. The green lines represent the shape function $r$ and the red curves denote the pressures $p$, which diverge at $r(l_{\text{div}})$ (zigzag lines). Lighter colors correspond to stars whose density is nearing $\rho_{\text{reg-p}}$ (the critical solution $\rho_{\rm{c}\text{-clas}}$ has the darkest colors). Any attempt of regularizing the pressure makes the star highly super-critical, causing $p'$ and $r'$ to diverge at $l=0$. The shape function of a regular star is drawn in blue for comparison.
    Right panel: series of semiclassical sub-critical super-Buchdahl stars approaching the critical solution. Again, the shape functions $r$ are shown in green, while the pressures $p$ are plotted in red. Lighter colors correspond to stars whose density is nearing $\rho_{\rm c}$. The critical solution is approached together with the regularization in the pressure. The vertical dashed lines represent the necks of the solutions and singularities are represented by vertical zigzag lines. We have drawn in blue the shape function of a hypothetical regular star. Note that regularity requires $r'(0)=1+\order{l^{2}}$. The classical super-critical configuration with finite pressure has $r'(0)\to+\infty$, whereas semiclassical configurations with small wormhole necks are \emph{$\epsilon$-strict} stellar spacetimes. Regularization of these profiles amounts to selecting a suitable regularization for the Polyakov RSET.}
    \label{fig:CriticalStars}
\end{figure*}

First of all, we have found that the set of strict stellar spacetimes for the semiclassical field equations analyzed here is almost coincident with the corresponding classical set.
That is, only for sub-Buchdahl stars we find strictly regular semiclassical solutions (with the exception of Planckian size stars for which we can attain much higher compactness). However, this should not be understood as showing that semiclassical gravity exhibits a Buchdahl limit essentially equal to that in classical general relativity.
This can be illustrated by taking a closer look at the super-Buchdahl non-regular solutions that satisfy our definition of \emph{$\epsilon$-strict} spacetime.

When analyzing super-Buchdahl ultracompact configurations of large size, there is a stark difference between the classical and semiclassical cases. In the classical case, for arbitrarily compact configurations it is not possible to define a small value of the radius $r_\epsilon$ so that the compactness remains small enough outside this radius, and at the same time the pressure is finite up to this radius. For ultracompact stars the pressure diverges very close to the surface, and the only way to tame the divergence of the pressure so that it is maintained finite up to $r_\epsilon$ is to become strongly supercritical in the density; this in turn leaves us outside the regime that we have denoted as \emph{$\epsilon$-strict}. In other words, in classical general relativity there are no \emph{$\epsilon$-strict} solutions with a compactness that is appreciably greater than the Buchdahl limit. In fact, the compactness of \emph{$\epsilon$-strict} solutions is bounded by the Buchdahl limit plus small $\mathcal{O}(\epsilon)$ corrections.

On the contrary, super-Buchdahl solutions around the critical solution in the semiclassical theory are just that in all super-critical solutions the pressure is finite at the origin. For the super-critical solution in Fig. \ref{fig:noncritsemi}, the compactness is within the \emph{$\epsilon$-strict} bound for $r_{\epsilon}>1.8 l_{\rm P}$. Going further into the supercritical regime the compactness diverges more strongly making the core to grow, thus making the geometry go outside the notion of \emph{$\epsilon$-strict} spacetime, which requires $\epsilon\ll1$. On the other hand, the sub-critical solutions close to the critical one are such that the compactness turns from negative to positive values, producing a wormhole neck. Hence, there are sub-critical solutions with cores in which the compactness remains very close to zero, while the pressure remains bounded. Therefore, close below and above criticality we have solutions which are \emph{$\epsilon$-strict} configurations, for any value of the compactness.

Figure \ref{fig:CriticalStars} shows a comparison between a critical super-Buchdahl star from the classical theory and a series of semiclassical super-Buchdahl stars that approach the critical solution from the sub-critical regime. Close-to-critical semiclassical solutions are far closer to attaining regularity than classical ones.

The lack of a compactness limit for \emph{$\epsilon$-strict} spacetimes in semiclassical gravity is the main physical result of this paper.  Conceptually, this result illustrates that the Polyakov approximation is successful in regularizing the super-Buchdahl classical stellar profiles in the regions of spacetime in which the approximation is expected to be reliable. It is clear that the only missing physical information to complete the picture is the behavior of the semiclassical source around $r=0$. There are different ways of regularizing the Polyakov approximation around $r=0$, and in this first work we have used the simplest one, also on the basis that it proved adequate for the analysis of vacuum spacetimes. However, our results here show that alternative regularizations must be studied in the presence of matter, as it is reasonable to think that there may exist a regularization in which the ultracompact \emph{$\epsilon$-strict} spacetimes discussed in this paper become regular. The possible existence of such a regularization, and related issues such as its physical interpretation, will be discussed elsewhere.

\section*{Acknowledgements}

The authors thanks Gerardo Garc\'{\i}a-Moreno and Valentin Boyanov for very useful discussions. Financial support was provided by the Spanish Government through the projects FIS2017- 86497-C2-1-P, FIS2017-86497-C2-2-P, PID2020-118159GB-C43/AEI/10.13039/501100011033, PID2020-118159GB-C44/AEI/10.13039/501100011033, PID2019-107847RB-C44/AEI/10.13039/501100011033, and by the Junta de Andalucía through the project FQM219. CB and JA acknowledges financial support from the State Agency for Research of the Spanish MCIU through the “Center of Excellence Severo Ochoa” award to the Instituto de Astrofísica de Andalucía (SEV-2017-0709).

\bibliographystyle{unsrt}
\bibliography{Constantdensity}

\begin{thebibliography}{10}

\bibitem{Wheeler1955}
John~Archibald Wheeler.
\newblock Geons.
\newblock {\em Phys. Rev.}, 97:511--536, Jan 1955.

\bibitem{Buchdahl1959}
H.~A. Buchdahl.
\newblock General relativistic fluid spheres.
\newblock {\em Phys. Rev.}, 116:1027--1034, Nov 1959.

\bibitem{Schwarzschild1916}
Karl Schwarzschild.
\newblock {On the gravitational field of a sphere of incompressible fluid
  according to Einstein's theory}.
\newblock {\em Sitzungsber. Preuss. Akad. Wiss. Berlin (Math. Phys.)},
  1916:424--434, 1916.

\bibitem{Eckart2017}
Andreas Eckart, Andreas H\"uttemann, Claus Kiefer, Silke Britzen, Michal
  Zaja\v{c}ek, Claus L\"ammerzahl, Manfred St\"ockler, Monica Valencia-S,
  Vladimir Karas, and Macarena Garc\'\i{}a-Mar\'\i{}n.
\newblock {The Milky Way\textquoteright{}s Supermassive Black Hole: How Good a
  Case Is It?}
\newblock {\em Found. Phys.}, 47(5):553--624, 2017.

\bibitem{Cardoso2019}
Vitor Cardoso and Paolo Pani.
\newblock {Testing the nature of dark compact objects: a status report}.
\newblock {\em Living Rev. Rel.}, 22(1):4, 2019.

\bibitem{Hawking1976}
S.~W. Hawking.
\newblock Breakdown of predictability in gravitational collapse.
\newblock {\em Phys. Rev. D}, 14:2460--2473, Nov 1976.

\bibitem{Colpi1986}
Monica Colpi, Stuart~L. Shapiro, and Ira Wasserman.
\newblock Boson stars: Gravitational equilibria of self-interacting scalar
  fields.
\newblock {\em Phys. Rev. Lett.}, 57:2485--2488, Nov 1986.

\bibitem{Raposo2018}
Guilherme Raposo, Paolo Pani, Miguel Bezares, Carlos Palenzuela, and Vitor
  Cardoso.
\newblock {Anisotropic stars as ultracompact objects in General Relativity}.
\newblock {\em Phys. Rev.}, D99(10):104072, 2019.

\bibitem{Freedman1978}
Barry Freedman and Larry McLerran.
\newblock Quark star phenomenology.
\newblock {\em Phys. Rev. D}, 17:1109--1122, Feb 1978.

\bibitem{Mazur2015}
Pawel~O. Mazur and Emil Mottola.
\newblock {Surface tension and negative pressure interior of a non-singular
  ‘black hole’}.
\newblock {\em Class. Quant. Grav.}, 32(21):215024, 2015.

\bibitem{Cattoen2005}
Celine Cattoen, Tristan Faber, and Matt Visser.
\newblock Gravastars must have anisotropic pressures.
\newblock {\em Classical and Quantum Gravity}, 22(20):4189–4202, Sep 2005.

\bibitem{Olmo2019}
Gonzalo~J. Olmo, Diego Rubiera-Garcia, and Aneta Wojnar.
\newblock {Stellar structure models in modified theories of gravity: Lessons
  and challenges}.
\newblock {\em Phys. Rept.}, 876:1--75, 2020.

\bibitem{Barcelo2008}
Carlos Barceló, Stefano Liberati, Sebastiano Sonego, and Matt Visser.
\newblock Fate of gravitational collapse in semiclassical gravity.
\newblock {\em Physical Review D}, 77(4), Feb 2008.

\bibitem{Mathur2005}
Samir~D. Mathur.
\newblock {The Fuzzball proposal for black holes: An Elementary review}.
\newblock {\em Fortsch. Phys.}, 53:793--827, 2005.

\bibitem{Carballo-Rubio2018}
Raúl Carballo-Rubio, Francesco Di~Filippo, Stefano Liberati, and Matt Visser.
\newblock {Phenomenological aspects of black holes beyond general relativity}.
\newblock {\em Phys. Rev.}, D98(12):124009, 2018.

\bibitem{Barcelo2009}
Carlos Barceló, Stefano Liberati, Sebastiano Sonego, and Matt Visser.
\newblock Black stars, not holes.
\newblock {\em Scientific American}, 301:38--45, 10 2009.

\bibitem{Visser2009}
Matt Visser, Carlos Barcelo, Stefano Liberati, and Sebastiano Sonego.
\newblock Small, dark, and heavy: But is it a black hole?, 2009.

\bibitem{Carballo-Rubio2018a}
Raúl Carballo-Rubio.
\newblock Stellar equilibrium in semiclassical gravity.
\newblock {\em Physical Review Letters}, 120(6), Feb 2018.

\bibitem{FullingDavies1977}
P.~C.~W. Davies and S.~A. Fulling.
\newblock Quantum vacuum energy in two dimensional space-times.
\newblock {\em Proceedings of the Royal Society of London. Series A,
  Mathematical and Physical Sciences}, 354(1676):59--77, 1977.

\bibitem{BirrellDavies1982}
N.~D. Birrell and P.~C.~W. Davies.
\newblock {\em {Quantum Fields in Curved Space}}.
\newblock Cambridge Monographs on Mathematical Physics. Cambridge Univ. Press,
  Cambridge, UK, 1984.

\bibitem{Brout1995}
R.~Brout, S.~Massar, R.~Parentani, and Ph. Spindel.
\newblock {A Primer for black hole quantum physics}.
\newblock {\em Phys. Rept.}, 260:329--454, 1995.

\bibitem{Hiscock1997}
William~A. Hiscock, Shane~L. Larson, and Paul~R. Anderson.
\newblock Semiclassical effects in black hole interiors.
\newblock {\em Phys. Rev. D}, 56:3571--3581, Sep 1997.

\bibitem{Chakraborty2015}
Sumanta Chakraborty, Suprit Singh, and T.~Padmanabhan.
\newblock {A quantum peek inside the black hole event horizon}.
\newblock {\em JHEP}, 06:192, 2015.

\bibitem{Fischetti1979}
M.~V. Fischetti, J.~B. Hartle, and B.~L. Hu.
\newblock {Quantum Effects in the Early Universe. 1. Influence of Trace
  Anomalies on Homogeneous, Isotropic, Classical Geometries}.
\newblock {\em Phys. Rev. D}, 20:1757--1771, 1979.

\bibitem{Grib1981}
A.~A. Grib, S.~G. Mamaev, and V.~M. Mostepanenko.
\newblock {Self-consistent treatment of vacuum quantum effects in isotropic
  cosmology}.
\newblock In {\em {Second Seminar on Quantum Gravity}}, pages 197--212, 1981.

\bibitem{Simon1992}
Jonathan~Z. Simon.
\newblock No starobinsky inflation from self-consistent semiclassical gravity.
\newblock {\em Phys. Rev. D}, 45:1953--1960, Mar 1992.

\bibitem{Parker1993}
Leonard Parker and Jonathan~Z. Simon.
\newblock Einstein equation with quantum corrections reduced to second order.
\newblock {\em Physical Review D}, 47(4):1339–1355, Feb 1993.

\bibitem{Banerjee2009}
Kinjal Banerjee and Aseem Paranjape.
\newblock {Semiclassical environment of collapsing shells}.
\newblock {\em Phys. Rev. D}, 80:124006, 2009.

\bibitem{Barcelo2019}
Carlos Barcel\'o, Valentin Boyanov, Ra\'ul Carballo-Rubio, and Luis~J. Garay.
\newblock {Semiclassical gravity effects near horizon formation}.
\newblock {\em Class. Quant. Grav.}, 36(16):165004, 2019.

\bibitem{Visser1996}
Matt Visser.
\newblock Gravitational vacuum polarization. ii. energy conditions in the
  boulware vacuum.
\newblock {\em Physical Review D}, 54(8):5116–5122, Oct 1996.

\bibitem{Barcelo2002}
Carlos Barcelo and Matt Visser.
\newblock {Twilight for the energy conditions?}
\newblock {\em Int. J. Mod. Phys. D}, 11:1553--1560, 2002.

\bibitem{Kontou2020}
Eleni-Alexandra Kontou and Ko~Sanders.
\newblock {Energy conditions in general relativity and quantum field theory}.
\newblock {\em Class. Quant. Grav.}, 37(19):193001, 2020.

\bibitem{Barcelo2015}
Carlos Barceló, Raúl Carballo-Rubio, and Luis~J. Garay.
\newblock {Where does the physics of extreme gravitational collapse reside?}
\newblock {\em Universe}, 2(2):7, 2016.

\bibitem{Harada2018}
Tomohiro Harada, Vitor Cardoso, and Daiki Miyata.
\newblock {Particle creation in gravitational collapse to a horizonless compact
  object}.
\newblock {\em Phys. Rev. D}, 99(4):044039, 2019.

\bibitem{Arrechea2020}
Julio Arrechea, Carlos Barcel\'o, Ra\'ul Carballo-Rubio, and Luis~J. Garay.
\newblock Schwarzschild geometry counterpart in semiclassical gravity.
\newblock {\em Phys. Rev. D}, 101:064059, Mar 2020.

\bibitem{Frolov1987}
V.~P. Frolov and A.~I. Zel'nikov.
\newblock Killing approximation for vacuum and thermal stress-energy tensor in
  static space-times.
\newblock {\em Phys. Rev. D}, 35:3031--3044, May 1987.

\bibitem{Anderson1995}
Paul~R. Anderson, William~A. Hiscock, and David~A. Samuel.
\newblock Stress-energy tensor of quantized scalar fields in static spherically
  symmetric spacetimes.
\newblock {\em Phys. Rev. D}, 51:4337--4358, Apr 1995.

\bibitem{Groves2002}
Peter~B. Groves, Paul~R. Anderson, and Eric~D. Carlson.
\newblock Method to compute the stress-energy tensor for the massless spin
  $\frac{1}{2}$ field in a general static spherically symmetric spacetime.
\newblock {\em Phys. Rev. D}, 66:124017, Dec 2002.

\bibitem{Hochberg1997}
David Hochberg, Arkadiy Popov, and Sergey~V. Sushkov.
\newblock Self-consistent wormhole solutions of semiclassical gravity.
\newblock {\em Physical Review Letters}, 78(11):2050–2053, Mar 1997.

\bibitem{Popov2003}
Arkady~A. Popov.
\newblock {Analytical approximation of the stress energy tensor of a quantized
  scalar field in static spherically symmetric space-times}.
\newblock {\em Phys. Rev. D}, 67:044021, 2003.

\bibitem{Simon1990}
Jonathan~Z. Simon.
\newblock {The Stability of flat space, semiclassical gravity, and higher
  derivatives}.
\newblock {\em Phys. Rev. D}, 43:3308--3316, 1991.

\bibitem{Polyakov1981}
Alexander~M. Polyakov.
\newblock {Quantum Geometry of Bosonic Strings}.
\newblock {\em Phys. Lett.}, B103:207--210, 1981.

\bibitem{Parentani1994}
Renaud Parentani and Tsvi Piran.
\newblock {The Internal geometry of an evaporating black hole}.
\newblock {\em Phys. Rev. Lett.}, 73:2805--2808, 1994.

\bibitem{Ayal1997}
Shai Ayal and Tsvi Piran.
\newblock {Spherical collapse of a massless scalar field with semiclassical
  corrections}.
\newblock {\em Phys. Rev. D}, 56:4768--4774, 1997.

\bibitem{Fabbri2005}
A.~Fabbri and J.~Navarro-Salas.
\newblock {\em Modeling Black Hole Evaporation}.
\newblock Imperial College Press, 2005.

\bibitem{Ho2017}
Pei-Ming Ho and Yoshinori Matsuo.
\newblock {Static Black Holes With Back Reaction From Vacuum Energy}.
\newblock {\em Class. Quant. Grav.}, 35(6):065012, 2018.

\bibitem{Fabbri2006}
A.~Fabbri, S.~Farese, J.~Navarro-Salas, G.~J. Olmo, and H.~Sanchis-Alepuz.
\newblock Semiclassical zero-temperature corrections to schwarzschild spacetime
  and holography.
\newblock {\em Physical Review D}, 73(10), May 2006.

\bibitem{Berthiere2017}
Clément Berthiere, Debajyoti Sarkar, and Sergey~N. Solodukhin.
\newblock {The fate of black hole horizons in semiclassical gravity}.
\newblock {\em Phys. Lett.}, B786:21--27, 2018.

\bibitem{Ho2018}
Pei-Ming Ho and Yoshinori Matsuo.
\newblock Static black hole and vacuum energy: thin shell and incompressible
  fluid.
\newblock {\em Journal of High Energy Physics}, 2018(3), Mar 2018.

\bibitem{Ensayo2021}
Julio Arrechea, Carlos Barceló, Valentin Boyanov, and Luis~J. Garay.
\newblock Vacuum semiclassical gravity does not leave space for safe
  singularities.
\newblock {\em Universe}, 7(8), 2021.

\bibitem{Arrechea2021}
Julio Arrechea, Carlos Barcel{\'{o}}, Ra{\'{u}}l Carballo-Rubio, and Luis~J
  Garay.
\newblock Reissner{\textendash}nordström geometry counterpart in semiclassical
  gravity.
\newblock {\em Classical and Quantum Gravity}, 38(11):115014, may 2021.

\bibitem{Balbinot2007}
R.~Balbinot, A.~Fabbri, S.~Farese, and R.~Parentani.
\newblock {Hawking radiation from extremal and non-extremal black holes}.
\newblock {\em Phys. Rev. D}, 76:124010, 2007.

\bibitem{Hiscock1988}
William~A. Hiscock.
\newblock Gravitational vacuum polarization around static spherical stars.
\newblock {\em Phys. Rev. D}, 37:2142--2150, Apr 1988.

\bibitem{Page1982}
Don~N. Page.
\newblock Thermal stress tensors in static einstein spaces.
\newblock {\em Phys. Rev. D}, 25:1499--1509, Mar 1982.

\bibitem{Brown1985}
M.~R. Brown and A.~C. Ottewill.
\newblock Effective actions and conformal transformations.
\newblock {\em Phys. Rev. D}, 31:2514--2520, May 1985.

\bibitem{Satz2004}
Alejandro Satz, Francisco~D. Mazzitelli, and Ezequiel Alvarez.
\newblock {Vacuum polarization around stars: Nonlocal approximation}.
\newblock {\em Phys. Rev.}, D71:064001, 2005.

\bibitem{Jensen1989}
B.~P. Jensen and Adrian Ottewill.
\newblock Renormalized electromagnetic stress tensor in schwarzschild
  spacetime.
\newblock {\em Phys. Rev. D}, 39:1130--1138, Feb 1989.

\bibitem{Jensen1991}
B.~P. Jensen, J.~G. McLaughlin, and A.~C. Ottewill.
\newblock Renormalized electromagnetic stress tensor for an evaporating black
  hole.
\newblock {\em Phys. Rev. D}, 43:4142--4144, Jun 1991.

\bibitem{Carlson2003}
Eric~D. Carlson, William~H. Hirsch, Benedikt Obermayer, Paul~R. Anderson, and
  Peter~B. Groves.
\newblock Stress-energy tensor for a massless spin $1/2$ field in static black
  hole spacetimes.
\newblock {\em Phys. Rev. Lett.}, 91:051301, Aug 2003.

\bibitem{Campanelli1996}
Manuela Campanelli and Carlos~O. Lousto.
\newblock {Semiclassical models for uniform density cosmic strings and
  relativistic stars}.
\newblock {\em Int. J. Mod. Phys. D}, 6:771--784, 1997.

\bibitem{Calmet2019}
Xavier Calmet, Roberto Casadio, and Folkert Kuipers.
\newblock {Quantum Gravitational Corrections to a Star Metric and the Black
  Hole Limit}.
\newblock {\em Phys. Rev. D}, 100(8):086010, 2019.

\bibitem{Volkmer2019}
Guilherme~L. Volkmer, Dimiter Hadjimichef, Moises Razeira, Benno Bodmann, and
  C\'esar A.~Zen Vasconcellos.
\newblock {Ultra-compact objects in semiclassical gravity}.
\newblock {\em Astron. Nachr.}, 340(9-10):914--919, 2019.

\bibitem{Prasetyo2021}
I.~Prasetyo, H.~S. Ramadhan, and A.~Sulaksono.
\newblock Ultra-compact objects from semi-classical gravity.
\newblock {\em Phys. Rev. D}, 103:123536, Jun 2021.

\bibitem{Lemaitre1997}
Abb{\'e}~Georges Lema{\^i}tre.
\newblock The expanding universe.
\newblock {\em General Relativity and Gravitation}, 29(5):641--680, May 1997.

\bibitem{Krasinski1997}
A.~Krasinski.
\newblock {Editor's Note: The Expanding Universe, by the Abbé Georges
  Lemaître}.
\newblock {\em Gen. Rel. and Grav.}, 29:637--640, 1997.

\bibitem{Misner1964}
Charles~W. Misner and David~H. Sharp.
\newblock {Relativistic equations for adiabatic, spherically symmetric
  gravitational collapse}.
\newblock {\em Phys. Rev.}, 136:B571--B576, 1964.

\bibitem{Hernandez1966}
Walter~C. Hernandez and Charles~W. Misner.
\newblock {Observer Time as a Coordinate in Relativistic Spherical
  Hydrodynamics}.
\newblock {\em Astrophys. J.}, 143:452, 1966.

\bibitem{Hayward1994}
Sean~A. Hayward.
\newblock {Gravitational energy in spherical symmetry}.
\newblock {\em Phys. Rev.}, D53:1938--1949, 1996.

\bibitem{Misner1974}
Charles~W. Misner, K.~S. Thorne, and J.~A. Wheeler.
\newblock {\em {Gravitation}}.
\newblock W. H. Freeman, San Francisco, 1973.

\bibitem{Urbano2018}
Alfredo Urbano and Hardi Veerm\"ae.
\newblock {On gravitational echoes from ultracompact exotic stars}.
\newblock {\em JCAP}, 04:011, 2019.

\bibitem{Tolman1939}
Richard~C. Tolman.
\newblock Static solutions of einstein's field equations for spheres of fluid.
\newblock {\em Phys. Rev.}, 55:364--373, Feb 1939.

\bibitem{Volkoff1939}
G.~M. Volkoff.
\newblock On the equilibrium of massive spheres.
\newblock {\em Phys. Rev.}, 55:413--413, Feb 1939.

\bibitem{Wyman1949}
Max Wyman.
\newblock Radially symmetric distributions of matter.
\newblock {\em Phys. Rev.}, 75:1930--1936, Jun 1949.

\bibitem{Oppenheimer1939}
J.~R. Oppenheimer and G.~M. Volkoff.
\newblock On massive neutron cores.
\newblock {\em Phys. Rev.}, 55:374--381, Feb 1939.

\bibitem{Mazur2004}
Pawel~O. Mazur and Emil Mottola.
\newblock {Gravitational vacuum condensate stars}.
\newblock {\em Proc. Nat. Acad. Sci.}, 101:9545--9550, 2004.

\bibitem{BarceloVolovik2004}
Carlos Barcelo and Grigory Volovik.
\newblock {A Stable Einstein universe}.
\newblock {\em JETP Lett.}, 80:209--213, 2004.

\bibitem{Tolman1987}
R.C. Tolman.
\newblock {\em Relativity, Thermodynamics, and Cosmology}.
\newblock Dover Books on Physics. Dover Publications, 1987.

\bibitem{Smoller1997}
J.~Smoller and B.~Temple.
\newblock {Solutions of the Oppenheimer--Volkoff Equations Inside $9/8^{ths}$
  of the Schwarzschild Radius}.
\newblock 3 1997.

\bibitem{Woszczyna2015}
Andrzej Woszczyna, Marek Kutschera, Sebastian Kubis, Wojciech Czaja, Piotr
  Plaszczyk, and Zdzis\l{}aw~A. Golda.
\newblock {Nakedly singular non-vacuum gravitating equilibrium states}.
\newblock {\em Gen. Rel. Grav.}, 48(1):5, 2016.

\bibitem{Bratek2019}
\L{}ukasz Bratek, Joanna Ja\l{}ocha, and Andrzej Woszczyna.
\newblock {Nakedly singular counterpart of Schwarzschild's incompressible star.
  A barotropic continuity condition in the center}.
\newblock {\em Gen. Rel. Grav.}, 51(11):142, 2019.

\bibitem{Anastopoulos2020}
Charis Anastopoulos and Ntina Savvidou.
\newblock {Classification theorem and properties of singular solutions to the
  Tolman-Oppenheimer-Volkoff equation}.
\newblock 10 2020.

\bibitem{Raychaudhuri1951}
Amal~Kumar Raychaudhuri.
\newblock Volkoff's massive spheres.
\newblock {\em Phys. Rev.}, 84:166--166, Oct 1951.

\end{thebibliography}

\end{document}